\newif\ifsingle

\ifsingle
\documentclass[12pt,draftclsnofoot, onecolumn]{IEEEtran}		
\else		
\documentclass[10pt,final, twocolumn]{IEEEtran}
\usepackage[a4paper, total={7.3in, 10.65in}]{geometry}
\fi

\usepackage{times}
\usepackage{amsmath,dsfont}
\usepackage{amssymb,amsthm}
\usepackage{epsfig,verbatim}
\usepackage{subfigure}
\usepackage{setspace}
\usepackage{color}
\usepackage{cite}
\usepackage{epstopdf}
\usepackage{enumitem}
\usepackage{graphics}
\usepackage{accents}
\usepackage{acronym}
\usepackage[bookmarks,colorlinks]{hyperref}
\usepackage{booktabs}
\usepackage{mathtools}
\usepackage{bbm}

\newtheorem{definition}{Definition}
\newtheorem{theorem}{Theorem}
\newtheorem{corollary}{Corollary}
\newtheorem{proposition}{Proposition}
\newtheorem{lemma}{Lemma}

\newtheorem{example}{Example}

\definecolor{NewColor}{rgb}{0,0,0}

\newcommand{\myVec}[1]{{\boldsymbol{#1}}}
\newcommand{\myMat}[1]{{\boldsymbol{#1}}}
\newcommand{\mySet}[1]{\mathcal{#1}}

\newcommand{\myDetVec}[1]{\myVec{\lowercase{#1}}}
\newcommand{\myRandVec}[1]{\myVec{\lowercase{#1}}}
\newcommand{\myDetMat}[1]{\myMat{\uppercase{#1}}}
\newcommand{\myRandMat}[1]{\myMat{\uppercase{#1}}}

\newcommand{\E}{\mathbb{E}}		 				
\newcommand{\myW}{{\myRandVec{W}}}			 		
\newcommand{\myY}{{\myRandVec{Y}}}			 		
\newcommand{\myX}{{\myRandVec{X}}}			 		
\newcommand{\myI}{{\myDetMat{i}}}			 		
\newcommand{\myA}{{\myDetMat{a}}}
\newcommand{\myB}{{\myDetMat{b}}}
\newcommand{\myAT}{\tilde{\myA}}
\newcommand{\myBT}{\tilde{\myB}}			 					 		
\newcommand{\myAB}{\bar{\myA}}
\newcommand{\myYmat}{{\myRandMat{Y}}}			 	
\newcommand{\myYvec}{\underline{\myY}}			 	
\newcommand{\mySmat}{{\myMat{\Theta}}}			 	
\newcommand{\myWmat}{{\myRandMat{W}}}			 	
\newcommand{\myWvec}{\underline{\myW}}			 	
\newcommand{\Gmat}{{\myRandMat{G}}}			 		
\newcommand{\Gvec}{\underline{\myVec{g}}}			 		
\newcommand{\Hmat}{{\myRandMat{H}}}			 		

\newcommand{\Dmat}{{\myDetMat{d}}}			 		
\newcommand{\Bmat}{{\myDetMat{f}}}			 		
\newcommand{\Phimat}{{\myMat{\Phi}}}			 		
\newcommand{\myTheta}{\theta}
\newcommand{\SigW}{\sigma_W^2}						
\newcommand{\Ncells}{n_c}							
\newcommand{\Nantennas}{\lenAsym}						
\newcommand{\Nusers}{\lenStag}							
\newcommand{\NcellsSet}{\mySet{N}_c}				
\newcommand{\NusersSet}{\mySet{K}}				
\newcommand{\dcoeff}{d}								
\newcommand{\bcoeff}{f}								
\newcommand{\phicoeff}{\phi}								
\newcommand{\Tpilots}{\lenXtag}						
\newcommand{\TpilotsSet}{\mySet{L}}						

\newcommand{\Pdf}[1]{f_{ #1}}
\newcommand{\Psd}[1]{s_{#1}}
\newcommand{\Acorr}[1]{c_{#1}}
\newcommand{\PSD}[1]{\myMat{S}_{#1}}

\newcommand{\CorrMat}[1][ ]{\myMat{C}_{#1}}
\newcommand{\CovMat}[1]{\myMat{\Sigma}_{#1}}			
\newcommand{\maxDiag}{\sigma^2_{l}}

\newcommand{\Rate}{R}
\newcommand{\Ratio}{r}
\newcommand{\AsymDist}{\mu}
\newcommand{\lenX}{\lenAsym\lenXtag}			 			
\newcommand{\lenS}{\lenAsym\lenStag}			 			
\newcommand{\lenAsym}{N}			 			
\newcommand{\lenZ}{P}			 			
\newcommand{\lenZT}{\tilde{\lenZ}}			 			
\newcommand{\lenZn}{\lenZ_q}
\newcommand{\lenZq}{\lenZ_r}
\newcommand{\Quan}[2]{Q_{{#1}}^{{#2}}}
\newcommand{\LmmseMat}{\myMat{\Gamma}}
\newcommand{\LmmseMatT}{\tilde{\LmmseMat}}

\newcommand{\DynRange}{\gamma}
\newcommand{\DynInt}[1][ ]{\Delta_{#1}}
\newcommand{\TilM}[1][ ]{\tilde{M}_{#1}}
\newcommand{\MyKappa}[1][]{\kappa_{#1}}

\newcommand{\Wlevel}{\zeta}
\newcommand{\myEta}{\eta}
\newcommand{\DistG}{D_{\rm G}}
\newcommand{\MMSE}{^{\rm MMSE}}
\newcommand{\Opt}{^{\rm Opt}}
\newcommand{\op}{^{\rm o}}
\newcommand{\Ign}{^{\rm Ign}}
\newcommand{\ADC}{^{\rm HL}}
\newcommand{\sADC}{^{\rm sHL}}
\newcommand{\myObs}{\underline{\myObstag}}
\newcommand{\mySOI}{\underline{\mySOItag}}
\newcommand{\mySOIEst}{\underline{\mySOIEsttag}}
\newcommand{\myQ}{\myVec{q}}
\newcommand{\lenXtag}{L}
\newcommand{\lenStag}{K}
\newcommand{\myObstag}{\myVec{y}}
\newcommand{\mySOItag}{\myVec{g}}
\newcommand{\mySOIEsttag}{\tilde{\mySOItag}}
\newcommand{\LmmseMattag}{{\LmmseMat}}
\newcommand{\eig}[1]{\lambda_{#1}}			
\newcommand{\eigT}[1]{\eig{#1}}
\newcommand{\CovYtag}{\CovMat{\myY_l}}
\newcommand{\myAtag}{\myA\op}
\newcommand{\myBtag}{\myB\op}
\newcommand{\Glevel}{\varphi}

\newcommand{\Plevel}{\Glevel}

\acrodef{bs}[BS]{base station}
\acrodef{mimo}[MIMO]{multiple-input multiple-output}
\acrodef{mac}[MAC]{multiple access channel}
\acrodef{tdd}[TDD]{time-division duplex}
\acrodef{ut}[UT]{user terminal}
\acrodef{cdf}[CDF]{cumulative distribution function}
\acrodef{pdf}[PDF]{probability density function}
\acrodef{ps}[PS]{pilot sequence}
\acrodef{se}[SE]{spectral efficiency}
\acrodef{mse}[MSE]{mean-squared error}
\acrodef{adc}[ADC]{analog-to-digital convertor}
\acrodef{dtft}[DTFT]{discrete-time Fourier transform}
\acrodef{dft}[DFT]{discrete Fourier transform}
\acrodef{nb}[NB]{narrowband}
\acrodef{dt}[DT]{discrete-time}
\acrodef{ct}[CT]{continuous-time}
\acrodef{evd}[EVD]{eigenvalue decomposition}
\acrodef{svd}[SVD]{singular valued decomposition}
\acrodef{soi}[SOI]{signal of interest}
\acrodef{awgn}[AWGN]{additive white Gaussian noise}
\acrodef{wss}[WSS]{wide-sense stationary}
\acrodef{mmse}[MMSE]{minimum \ac{mse}}
\acrodef{mi}[MI]{mutual information}
\acrodef{lmmse}[LMMSE]{linear MMSE}
\acrodef{map}[MAP]{maximum a-posteriori probability}
\acrodef{mi}[MI]{mutual information}
\acrodef{isi}[ISI]{intersymbol interference}
\acrodef{snr}[SNR]{signal-to-noise ratio}
\acrodef{pc}[PC]{proper-complex}
\acrodef{psd}[PSD]{power spectral density}
\acrodef{ptp}[PtP]{point-to-point}
\acrodef{sinr}[SINR]{signal-to-interference-and-noise ratio}
\acrodef{pdf}[PDF]{probability density function}
\acrodef{rv}[RV]{random variable}
\acrodef{csi}[CSI]{channel state information}

\ifsingle
\newcommand{\includefig}[1]{\includegraphics[width = 0.75\columnwidth]{#1} 	\vspace{-0.4cm}}
\else
\newcommand{\includefig}[1]{\includegraphics[width = \columnwidth]{#1} 	\vspace{-0.6cm}}
\fi 

\long\def\symbolfootnote[#1]#2{\begingroup\def\thefootnote{\fnsymbol{footnote}}\footnote[#1]{#2}\endgroup}

 \IEEEoverridecommandlockouts
\title{Asymptotic Task-Based Quantization with Application to Massive MIMO
}

\vspace{-0.75cm}

\author{
	\IEEEauthorblockN{ Nir Shlezinger, Yonina C. Eldar, and Miguel R. D. Rodrigues\\
	}		
	\thanks{Parts of this work were presented in the 2019 IEEE International Conference on Acoustics, Speech, and Signal Processing (ICASSP), Brighton, UK.}
	\thanks{This project has received funding from the European Union’s Horizon 2020 research and innovation program under grant No. 646804-ERC-COG-BNYQ, from the Israel Science Foundation under grant No. 0100101, and from the Royal Society International Exchange scheme IE 160348.
	}
	\thanks{N. Shlezinger  and Y. C. Eldar are with the faculty of Math and CS, Weizmann Institute of Science, Rehovot, Israel (e-mail: nirshlezinger1@gmail.com; yonina@weizmann.ac.il). 	
	}
	\thanks{M. R. D. Rodrigues  is with the department of EE,  University College, London, UK (e-mail: m.rodrigues@ucl.ac.uk).  	
	}
	
	\vspace{-1.0cm}
	
}

\vspace{-0.5cm}

\begin{document}

\maketitle
\pagestyle{plain}
\thispagestyle{plain}
\begin{abstract}
Quantizers take part in nearly every digital signal processing system which operates on physical signals. They are commonly designed to accurately represent the underlying signal, regardless of the specific task to be performed on the quantized data. In systems working with high-dimensional signals, such as massive multiple-input multiple-output (MIMO) systems, it is beneficial to utilize low-resolution quantizers, due to cost, power, and memory constraints. In this work we study quantization of high-dimensional inputs, aiming at improving performance under resolution constraints by accounting for the system task in the quantizers design. We focus on the task of recovering a desired signal statistically related to the high-dimensional input, and analyze two quantization approaches: 
We first consider vector quantization,  which is typically computationally infeasible, and characterize the optimal performance achievable with this approach. Next, we focus on practical systems which utilize hardware-limited scalar uniform analog-to-digital converters (ADCs), and design a task-based quantizer under this model. 
The resulting system accounts for the task by linearly combining the observed signal into a lower dimension prior to quantization. 
We then apply our proposed technique to channel estimation in massive MIMO networks. Our results demonstrate that a system utilizing low-resolution scalar ADCs can approach the optimal channel estimation performance by properly accounting for the task in the system design.  
\end{abstract}

\vspace{-0.4cm}
\section{Introduction}
\label{sec:Intro}
\vspace{-0.15cm}
Digital signal processing and communications systems use quantized representations of continuous-amplitude physical quantities \cite{Eldar:15}. These digital representations are typically designed to accurately match the original analog signal, by minimizing some distortion measure between the analog signal and the digital representation \cite{Gray:98}, regardless of the task of the system. 
Nonetheless, in many cases, the system task is not to recover the analog signal, but to extract some other information from its quantized representation \cite{Rodrigues:17}. 
It is therefore possible that in such systems -- which we refer to as {\em task-based quantizers} -- one can obtain further performance improvements in terms of the quantization rate necessary to achieve a certain performance.

Practical quantizers typically utilize scalar uniform \acp{adc} \cite{Eldar:15}.
Recent years have witnessed a growing interest in systems operating with quantized large-scale vectors obtained using low-resolution scalar \acp{adc}. One of the main applications considered is massive \ac{mimo} communications \cite{Mo:18,Li:17,Choi:16,Choi:17,Choi:18,Mo:17,Zhang:16,Jacobsson:17,Roth:18,Pirzadeh:18,Ordonez:16,Studer:16,Mollen:17,Zhang:18,Ding:18}, which is a key technology for the realization of next generation wireless networks \cite{Andrews:14}. In such systems, a wireless \ac{bs} is equipped with a large number of antennas \cite{Marzetta:10,Hoydis:13, Shlezinger:17}. 
	The \ac{bs} first quantizes the received signal using a set of \acp{adc}, commonly implementing scalar uniform quantization. Then, the quantized representation is used to estimate the underlying channel \cite{Mo:18, Li:17, Choi:16, Jacobsson:17, Pirzadeh:18,Mollen:17,Studer:16} and/or recover the transmitted messages \cite{Choi:18, Choi:17, Mo:17, Li:17, Choi:16, Zhang:16, Roth:18, Jacobsson:17, Pirzadeh:18,Mollen:17,Studer:16, Zhang:18,Ordonez:16}. For large-scale inputs, i.e., large number of \ac{bs} antennas, accurate quantizers become costly in terms of power and memory usage, particularly when utilizing a large bandwidth, making low-resolution  quantization essential for realizing massive \ac{mimo} systems \cite{Andrews:14}. As the task in massive \ac{mimo} is not to recover the input signal, but to estimate the  channel or decode the transmitted message, reasonable performance with low-resolution scalar quantizers has been observed \cite{Mo:18, Li:17, Choi:16, Choi:17, Choi:18, Mo:17, Zhang:16, Jacobsson:17, Roth:18, Pirzadeh:18,Mollen:17,Studer:16, Zhang:18,Ordonez:16,Ding:18}. 
	However, most prior works  assume that the quantizers are fixed, commonly assuming one-bit sign quantization \cite{Zhang:18, Mollen:17, Choi:16,Li:17}. Thus, they do not characterize the achievable performance  when the quantizers are designed to account for the system task.

In the presence of multivariate inputs, joint (vector) quantization is known to outperform scalar quantization \cite[Ch. 10]{Cover:06}. Task-based vector quantization can be considered as an indirect lossy-source coding setup \cite{Gray:98}. In such scenarios, one wishes to recover a desired source based on a discrete representation of its noisy version, in the sense of minimizing a given distortion measure \cite{Witsenhausen:80}. For the \ac{mse} distortion, it was shown in \cite{Wolf:70} that the optimal system which achieves the rate-distortion curve, namely, uses the minimal number of bits per input sample required to achieve a fixed distortion, applies vector quantization to the \ac{mmse} estimate of the desired source. This observation was used in \cite{Kipnis:18, Kipnis:18a} to study sampling and vector quantization of continuous-time signals. Nonetheless, in the presence of high-dimensional inputs, vector quantization becomes infeasible, so that practical task-based quantization approaches are required. 

  Task-based quantization with scalar uniform \acp{adc}, referred to as {\em hardware-limited task-based quantization},
 can be realized by allowing analog linear processing prior to quantization \cite{Shlezinger:18}. \ac{mimo} communications systems utilizing both analog and digital processing are known as {\em hybrid architectures}  \cite{Mo:17, Rial:16,Kim:15}, and are the focus of a large amount of recent works. 
 In particular, \cite{Cuba:17} compared the achievable-rate versus power efficiency tradeoff for various analog combining systems, 
 \cite{Mo:17} and \cite{Stein:17} designed hybrid architectures aimed at maximizing the achievable rate and signal recovery \ac{mse}, respectively, with full \ac{csi}, 
 while  \cite{Choi:17} studied  bit allocation  for minimizing the quantization error when the analog combining is set to the largest channel eigenmodes, using high rate quantization analysis. 
 Additionally, \cite{Roth:18} studied the achievable rate with imperfect \ac{csi} when distinct sets of inputs are each combined in the analog domain to maximize the receive power,  
 while \cite{Rini:17} characterized bounds on the capacity of \ac{mimo} communications with analog combining and one-bit quantizers. 
Most previous works which designed hybrid \ac{mimo} receivers, e.g., \cite{Mo:17,Choi:17,Stein:17}, considered finite-size inputs and required \ac{csi} in their design, and thus cannot be utilized for massive \ac{mimo} channel estimation. 
 Specifically, the joint design of analog combining, quantization rule, and digital processing, to optimize the accuracy of massive \ac{mimo} channel estimation with scalar  \acp{adc}  has not yet been studied, to the best of our knowledge. 
 
In this work we study task-based quantization for channel estimation in massive \ac{mimo} systems operating with scalar \acp{adc}. 
Our analysis is based on an extension of the hardware-limited task-based quantization framework proposed in our previous work  \cite{Shlezinger:18}, which studied parameter estimation from a finite-sized quantized observed signal. The work \cite{Shlezinger:18} proposed to jointly optimize the analog combining, quantization rule, and digital processing, to minimize the \ac{mse} in recovering the desired finite-sized vector. Here, we extend the study of \cite{Shlezinger:18} 
to account for asymptotically large data, developing a framework for task-based quantization with high-dimensional inputs, and then apply the resulting analysis to massive \ac{mimo} systems, which are commonly studied in the asymptotic number of antennas regime \cite{Marzetta:10, Hoydis:13}.  
In particular, we focus on massive \ac{mimo} channel estimation, carried out in a \ac{tdd} manner \cite{Marzetta:10, Hoydis:13, Shlezinger:17}.  
Unlike previous works on hybrid architectures optimization with low-resolution quantization, e.g., \cite{Mo:17,Choi:17,Stein:17}, our work does not require knowledge of the channel. In fact, in the presence of adjustable analog combining hardware, such as dynamic metasurface antennas \cite{Shlezinger:DMA}, our analysis can be combined with previously proposed  hybrid systems by reconfiguring the analog combining hardware once the channel is accurately estimated. 
We also note that our  analysis can be applied to different tasks, such as signal recovery and noise mitigation.

We begin by studying task-based vector quantization using indirect lossy source coding theory. We characterize the minimal achievable average \ac{mse} for any quantization system operating with a fixed quantization rate, namely, a fixed number of bits per input sample. Then, we study the performance when vector quantization is carried out independently from the task, referred to as {\em task-ignorant vector quantization}. 
Since the input dimensionality here is asymptotically large, we are able to explicitly obtain the achievable performance, unlike \cite{Shlezinger:18}, using indirect rate-distortion theory. 
Studying vector quantizers allows us to quantify the performance bounds of task-based quantization with large-scale inputs, and in particular, understand the fundamental limits of massive \ac{mimo} channel estimation. 

Next, we study task-based quantization with scalar uniform \acp{adc},  allowing analog combining prior to quantization.  While analog combining can contribute in aspects other than improving the performance with  finite-resolution quantizers, e.g., reducing the number of costly RF chains in massive \ac{mimo} systems \cite{Stein:17}, we focus here on the achievable performance for a given quantization rate. 
For this setup we propose a task-based quantization system with linear analog and digital processing which minimizes the average \ac{mse} under such hardware-limited structure constraints. 
We show that, unlike in the fixed size regime studied in \cite{Shlezinger:18}, for large-scale inputs an important parameter which greatly affects the system performance is the {\em analog combining ratio}, which determines how the number of scalar quantizers grows as the input size tends to infinity.

 Then, we focus on massive \ac{mimo} systems, and show how the proposed task-based quantization system can be applied to channel estimation from quantized measurements. 
 We note that in this scenario the inputs are gathered over different antennas as well as over different time instances. Since in some cases, it may be desirable to combine only samples received at the same time instance, to avoid introducing delays in the analog domain, we also derive the system which minimizes the average \ac{mse} subject to the constraint that only inputs taken at the same time instance can be combined. This constraint reduces the complexity of the resulting system at the cost of degraded \ac{mse} performance. 
In our numerical study, we illustrate the fundamental performance limits of massive \ac{mimo} channel estimation achievable using vector quantizers, and compare these limits to our proposed task-based quantization systems with scalar \acp{adc}, and to massive \ac{mimo} channel estimators which operate only in the digital domain. Our results demonstrate that the proposed quantizers, which utilize practical low-resolution scalar \acp{adc}, are capable of approaching the optimal performance, achievable using vector quantizers, and outperform previously proposed estimators.

The rest of this paper is organized as follows: 
Section~\ref{sec:Preliminaries} reviews some basics in quantization theory.
Section~\ref{sec:MSE} extends the results of \cite{Shlezinger:18} to large-scale data, and Section \ref{sec:MIMO} applies them to massive \ac{mimo} channel estimation.
Section~\ref{sec:Simulations} provides simulation examples.
Finally,  Section~\ref{sec:Conclusions}  concludes the paper.

Throughout the paper, we use boldface lower-case letters for vectors, e.g., ${\myDetVec{x}}$;
the $i$th element of ${\myDetVec{x}}$ is written as $(\myDetVec{x})_i$. 
Matrices are denoted with boldface upper-case letters,  e.g., 
$\myDetMat{m}$, and we use $(\myDetMat{m})_{i,j}$ to denote its $(i,j)$th element. We use $\myI_{n}$ to denote the $n \times n$ identity matrix.  
Sets are expressed with calligraphic letters, e.g., $\mySet{X}$, and $\mySet{X}^n$ is the $n$th order Cartesian power of $\mySet{X}$. 
Hermitian transpose, transpose, complex conjugate, stochastic expectation,  and mutual information are written as $(\cdot)^H$, $(\cdot)^T$, $(\cdot)^*$,  $\E\{ \cdot \}$,  and $I\left( \cdot ~ ; \cdot \right)$, respectively.  
For a real number $a$, we use $a^+ \triangleq \max(a,0)$;
$\left\langle \cdot \right \rangle$ denotes the integer divisor (plus one) of the value in the brackets (minus one), namely, $\left\langle n \right \rangle_m \triangleq \lfloor\frac{n-1}{m} \rfloor + 1$. 
We use ${\rm {Tr}}\left(\cdot\right)$ to denote the trace operator, 
$\delta_{(\cdot)}$ is the indicator function, $\otimes$ is the Kronecker product, $\mySet{R}$  and $\mySet{C}$ are the sets of real and complex numbers, respectively. 
All logarithms are taken to base-2. 
Finally, for an $n \! \times \! n$ matrix $\myMat{X}$,  $\myVec{x}\! =\! {\rm vec}\left(\myMat{X}\right)$ is the $n^2 \! \times \! 1$  vector obtained by stacking the columns of $\myMat{X}$. 

\vspace{-0.2cm}
\section{Preliminaries in Quantization Theory}
\label{sec:Preliminaries}
\vspace{-0.1cm}
To formulate the task-based quantization setup. we first briefly review standard quantization notions. 
While parts of this review also appear in our previous work \cite{Shlezinger:18}, it is included for completeness.
We begin with the definition of a quantizer:
\begin{definition}[Quantizer]
	\label{def:Quantizer}
	A quantizer $\Quan{M}{\lenAsym,\lenStag}\left(\cdot \right)$ with $\log M$ bits, input size $\lenAsym$, input alphabet $\mySet{X}$, output size $\lenStag$, and output alphabet $\hat{\mySet{X}}$, consists of: 
	{\em 1)} An  encoding function $g_\lenAsym^{\rm e}: \mySet{X}^\lenAsym \mapsto \{1,2,\ldots,M\} \triangleq \mySet{M}$ which maps the input from $\mySet{X}^\lenAsym$ into a discrete index $i \in \mySet{M}$.
	{\em 2)} A decoding function  $g_\lenStag^{\rm d}: \mySet{M} \mapsto \hat{\mySet{X}}^\lenStag$ which maps each index $i \in \mySet{M}$ into a codeword $\myVec{q}_i \in  \hat{\mySet{X}}^\lenStag$. 
\end{definition}
The quantizer output for input $\myX^\lenAsym = \{\myX_i\}_{i=1}^{\lenAsym} \in \mySet{X}^\lenAsym$ is $\hat{\myX}^\lenStag = g_\lenStag^{\rm d}\left( g_\lenAsym^{\rm e}\left( \myX^\lenAsym\right) \right) \triangleq \Quan{M}{\lenAsym,\lenStag}\left( \myX\right)$. 
{\em Scalar quantizers} operate on a scalar input, i.e., $\lenAsym=1$ and $\mySet{X}$ is a scalar space, while {\em vector quantizers} have a multivariate input. Note that when $\mySet{X}$ is a vector space, then each $\myX_i$ is a random vector. 
When the input size and  output size are equal, namely, $\lenAsym=\lenStag$, we write $\Quan{M}{\lenAsym}\left(\cdot \right) \triangleq \Quan{M}{\lenAsym,\lenAsym}\left(\cdot \right)$.

In the standard quantization problem,  a $\Quan{M}{\lenAsym}\left(\cdot \right)$ quantizer is designed to minimize some distortion measure  $d_\lenAsym:\mySet{X}^\lenAsym\times\hat{\mySet{X}}^\lenAsym \mapsto \mySet{R}^+$  between its input and its output. 
The performance of a quantizer is therefore characterized using two measures: The quantization rate, defined as $\Rate \triangleq \frac{1}{\lenAsym}\log M$, and the expected distortion $\E\{d_\lenAsym\left(\myX^\lenAsym, \hat{\myX}^\lenAsym \right)\}$. For a fixed input size $\lenAsym$ and codebook size $M$, the optimal quantizer is given by  
\begin{equation}
\label{eqn:OptQuantizer}
\Quan{M}{\lenAsym, {\rm opt}}\left(\cdot \right) = \mathop{ \min}\limits_{\Quan{M}{\lenAsym}\left(\cdot \right)} \E \left\{d_\lenAsym\left(\myX^\lenAsym, \Quan{M}{\lenAsym}\left( {\myX}^\lenAsym \right)\right)   \right\}.
\end{equation}

Characterizing the optimal quantizer via \eqref{eqn:OptQuantizer} and the optimal tradeoff between distortion and quantization rate is in general a very difficult task. Consequently, optimal quantizers are typically studied assuming either high quantization rate, i.e., $\Rate \rightarrow \infty$, see, e.g., \cite{Li:99}, or asymptotically large input size, namely, $\lenAsym \rightarrow \infty$, typically with stationary inputs, via rate-distortion theory \cite[Ch. 10]{Cover:06}. 
For example, when the quantizer input represents a stationary source, and the distortion measure is {subadditive, i.e., for any $\lenAsym_1$, $\lenAsym_2$, $\myX^{\lenAsym_1}\in \mySet{X}^{\lenAsym_1}$, $\hat{\myX}^{\lenAsym_1} \in \mySet{X}^{\lenAsym_1}$,  $\myX^{\lenAsym_2}\in \mySet{X}^{\lenAsym_2}$, $\hat{\myX}^{\lenAsym_2} \in \mySet{X}^{\lenAsym_2}$, it holds that 
$d_{\lenAsym_1+\lenAsym_2}\left(\{\myX^{\lenAsym_1}, \myX^{\lenAsym_2}\}, \{\hat{\myX}^{\lenAsym_1}, \hat{\myX}^{\lenAsym_2}\} \right) \le d_{\lenAsym_1}\left(\myX^{\lenAsym_1}, \hat{\myX}^{\lenAsym_1} \right) + d_{\lenAsym_2}\left(\myX^{\lenAsym_2}, \hat{\myX}^{\lenAsym_2} \right)$. Then, by  \cite[Thm. 5.9.1]{Han:03}} the optimal distortion in the limit $\lenAsym \rightarrow \infty$ for a fixed rate $\Rate$ is given by the distortion-rate function:
\begin{definition}[Distortion-rate function]
	\label{def:DistRateFunction}
	The distortion-rate function for  	a stationary source $\{\myX_i\}_{i=1}^{\infty}$ with respect to the subadditive distortion measure $d_{\lenAsym}$  is defined as
	\vspace{-0.1cm}
	\begin{equation}
	\label{eqn:DistRateFunction}
	\!\!D_{\myX}\!\left( R \right)\! =\! \mathop {\lim }\limits_{\lenAsym \rightarrow \infty}\mathop {\min }\limits_{{\Pdf{\hat{\myX}^\lenAsym|\myX^\lenAsym}}:\frac{1}{\lenAsym}I\left( \hat{\myX}^\lenAsym;\myX^\lenAsym \right)\!\le\! R}\frac{1}{\lenAsym} \E\left\{ {d_\lenAsym\!\left( \hat{\myX}^\lenAsym,\myX^\lenAsym \right)} \!\right\}.
	\vspace{-0.1cm}
	\end{equation} 
\end{definition}
The minimization in \eqref{eqn:DistRateFunction} is carried out over all conditional distributions $\Pdf{\hat{\myX}^{\lenAsym}|\myX^{\lenAsym}}$ which satisfy the given constraint on the resulting mutual information. 
The marginal output distribution of $\{\hat{\myX}_i\}$ which obtains the minima in \eqref{eqn:DistRateFunction} is referred to henceforth as the {\em optimal marginal distortion-rate distribution}. 
One scenario where $D_{\myX}\left( R \right)$ is given in closed-form is when each $\myX_i$ is a zero-mean $\lenXtag\times 1$ proper-complex Gaussian \ac{rv} \cite[Def. 1]{Massey:93}, i.e., $\mySet{X} = \mySet{C}^{\lenXtag}$,   such that for each $l \in \{1,2,\ldots,\lenXtag\}$, the source $\{(\myX_i)_l\}_{i=1}^{\infty}$ is stationary\footnote{ {Following \cite{Han:03}, we use the term {\em stationary source} for stationary and ergodic signals with time index $i =\{1,2,\ldots\}$.}} with scalar \ac{psd} $\Psd{\myX}:[0,2\pi)\mapsto\mySet{R}^+$, thus its multivariate \ac{psd} is $\PSD{\myX}(\cdot) = \E\{\myX_i\myX_i^H\}\Psd{\myX}(\cdot)$. The distortion-rate function for this scenario is given in the following example:
\begin{example}
	\label{exm:Gaussian}
	Let $\{\myX_i\}_{i=1}^{\infty}$ be  zero-mean proper-complex $\lenXtag\times 1$ Gaussian source with multivariate \ac{psd} $\PSD{\myX}(\omega) = \CovMat{\myX}\Psd{\myX}(\cdot)$, and let the eigenvalue decomposition of $\CovMat{\myX} \in \mySet{C}^{\lenXtag \times \lenXtag}$ be given by $\CovMat{\myX} = \myMat{U}_{\myX} \myMat{\Lambda}_{\myX}\myMat{U}_{\myX}^H$. 
	The distortion-rate function for $\myX$ with the \ac{mse} distortion is \cite[Cor. 1]{Gutierrez:18}
	\begin{subequations}
		\label{eqn:exmGaussian}
			\vspace{-0.1cm}
	\begin{equation}
	\label{eqn:exmGaussian1}
	\hspace{-0.2cm}
	\DistG(R, \CovMat{\myX}, \Psd{\myX} ) \!=\! \frac{1}{2\pi} \!\int_{0}^{2\pi}\!\sum\limits_{i=1}^{\lenXtag}\! \min \left(\zeta, \left(\myMat{\Lambda}_{\myX} \right)_{i,i}\!\Psd{\myX}(\omega)  \right)d\omega, 
		\vspace{-0.1cm}
	\end{equation}
	where $\zeta > 0 $ is the solution to 
		\vspace{-0.1cm}
	\begin{equation}
	\label{eqn:exmGaussian2}
	R = \frac{1}{2\pi} \int_{0}^{2\pi}\sum\limits_{i=1}^{\lenXtag} \left( \log \frac{\left(\myMat{\Lambda}_{\myX} \right)_{i,i}\Psd{\myX}(\omega)  }{\zeta}\right)^+ d\omega. 
		\vspace{-0.1cm}
	\end{equation}
	The optimal marginal distribution for this setup is a zero-mean proper-complex  multivariate Gaussian distribution with \ac{psd} $\PSD{\hat{\myX}}(\omega) = \myMat{U}_{\myX} \myMat{\Lambda}_{\hat{\myX}}(\omega)\myMat{U}_{\myX}^H$, where $\myMat{\Lambda}_{\hat{\myX}}(\omega)$ is a diagonal matrix with diagonal entries $\left( \myMat{\Lambda}_{\hat{\myX}}(\omega)\right)_{i,i} =  \big( \left(\myMat{\Lambda}_{\myX} \right)_{i,i}\Psd{\myX}(\omega) - \zeta\big)^+$.  
	\end{subequations}
\end{example}

Comparing high  rate analysis for scalar quantizers and rate-distortion theory for vector quantizers demonstrates the sub-optimality of serial scalar quantization. For example, for quantizing a large-scale real-valued Gaussian random vector with i.i.d. entries and sufficiently large quantization rate $\Rate$, where one would imagine there is little benefit in quantizing the entries jointly over quantizing each entry independently, vector quantization notably outperforms serial scalar quantization \cite[Ch. 23.2]{Polyanskiy:15}. 

Finally, we introduce the notion of {\em dithered quantization}, which will be frequently used in our analysis of hardware-limited task-based quantization systems:
 \begin{definition}[Dithered quantizer]
 	\label{def:DithQuant}
	A scalar quantizer $\Quan{M}{1}$ implements serial non-subtractive uniform dithered quantization \cite{Gray:93}, referred to henceforth as {\em dithered quantization}, with support $\DynRange$ and quantization spacing $\DynInt[] = \frac{2\DynRange}{M}$, if its output for an input sequence  $y_1, y_2, \ldots, y_\lenZ$ can be written as $\Quan{M}{1}\left( y_i\right)  =  q\left({\rm Re}\left\{y_i + z_i\right\}  \right) +j \cdot q\left({\rm Im}\left\{y_i + z_i\right\}   \right)$. Here, $z_1, \ldots, z_\lenZ $ are i.i.d. \acp{rv} with i.i.d. real and imaginary parts uniformly distributed over $\left[-\frac{\DynInt[]}{2},\frac{\DynInt[]}{2} \right]$, mutually independent of the input, and  $q(\cdot)$ implements  uniform quantization defined as
	\ifsingle
	\begin{equation*}
	q(y) = \begin{cases}
	-\DynRange + \DynInt\left(l +\frac{1}{2} \right)   & 
	\alpha - l \cdot \DynInt + \DynRange \in  \left[0,\DynInt \right], \quad l \in \{0,1, \ldots, M - 1 \}  \\
	{\rm sign}\left( \alpha\right) \left( \DynRange - \frac{\DynInt}{2}\right)    & |\alpha| > \DynRange.
	\end{cases}
	\end{equation*}  
	\else
	\begin{equation*}
	q(\alpha) = \begin{cases}
	-\DynRange + \DynInt\left(l + \frac{1}{2} \right)   & \begin{array}{c}
	\alpha - l \cdot \DynInt + \DynRange \in  \left[0,\DynInt \right] \\ l \in \{0,1, \ldots, M - 1 \}  \end{array} \\
	{\rm sign}\left( \alpha\right) \left( \DynRange - \frac{\DynInt}{2}\right)    & |\alpha| > \DynRange.
	\end{cases}
	\end{equation*}  
	\fi 
 \end{definition}
 	Note that when $M = 2$, the uniform quantizer $q(y)$ is a standard one-bit sign quantizer of the form $q(\alpha) = c \cdot {\rm sign}(\alpha)$, where the $c >0$ is determined by the support $\DynRange$.

 In the following we study hardware-limited systems assuming dithered quantizers. Our motivation for using dithered quantizers stems from the fact that conventional analysis of uniform quantizers, e.g., \cite{Max:60}, does not lead to a tractable model for the quantizer output, nor does it extend to the task-based setup. However, when using dithered quantizers, the digital representation of an input which is in the support of the quantizer  can be written as the sum of the quantizer input and an additive uncorrelated white noise signal \cite{Gray:93}. This significantly facilitates our analysis and allows to characterize the system which minimizes the \ac{mse}. 
 Nonetheless, it is emphasized that this property of dithered quantizers is also {\em approximately} satisfied  in uniform quantization  {\em without dithering} for  various input distributions, including Gaussian inputs\footnote{For a Gaussian input with magnitude smaller than $\DynRange$ with sufficiently high probability, if the quantization spacing is in the order of the input standard deviation (or smaller), then the  output can be modeled as the input corrupted by additive uncorrelated white noise, even without dithering \cite[Sec. VII]{Widrow:96}.} \cite{Widrow:96}. 
 Therefore, the rigorous analysis which follows from considering dithered quantization, also holds approximately when using standard uniform quantizers without dithering, as demonstrated in \cite{Shlezinger:18}.


\vspace{-0.2cm}
\section{Task-Based Quantization of Large-Scale Data}
\label{sec:MSE}
\vspace{-0.1cm}
We now extend the analysis of task-based quantization carried out in our previous work  \cite{Shlezinger:18}, which considered fixed-size signals, to asymptotically large input signals. 
The motivation of this extension stems from the need to properly design and characterize quantizers for massive \ac{mimo} systems, which is our main target application discussed in Section \ref{sec:MIMO}. 
To that aim, we first present the problem formulation  in Subsection \ref{subsec:Pre_Problem}, and derive the achievable \ac{mse} without quantization constraints in Subsection \ref{subsec:MSE_NoQuant}. Then, we study task-based quantization with vector quantizers in Subsection \ref{subsec:MSE_VecQuant} and with hardware-limited quantizers in Subsection \ref{subsec:MSE_ScaQuant}. 
Focusing on the asymptotic regime allows us to rigorously characterize the achievable performance of vector quantizers, for which we were only able to obtain bounds in the finite horizon case studied in \cite{Shlezinger:18}. For the hardware-limited case, we formulate the dependency of task-based quantization systems on how  the system parameters grow proportionally with the size of the input signal, i.e., the quantization rate and the  analog combining ratio.

\vspace{-0.2cm}
\subsection{Problem Formulation}
\label{subsec:Pre_Problem}
\vspace{-0.1cm}
We study task-based quantization with asymptotically large observations and a proportionally large desired signal. 
The design objective of the quantizer is to quantize the observations such that the desired signal can be accurately recovered from the  quantized observations in the sense of minimizing the \ac{mse}.  
The desired signal consists of $\lenAsym$ zero-mean $\lenStag \times 1$ random vectors $\{\mySOItag_i\}_{i=1}^{\lenAsym}$, sampled from a stationary source with multivariate autocorrelation function $\E\{\mySOItag_{i+l}\mySOItag_i^H\} = \CovMat{\mySOItag} \Acorr{}[l]$, 
where  $\CovMat{\mySOItag} \in \mySet{C}^{\lenStag \times \lenStag}$ is Hermitian and positive semi-definite, while $\Acorr{}[\cdot]$ is an absolutely summable scalar autocorrelation function satisfying $\Acorr{}[0]=1$. By letting  $\Psd{}(\cdot)$ be the \ac{dtft} of $\Acorr{}[\cdot]$, the corresponding multivariate \ac{psd} is given by  $ \CovMat{\mySOItag} \Psd{}(\cdot)$. 
 The observations are a set of  $\lenXtag \times 1$ random vectors $\{\myObstag_i\}_{i=1}^{\lenAsym}$ with multivariate \ac{psd} $ \CovMat{\myObstag} \Psd{}(\cdot)$,  where  $\lenXtag \ge \lenStag$, and each vector $\myObstag_i$ is related to its corresponding $\mySOItag_i$ via the same conditional probability measure, denoted $\Pdf{\myObstag | \mySOItag}$. 
 \label{txt:VecSize}
 The model assumption that the size of the desired signal is not larger than that of the observed signal allows us to clearly demonstrate the benefits of task-based quantization as noted in \cite{Shlezinger:18}, and faithfully represent our main target application of channel estimation in massive \ac{mimo} systems discussed in Section \ref{sec:MIMO}.
 
We assume that the \ac{mmse} estimator which stems from $\Pdf{\myObstag | \mySOItag}$ is linear, i.e., there exists $ \LmmseMattag \in \mySet{C}^{\lenStag \times \lenXtag}$ such that the \ac{mmse} estimate of $ \mySOItag_i$ from $\{\myObstag_{i'}\} $ can be written as $\mySOIEsttag_i = \LmmseMattag \myObstag_i$, for each $i \in \{1,\ldots, \lenAsym\}$.
Since we focus on large-scale data, $\lenAsym$ is arbitrarily large.
Clearly, this setup specializes to the  case in which the desired signal and the observed signal consist of i.i.d. elements.
Such scenarios arise, for example, in signal recovery over memoryless channels, where $\mySOItag_i$ is the channel input at time index $i$ and $\myObstag_i$ is the corresponding channel output, or alternatively, in the estimation of fast fading memoryless channels, in which $\mySOItag_i$ is the unknown channel at time index $i$ and $\mySOItag_i$  is the channel output. 
 Furthermore, in  Section \ref{sec:MIMO} we show that this model can also represent  channel estimation in massive \ac{mimo} systems with correlated antennas.

\begin{figure}
		\centering
		\scalebox{0.85}
		{\includegraphics[trim={0.2cm 0 0 0},clip]{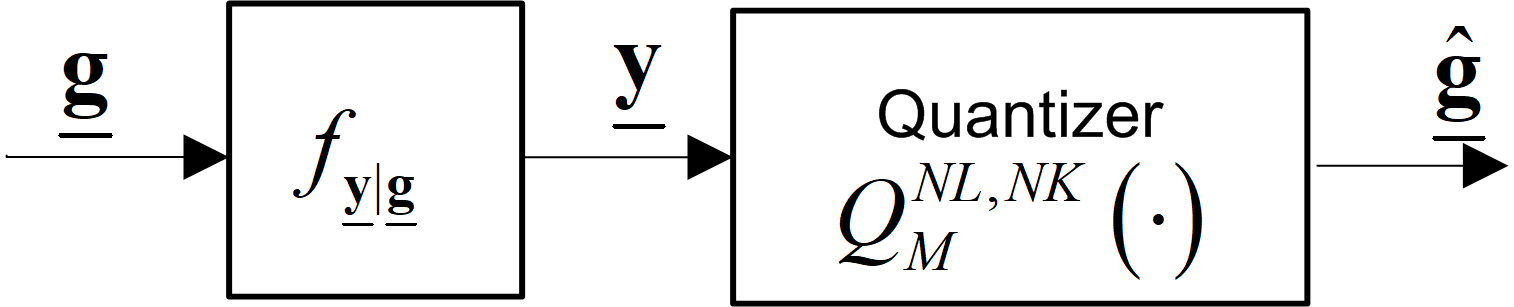}}
		\caption{Task-based quantization system.}
		\label{fig:TB_System}	
		\vspace{-0.4cm}	
	\end{figure}

We write the desired vector and the observed vector as $\mySOI = {\rm vec}\big([\mySOItag_1, \ldots, \mySOItag_{\lenAsym}]^T\big)$ and $\myObs = {\rm vec}\big([\myObstag_1, \ldots, \myObstag_{\lenAsym}]^T\big)$, respectively. By letting $\CorrMat$ be a $\lenAsym \times \lenAsym$ Toeplitz matrix whose entries are given by $\left( \CorrMat\right)_{i_1,i_2} = \Acorr{}[i_1 - i_2]$ for each $i_1, i_2 \in \{1,\ldots, \lenAsym\}$, it holds that the covariance matrices of $\mySOI$ and $\myObs$ are equal to $\CovMat{\mySOItag} \otimes \CorrMat$ and $\CovMat{\myObstag} \otimes \CorrMat$, respectively.  
%
The main model notations along with their meaning in the massive \ac{mimo} setup considered in Section \ref{sec:MIMO} are summarized in Table \ref{tbl:notation}. 
The proposed system forms a quantized representation of  $\mySOI$ based on the observed  $\myObs$,  using up to $\log M$ bits, where the quantization rate $\Rate \triangleq \frac{1}{\lenS} \log M$ is fixed. 
An illustration of such a system  is depicted in Fig. \ref{fig:TB_System}. 
\begin{table*}[t]
	\centering
	\caption{Main model notations.}
	\vspace{-0.2cm}
	\label{tbl:notation}
\ifsingle
	\small
\fi 
	\begin{tabular}{|c|c|c|c|}
		\hline
		Notation & Type & General Setup (Section \ref{sec:MSE}) & Massive \ac{mimo} Setup (Section \ref{sec:MIMO}) \\ \hline\hline 
		$\lenAsym$ & Large integer & Number of observations & Number of antennas \\ \hline
		$\lenStag$ & Integer & Size of desired signal samples & Number of users in cell \\ \hline
		$\lenXtag$ & Integer & Size of observation samples & Number of pilot symbols \\ \hline
		$\mySOI_{\vphantom{A}}^{\vphantom{A}}$ & $\lenS \times 1$ complex vector & Desired signal & Channel coefficients in vector form \\ \hline
		$\myObs_{\vphantom{A}}^{\vphantom{A}}$ & $\lenX \times 1$ complex vector & Observations & Channel outputs in vector form \\ \hline
		$\LmmseMattag$ & $\lenStag \times \lenXtag$ complex matrix & Linear \ac{mmse} matrix & Linear \ac{mmse} matrix \\ \hline
		$\CovMat{\myObstag}$ & $\lenXtag \times \lenXtag$ complex matrix & Covariance of each observed sample & Temporal covariance of channel outputs  \\ \hline
		$\{\phicoeff_i\}$ &  $\lenStag$ real numbers & Singular values of $\LmmseMattag\CovMat{\myObstag}^{1/2}$ & Singular values of $\LmmseMattag\CovMat{\myObstag}^{1/2}$  \\ \hline
		$\Acorr{}[\cdot]_{\vphantom{A}}^{\vphantom{A}}$ & Mapping $\mySet{Z}\mapsto \mySet{R}$ & Entry-wise correlation & Spatial correlation between antennas \\ \hline
		$\CorrMat$ & $\lenAsym \times \lenAsym$ complex matrix & Toeplitz matrix constructed from $\Acorr{}[\cdot]$ & Spatial correlation matrix \\ \hline
		$\Psd{}(\cdot)_{\vphantom{A}}^{\vphantom{A}}$ & Mapping $[0,2\pi]\mapsto \mySet{R}^+$ & \ac{dtft} of $\Acorr{}[\cdot]$ & Spatial \ac{psd} of each channel output \\ \hline
		$\Rate$ & Real number & Quantization rate  & Quantization rate \\ \hline
		$\lenZ$ & Integer & Number of scalar quantizers & Number of scalar quantizers \\ \hline
		$\Ratio$ & Real number & Analog combining ratio  & Analog combining ratio \\ \hline
		$\TilM^{\vphantom{\left( \right) ^A}}$ & Integer & Number of scalar quantization regions  & Number of scalar quantization regions \\ \hline
		$\DynRange$ & Real number & Scalar quantizer support  & Scalar quantizer support \\ \hline
		\hline
	\end{tabular} 
	\vspace{-0.4cm}
\end{table*}
The distortion measure for a quantized representation $\hat{\mySOI}$ is the average \ac{mse},  defined as 
\vspace{-0.15cm}
\begin{equation}
\label{eqn:MSEDef}
\AsymDist \triangleq \mathop{\lim}\limits_{\lenAsym \rightarrow \infty} \frac{1}{\lenS} \E \{\|\mySOI- \hat{\mySOI}  \|^2 \}.
\vspace{-0.15cm}
\end{equation}

\begin{figure}
	\centering
	\scalebox{0.85}
	{\includegraphics[trim={0.2cm 0 0 0},clip]{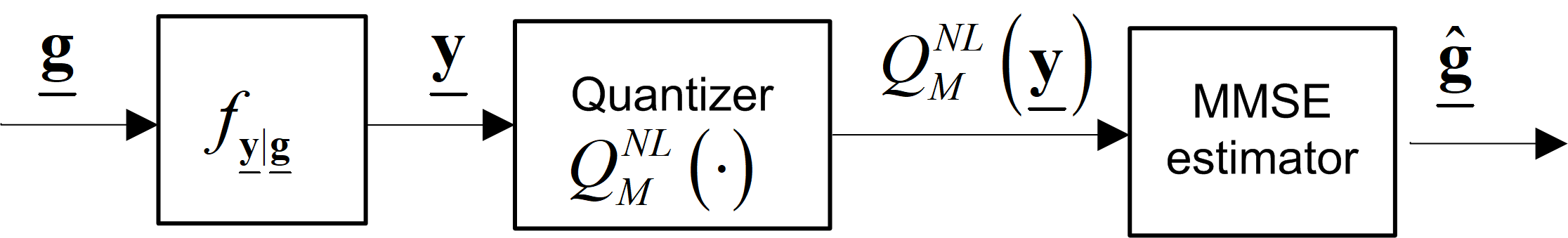}}
	\caption{Task-ignorant quantizer.}
	\label{fig:TaskIgn_System}
	\vspace{-0.4cm}
\end{figure}

We consider vector quantizers as well  as {hardware-limited quantizers}. In the following we elaborate on these systems:

\smallskip
\noindent
{\bf Vector Quantizers:} 
Joint (vector) quantization is known to be superior to separate (scalar) quantization \cite[Ch. 23]{Polyanskiy:15}. Thus, analyzing systems utilizing vector quantizers provides the fundamental limits of task-based quantization with large-scale inputs. 
We consider two different vector quantization systems: 
\begin{enumerate}
	\item {\bf Task-based optimal vector quantization} - in the optimal quantization system, the quantizer $\Quan{M}{\lenX,\lenS}(\cdot)$ 	in Fig. \ref{fig:TB_System} is the vector quantizer which minimizes the distortion between the quantized representation $\hat{\mySOI}$ and $\mySOI$. The performance of this system represents the optimal distortion achievable with any quantization system operating at rate $\Rate$. 
	\item {\bf Task-ignorant vector quantization} -  here, the quantizer is designed to recover
	the observed  $\myObs$ separately from the task, using the optimal vector quantizer for representing $\myObs$,  { namely, the quantizer here is ignorant of the task and is designed to accurately represent the observations.} The desired vector $\mySOI$ is estimated from the quantized representation using the \ac{mmse} estimator, as illustrated in  Fig. \ref{fig:TaskIgn_System}. This is a plausible system when the quantizer is ignorant of the task. 
\end{enumerate}

\begin{figure}
	\centering
	{\includefig{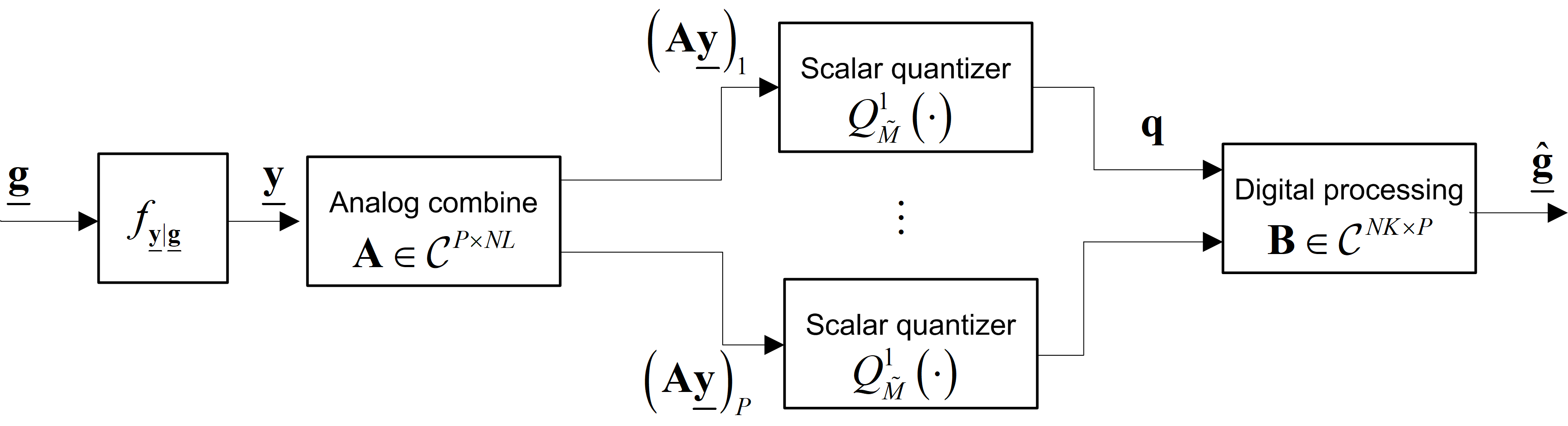}}
	\caption{Hardware-limited task-based quantization system.}
	\label{fig:GenericStructure1}		
	\vspace{-0.6cm}
\end{figure}

\smallskip
\noindent
{\bf Hardware-Limited Quantizers:} 
Vector quantization may be difficult to implement, especially for large input sizes. Consequently, systems utilizing vector quantizers may not be feasible in practice. 
As discussed in the introduction, practical systems typically implement quantization using  scalar \acp{adc}. In such systems, each continuous-amplitude element is converted into a discrete representation using a single quantization rule, which commonly corresponds to uniform quantization. This operation can be modeled using identical scalar uniform quantizers.  
In particular, we consider the system depicted in Fig. \ref{fig:GenericStructure1}. 
The observed vector $\myObs$, is projected into $\mySet{C}^{\lenZ}$, where 
$\lenZ \le \lenX$, using some pre-quantization processing carried out in the analog domain. 
As arbitrary processing may be difficult to implement in analog, we henceforth restrict our attention to linear pre-quantization processing only. This {\em analog combining} is modeled via the matrix $\myA \in   \mySet{C}^{\lenZ \times \lenX}$. 
We write the number of scalar quantizers $\lenZ$ in terms of its integer quotient and remainder with respect to $\lenAsym$, denoted $\lenZn$ and $\lenZq$, respectively,  i.e.,
\vspace{-0.1cm}
\begin{equation}
\label{eqn:lenZDefa}
\lenZ = \lenZn \cdot \lenAsym + \lenZq, \quad 0 < \lenZq < \lenAsym.
\vspace{-0.1cm}
\end{equation} 
The motivation for expressing $\lenZ$ using $\lenAsym$ in \eqref{eqn:lenZDefa} stems from the fact that for large-scale inputs, $\lenAsym$ tends to infinity, and thus $\lenZn$ and $\lenZq$ represent how $\lenZ$ scales accordingly. These scaling parameters play an important role  when analyzing the performance of hardware-limited task-based quantizers, as shown in Subsection \ref{subsec:MSE_ScaQuant}.

The real and imaginary parts of each entry of $\myA\myObs$ are quantized using the same scalar  quantizer with resolution $\TilM \triangleq \lfloor M^{1/2\lenZ} \rfloor$, denoted $\Quan{\TilM}{1}(\cdot)$. 
Define the {\em analog combining ratio}
\vspace{-0.15cm}
\begin{equation}
\label{eqn:RatioDefa}
{\Ratio} \triangleq \frac{\lenZ}{\lenX} =  \frac{\lenZn}{\lenXtag} + \frac{\lenZq}{\lenX}.
\vspace{-0.15cm}
\end{equation}
Note that  $\TilM = \lfloor 2^{\frac{ \Rate}{2 \cdot \Ratio}} \rfloor$.
The overall quantization rate is $\frac{2\cdot \lenZ}{\lenX} \log \big(\TilM\big)  \le\frac{1}{\lenX} \log M = \Rate$. 
The identical scalar quantizers $\Quan{\TilM}{1}$ implement dithered quantization, as defined in Def. \ref{def:DithQuant}.  
The quantizer is designed to operate within the support $\DynRange$, namely,  the amplitude of the input is not larger than $\DynRange$ with sufficiently large probability. To guarantee this, we fix $\DynRange$ to be some multiple $\myEta$ of the maximal standard deviation of the input. For example, for  proper-complex Gaussian  inputs, when $\myEta \ge  \sqrt{2}$   the amplitude of both the real and imaginary parts of the input are smaller than the support with probability over $94 \%$. We assume that $\myEta < \sqrt{3/2} \TilM$, such that the variable $\MyKappa \triangleq \myEta^2 \big(1 - \frac{ 2\myEta^2 }{3\TilM^2}\big)^{-1}$ is strictly positive. Note that $\eta = 2$ satisfies this requirement for any $\TilM \ge 2$, i.e., the \ac{adc} is implemented using scalar quantizers with at least one bit. 

Finally, in the digital domain, the system approximates the linear \ac{mmse}  estimate based on the output of the \ac{adc}, denoted $\myQ \in \mySet{C}^{\lenZ}$, where $(\myQ)_i = \Quan{\TilM}{1}\big((\myA \myObs )_i \big)$. Consequently, the estimate can be written as $\hat{\mySOI}= \myB \myQ$ for some $\myB \in \mySet{C}^{\lenS \times \lenZ}$.  We focus on linear digital processing to keep the analysis tractable, and since linear  estimators are commonly used in our main application, massive \ac{mimo} channel estimation with quantized outputs \cite{Li:17,Jacobsson:17}. This restriction is not expected to notably affect the overall performance, especially when the error due to quantization is small, as the \ac{mmse} estimator in the considered setup is linear. 

\vspace{-0.2cm}
\subsection{No Quantization Constraints}
\label{subsec:MSE_NoQuant}
\vspace{-0.1cm} 
As a preliminary step, we note that the \ac{mmse} estimate of $\mySOI$ from $\myObs$, denoted $\mySOIEst$  consists of the $\lenStag \times 1$ random vectors $\{\mySOIEsttag_i \}_{i=1}^{\lenAsym}$, sampled from a stationary source with multivariate \ac{psd} $\PSD{\mySOIEsttag}(\cdot) = \LmmseMattag\CovMat{\myObstag}\LmmseMattag^H \Psd{}(\cdot)$. Since $\frac{1}{2\pi}\int_{0}^{2\pi}\Psd{}(\omega) d\omega = \Acorr{}[0]=1$, the  average \ac{mmse} can  be written as 
\vspace{-0.15cm}
\begin{align}
\AsymDist\MMSE =  \frac{1}{\lenStag}{\rm Tr}\left(\CovMat{\mySOItag} - \LmmseMattag\CovMat{\myObstag}\LmmseMattag^H \right).
\label{eqn:MMSE1}  
\vspace{-0.15cm}
\end{align} 
The \ac{mmse} in \eqref{eqn:MMSE1} is achievable without quantization, and thus serves as a lower bound on the achievable distortion of the quantization systems discussed in the following subsections.

\vspace{-0.2cm}
\subsection{Vector Quantization}
\label{subsec:MSE_VecQuant}
\vspace{-0.1cm} 
We now study the  average \ac{mse} achievable of the vector quantization systems detailed in Subsection \ref{subsec:Pre_Problem}. 
We note that  for fixed size inputs, the achievable performance of vector quantizers can only be obtained in terms of upper and lower bounds, see \cite[Prop. 1]{Shlezinger:18}. However, as we show next, for large-scale data, we  explicitly characterize  the minimal achievable average \ac{mse} for each system using indirect rate-distortion theory analysis, which considers asymptotically large inputs. 
\subsubsection{Optimal Vector Quantizer} 
The optimal vector quantizer minimizes the \ac{mse} between the unknown desired vector and the system output.  
Recovering the desired signal $\mySOI$ from quantized observations is a special case of  indirect lossy source coding  \cite{Witsenhausen:80}. For the \ac{mse} distortion, it follows from \cite{Wolf:70} that the optimal vector quantizer first recovers the \ac{mmse} estimate   $\mySOIEst$, and then uses a vector quantizer to represent $\mySOIEst$. The resulting \ac{mse} is given in the following theorem:
\begin{theorem}
	\label{thm:OptVecQuant}
	The \ac{mse} of the optimal vector quantizer is 
	\begin{equation}
	\label{eqn:OptVecQuant}
	\AsymDist \Opt = \AsymDist \MMSE + \frac{1}{\lenStag} D_{\mySOIEst}\left(\frac{\lenXtag}{\lenStag}\cdot \Rate \right), 
	\end{equation}
	where $D_{\mySOIEsttag}(\cdot)$ is the distortion-rate function, given in Def. \ref{def:DistRateFunction}, of the random vector $\mySOIEst$ with the \ac{mse} distortion.
\end{theorem}

{\em Proof:}
See Appendix \ref{app:ProofVecAsym}.

\smallskip
Theorem \ref{thm:OptVecQuant} holds since the \ac{mmse} estimate $\mySOIEst$ represents a stationary source, thus, in the limit $\lenAsym \rightarrow \infty$, the minimal \ac{mse} for a fixed quantization rate is given by the distortion-rate function. The achievable average \ac{mse} in \eqref{eqn:OptVecQuant} constitutes the minimal achievable distortion of any system which recovers $\mySOI$ from $\myObs$ using up to $\Rate$ bits per input sample.

\subsubsection{Task-Ignorant Vector Quantizer}  
In task-ignorant quantization, 
 the desired signal is estimated from the quantized observations, which are in turn designed to yield an accurate representation of the input signal. 
The resulting quantization system, depicted in Fig. \ref{fig:TaskIgn_System}, first quantizes  $\myObs$ via a quantizer $\Quan{M}{\lenX}(\cdot)$, which minimizes the \ac{mse} between its output and $\myObs$. Then, $\mySOI$ is estimated from the output of the quantizer using the \ac{mmse} estimator.  
Characterizing the average \ac{mse} of such systems is in general a challenging task, due to difficulty in formulating the conditional distribution of the desired signal given the output of the quantizer  $\Quan{M}{\lenX}(\cdot)$. However, in the special case where the signals are i.i.d., and thus $\Psd{}(\omega) = 1$, the resulting average \ac{mse} is given in the following theorem:
\begin{theorem}
	\label{thm:TIVecQuant}
	When $\{ \myObstag_i\}$ are i.i.d. the average \ac{mse} of the task-ignorant vector quantizer is given by
	\begin{equation}
	\label{eqn:TIVecQuant}
	\AsymDist\Ign = \AsymDist\MMSE + \frac{1}{  \lenStag} {\rm Tr}\left( \left( \LmmseMattag\right)^H \LmmseMattag\left( \CovMat{\myObstag} - \CovMat{\myObstag, D}(\Rate)  \right)  \right). 
	\end{equation}
	Here,  $\CovMat{\myObstag, D}(\Rate)$ is the covariance matrix of the optimal marginal distribution which achieves the distortion-rate function $D_{\myObstag}\left(\Rate\right)$ with the \ac{mse} distortion, given in Def. \ref{def:DistRateFunction}.
\end{theorem}

{\em Proof:}
See Appendix \ref{app:ProofTIAsym}.

\smallskip
Theorem \ref{thm:TIVecQuant} exploits the fact that when $\myObs$ consists of $\lenAsym$ i.i.d. $\lenXtag \times 1$ vectors, then, as $\lenAsym$ grows arbitrarily, the output of the optimal quantizer for representing $\myObs$ converges to a set of $\lenAsym$ i.i.d. vectors, each distributed via the optimal marginal distribution which achieves $D_{\myObstag}\left(\Rate\right)$. In our numerical study in Section \ref{sec:Simulations} it is illustrated that for relatively small quantization rates, there is a notable gap between the performance of task-ignorant quantization and the optimal average \ac{mse} in \eqref{eqn:OptVecQuant}. 

\vspace{-0.2cm}
\subsection{Hardware-Limited Quantization}
\label{subsec:MSE_ScaQuant}
\vspace{-0.1cm} 
We now characterize the optimal hardware-limited task-based quantization system, using the setup depicted in Fig. \ref{fig:GenericStructure1}.  We derive the analog combining matrix and digital processing matrix which minimize the average \ac{mse}, denoted $\myA\op$ and $\myB\op$, respectively, and the corresponding support $\DynRange$. 

To formulate the proposed system, define the $\lenStag \times \lenXtag$ matrix $\LmmseMatT \triangleq  \LmmseMattag \CovMat{\myObstag}^{1/2}$, and let $\{ \phicoeff_i \}$ be its singular values arranged in descending order. Note that for $i > {\rm rank} \big( \LmmseMatT\big)$, $\phicoeff_i  = 0$.   
Let $\{\eigT{i} \}$ be the singular values of $\LmmseMatT\otimes \CorrMat$ arranged in descending order, and define the function $\Glevel(\alpha) \triangleq \big( \alpha   - 1 \big)^ +$,  $\alpha \in \mySet{R}^+$.  Recall that $\MyKappa$ is defined as $\MyKappa = \myEta^2 \big(1 - \frac{ 2\myEta^2 }{3\TilM^2}\big)^{-1}$, where $\myEta$ is the ratio of the quantizer support to the maximal input standard deviation.
The hardware-limited quantization system which minimizes the average \ac{mse} is stated in the following theorem:
\begin{theorem}
	\label{thm:OptimalDes}
	\begin{subequations}
		\label{eqn:OptimalDes}
		In the hardware-limited quantization system which minimizes the average \ac{mse}, 	the analog combining matrix $\myA\op$ is given by $\myA \op = \myMat{U}_{\myA} \myMat{\Lambda}_{\myA} \left( \myMat{V}_{\myA}^H \CovMat{\myObstag}^{-1/2} \otimes  \CorrMat^{-1/2}\right) $, where
		\begin{itemize}
			\item $\myMat{V}_{\myA} \in \mySet{C}^{\lenXtag \times \lenXtag}$ is the right singular vectors matrix of  $\LmmseMatT$.
			\item  $\myMat{\Lambda}_{\myA} \in \mySet{C}^{\lenZ \times \lenX}$ is a diagonal matrix   with diagonal entries  
			\vspace{-0.1cm}
			\begin{equation}
			\label{eqn:OptimalDesA}
			\left( \myMat{\Lambda}_{\myA} \right)_{l,l}^2 = \frac{4 \MyKappa}{{3\TilM^2} \cdot \Ratio}  \Glevel(\Wlevel\cdot \eigT{l}),
			\vspace{-0.1cm}
			\end{equation} 
			where $\Wlevel$ is set such that $\frac{4\MyKappa}{3 \TilM^2 \cdot \lenZ} \sum\limits_{l=1}^{\lenZ} \Glevel(\Wlevel\cdot \eigT{l}) = 1$, $\Ratio$ is defined in \eqref{eqn:RatioDefa}, and $\TilM =   \lfloor 2^{\frac{ \Rate}{2 \cdot \Ratio}} \rfloor$.
			\item $\myMat{U}_{\myA} \in \mySet{C}^{\lenZ \times \lenZ}$ is a unitary matrix which guarantees that  $\myMat{U}_{\myA}\myMat{\Lambda}_{\myA}\myMat{\Lambda}_{\myA}^H\myMat{U}_{\myA}^H$ has identical diagonal entries, which can be  obtained
			via \cite[Alg. 2.2]{Palomar:07}. 
		\end{itemize}
		
		The support of the \ac{adc} is given by $	\DynRange^2   =  \frac{\MyKappa}{\Ratio}$, 
		and the digital processing matrix is 
		\ifsingle		
		\begin{equation}
		\label{eqn:OptimalDesB}
		\myB \op  =  \left( \LmmseMattag\CovMat{\myObstag} \otimes\CorrMat\right)    \left( \myA \op\right) ^H\left( \myA \op \left( \CovMat{\myObstag} \otimes \CorrMat \right)  \left( \myA \op\right) ^H + \frac{{4{\DynRange^2}}}{{3\TilM^2}}{\myI_\lenZ} \right)^{ - 1}.
		\end{equation}
		\else
		\begin{align}
		\myB \op  &=  \left( \LmmseMattag\CovMat{\myObstag} \otimes\CorrMat\right) \left( \myA \op\right) ^H   \notag \\
		& \quad \times \left( \myA \op \left( \CovMat{\myObstag} \otimes\CorrMat\right)  \left( \myA \op\right) ^H \!+\! \frac{{4{\DynRange^2}}}{{3\TilM^2}}{\myI_\lenZ} \right)^{ - 1}\!\!.
		\label{eqn:OptimalDesB}
		\end{align}
		\fi 
		The corresponding achievable average \ac{mse} at the limit $\lenAsym\rightarrow \infty$ when $\lenZn \ge {\rm rank}(\LmmseMattag\CovMat{\myObstag}\LmmseMattag^H)$ is given by
		\begin{align}
		&\hspace{-0.2cm}\AsymDist \ADC
		\!=\!  \AsymDist \MMSE\! +\!  \frac{1}{2\pi}\int_{0}^{2\pi} \frac{1}{\lenStag } \sum\limits_{i=1}^{\lenStag }\!   \frac{  \phicoeff_i^2{\Psd{}(\omega)}  } { \Glevel(\Wlevel\cdot \phicoeff_i\sqrt{\Psd{}(\omega)}) \! + \! 1} d\omega.
		\vspace{-0.2cm}
		\label{eqn:OptimalMSE2}
		\end{align}
		Furthermore, when the signals consists of uncorrelated vectors, i.e., $\Acorr{}[{\tau}]=\delta_{{\tau}}$, the asymptotic average \ac{mse} for any $\lenZn \ge 0$ reduces to
\ifsingle		
		\begin{align}
		\AsymDist \ADC &= \AsymDist \MMSE + \frac{1}{\lenStag}\sum\limits_{i=1}^{\min(\lenStag, \lenZn)} \frac{ \phicoeff_i^2}{\Glevel(\Wlevel\cdot \phicoeff_i) +1} \notag \\
		&+ \delta_{(\lenZn < \lenStag)}\left(\frac{1}{\lenStag  } \sum\limits_{i=\lenZn + 1 }^{\lenStag } \phicoeff_i^2 
		- \left(\Ratio \cdot \lenXtag - \lenZn \right)  \frac{\phicoeff_{\lenZn+1}^2 \Glevel(\Wlevel\cdot \phicoeff_{\lenZn+1}) } {\Glevel(\Wlevel\cdot \phicoeff_{\lenZn+1}) + 1} \right).
		\label{eqn:OptimalMSE}
		\end{align}
\else
\vspace{-0.1cm}
		\begin{align}
&\AsymDist \ADC = \AsymDist \MMSE + \frac{1}{\lenStag}\sum\limits_{i=1}^{\min(\lenStag, \lenZn)} \frac{ \phicoeff_i^2}{\Glevel(\Wlevel\cdot \phicoeff_i) +1} 
+ \delta_{(\lenZn < \lenStag)} \notag \\
& \times \Bigg(\frac{1}{\lenStag  }\! \sum\limits_{i=\lenZn \! + \! 1 }^{\lenStag }\! \phicoeff_{ i}^2 
\! -  \! \left(\Ratio \cdot \lenXtag \! -  \! \lenZn \right)  \frac{\phicoeff_{\lenZn\! + \!1}^2 \Glevel(\Wlevel\cdot \phicoeff_{\lenZn+1}) }{\Glevel(\Wlevel\cdot \phicoeff_{\lenZn+1})  \! + \! 1} \Bigg). 
\label{eqn:OptimalMSE}
\vspace{-0.1cm}
\end{align}
\fi 
		%
	\end{subequations}
\end{theorem}   

{\em Proof:}
See Appendix \ref{app:ProofThmDes}.

Theorem \ref{thm:OptimalDes} extends \cite[Thm. 1]{Shlezinger:18} to asymptotically large complex-valued inputs. A notable  difference between Theorem~\ref{thm:OptimalDes} and \cite[Thm. 1]{Shlezinger:18} is in the performance expression in \eqref{eqn:OptimalMSE2}-\eqref{eqn:OptimalMSE}: While \cite[Thm. 1]{Shlezinger:18} studied the \ac{mse} with finite-size inputs, here we consider the asymptotic average \ac{mse}. Thus \eqref{eqn:OptimalMSE2}-\eqref{eqn:OptimalMSE} depend on how the number of scalar quantizers grow with  the input size, and not on the exact number of inputs and scalar quantizers.  

\smallskip
Note that when $\lenZq$ in \eqref{eqn:lenZDefa} does not grow proportionally with $\lenAsym$, i.e., $\mathop{\lim}\limits_{\lenAsym \rightarrow \infty} \frac{\lenZq}{\lenAsym} = 0$, then by \eqref{eqn:lenZDefa}, $\Ratio \cdot \lenXtag = \lenZn$, and the last summand in \eqref{eqn:OptimalMSE}  vanishes. 
When $\lenZq$ equals zero, i.e., $\lenZ$ is an integer multiple of $\lenAsym$, and $\Acorr{ }[{\tau}] = \delta_{\tau}$, the 
optimal system processes $\myObstag_i$ using the same transformation for each $i \in \{1,\ldots,\lenAsym\}$ separately, as stated in the following corollary:
\begin{corollary}
	\label{cor:OptimalCor}
	When $\lenZq = 0$  and $\Acorr{ }[\tau] = \delta_{\tau}$, 
	the hardware-limited system which minimizes the \ac{mse} applies the same mapping to each $\myObstag_i$ separately. This mapping includes  analog combining via the matrix $\myAtag$, scalar quantizers with support   $	\DynRange^2   =  \frac{\MyKappa}{\Ratio}$, and digital processing with matrix $\myBtag$. In particular, 
	$\myAtag=   \myMat{U}_{\myA} \myMat{\Lambda}_{\myA}\myMat{V}_{\myA}^H \CovMat{\myObstag}^{-1/2}  $, where
	\begin{itemize}
		\item $\myMat{V}_{\myA} \in \mySet{C}^{\lenXtag \times \lenXtag}$ is the right singular vectors matrix of  $\LmmseMatT$.
		\item  $\myMat{\Lambda}_{\myA} \in \mySet{C}^{\lenZn \times \lenXtag}$ is  diagonal   with  entries  
		$\left( \myMat{\Lambda}_{\myA} \right)_{i,i}^2 = \frac{4 \MyKappa \cdot \Glevel(\Wlevel\cdot \phicoeff_i)}{{3\TilM^2}\cdot \lenZn }  $, where   $\Wlevel$ is set such that $		 \frac{4\MyKappa}{3 \TilM^2 \cdot \lenZn}\sum\limits_{i=1}^{\lenZn}  \Glevel(\Wlevel\cdot \phicoeff_i)  = 1$.  
		\item $\myMat{U}_{\myA} \in \mySet{C}^{\lenZn \times \lenZn}$ is a unitary matrix for which $\myMat{U}_{\myA}\myMat{\Lambda}_{\myA}\myMat{\Lambda}_{\myA} ^H \myMat{U}_{\myA}^H$ has identical diagonal entries. 
	\end{itemize}
	The matrix 
\ifsingle
\begin{equation*}
\myBtag \!= \! \LmmseMatT \myMat{V}_{\myA} \myMat{\Lambda}_{\myA}^H   \Big( \myMat{\Lambda}_{\myA} \myMat{\Lambda}_{\myA}^H \!+\! \frac{{4{\DynRange^2}}}{{3\TilM^2}}{\myI_{\lenZn}} \Big)^{ - 1}\!\!\! \myMat{U}_{\myA}^H
\end{equation*}
\else
	$
	\myBtag \!= \! \LmmseMatT \myMat{V}_{\myA} \myMat{\Lambda}_{\myA}^H   \Big( \myMat{\Lambda}_{\myA} \myMat{\Lambda}_{\myA}^H \!+\! \frac{{4{\DynRange^2}}}{{3\TilM^2}}{\myI_{\lenZn}} \Big)^{ - 1}\!\!\! \myMat{U}_{\myA}^H$
\fi	
represents the digital processing. 
	The  achievable average \ac{mse}  is given by:
	\ifsingle	
		\begin{equation}
		\AsymDist \ADC = \AsymDist \MMSE + \frac{1}{\lenStag}\sum\limits_{i=1}^{\min(\lenStag, \lenZn)} \frac{ \phicoeff_i^2}{ \Glevel(\Wlevel\cdot \phicoeff_i)+1} + \frac{\delta_{(\lenZn < \lenStag)}}{\lenStag  } \sum\limits_{i=\lenZn + 1 }^{\lenStag } \phicoeff_{ i}^2.
		\label{eqn:OptimalCorMSE}
		\end{equation}
	\else
		\begin{align}
		\AsymDist \ADC = \AsymDist \MMSE + &\frac{1}{\lenStag}\sum\limits_{i=1}^{\min(\lenStag, \lenZn)}  \frac{ \phicoeff_i^2}{ \Glevel(\Wlevel\cdot \phicoeff_i) +1} \notag \\
		&\qquad + \frac{\delta_{(\lenZn < \lenStag)}}{\lenStag  } \sum\limits_{i=\lenZn + 1 }^{\lenStag } \phicoeff_{ i}^2. 
		\label{eqn:OptimalCorMSE}
		\end{align}
	\fi 
	
%
%
%
\end{corollary}

\begin{IEEEproof}
	The corollary follows directly from Theorem \ref{thm:OptimalDes}. In particular, \eqref{eqn:OptimalCorMSE} and 	the requirement on $\Wlevel$ are obtained from Theorem \ref{thm:OptimalDes} since  $\Ratio \cdot \lenXtag = \lenZn$ when $\lenZ = \lenZn \cdot \lenAsym$. The resulting $\myAtag$ is a special case of $\myA \op $ in Theorem \ref{thm:OptimalDes} for $\lenZ = \lenZn \cdot \lenAsym$, and $\myBtag$ is obtained by plugging $\myAtag \otimes \myI_{\lenAsym}$  into \eqref{eqn:OptimalDesB}. 
\end{IEEEproof}
	
Corollary \ref{cor:OptimalCor} is quite surprising in light of known results in vector quantization. It is well-known that with unrestricted vector quantizers, jointly processing a set of \acp{rv} is beneficial even if they are i.i.d. \cite[Ch. 23]{Polyanskiy:15}.  However, Corollary \ref{cor:OptimalCor} indicates that in the presence of scalar \acp{adc}, if it is possible to process i.i.d. \acp{rv} using the same mapping separately, i.e.,  when $\lenZq = 0$ and the same number of scalar quantizers can be assigned to each $\myObstag_i$, 	then this strategy minimizes the \ac{mse}.

Theorem \ref{thm:OptimalDes} and Corollary \ref{cor:OptimalCor} indicate that the analog combining ratio $\Ratio$, and particularly the value of $\lenZn$, play an important part in the performance of hardware-limited systems. Guidelines for setting these values are stated 	
in the following corollary:
\begin{corollary}
	\label{cor:OptimalP}
	In order to minimize the average \ac{mse}, $\lenZn$ must not be larger than the rank of $\LmmseMatT\CovMat{\myObstag} \LmmseMatT^H$. 
\end{corollary}
\begin{IEEEproof}
	The proof is obtained by repeating the arguments in \cite[Appendix D]{Shlezinger:18}, and is thus omitted for brevity.
\end{IEEEproof}

In order to compare the achievable average \ac{mse} in Theorem~\ref{thm:OptimalDes} to the fundamental limit in Theorem~\ref{thm:OptVecQuant}, one must specify the distribution of the observations, as we do in the following example:
\begin{example}
	\label{exm:IIDGtilde}
	\begin{subequations}
		\label{eqn:IIDGtilde}
	Consider the case where the \ac{mmse} estimate $\mySOIEst$ has i.i.d. proper-complex Gaussian entries with variance $\sigma_{\tilde{g}}^2$. Here, the excess average \ac{mse} of the optimal vector quantizer of Theorem \ref{thm:OptVecQuant} is
	\vspace{-0.1cm}
	\begin{align}
	\AsymDist \Opt \!-\! \AsymDist \MMSE\! &= \!\frac{1}{\lenStag} D_{G}\left(\frac{\lenXtag}{\lenStag} \Rate, \sigma_{\tilde{g}}^2 \myI_{\lenStag}, 1 \right) 
\!	\stackrel{(a)}{=}\! \sigma_{\tilde{g}}^2 2^{-\frac{\lenXtag}{\lenStag} \Rate},
	\label{eqn:example1}
		\vspace{-0.1cm}
	\end{align}
	where $D_G(\cdot)$ is defined in \eqref{eqn:exmGaussian}, and $(a)$ follows from the distortion-rate function of Gaussian \acp{rv} \cite[Ch. 23]{Polyanskiy:15}.
	Next, we compute the excess average \ac{mse} of a hardware-limited quantizer with analog combining ratio $\Ratio = \frac{\lenXtag}{\lenStag}$, namely, $\lenZq = 0$ and $\lenZn = \lenStag$. By noting that  $ \phicoeff_i^2 = \sigma_{\tilde{g}}^2 $ for each $i$, it follows from Corollary \ref{cor:OptimalCor} that 
		\vspace{-0.1cm}
	\begin{align}
	\AsymDist \ADC \!-\! \AsymDist \MMSE 
	&=   \frac{ \sigma_{\tilde{g}}^2}{\frac{3}{4 \MyKappa}\TilM^2  \!+\! 1} 
	 \stackrel{(a)}{=} \frac{ \sigma_{\tilde{g}}^2}{\frac{3}{4 \MyKappa}\lfloor 2^{-\frac{\lenXtag}{2\lenStag}\cdot \Rate}\rfloor ^2  \!+ \!1},
	 \label{eqn:example2}
	 	\vspace{-0.1cm}
	\end{align} 
	where $(a)$ holds as $\Ratio = \frac{\lenXtag}{\lenStag}$. 
	Note that \eqref{eqn:example1}-\eqref{eqn:example2} imply that as $\Rate$ increases, the ratio of the excess average \acp{mse} satisfies
		\vspace{-0.1cm}
	\begin{equation}
	\frac{\AsymDist \ADC - \AsymDist \MMSE}{\AsymDist \Opt - \AsymDist \MMSE } \approxeq \frac{4\MyKappa}{3} = \frac{4 \myEta^2}{3 - \frac{2\myEta^2}{\TilM^2}}.
	\label{eqn:example3}
		\vspace{-0.1cm}
	\end{equation} 
	\end{subequations}	
	As we assume that the quantized input is within the support and each scalar quantizer uses at least one bit, i.e., $\myEta \ge 2$ and $\TilM \ge 2$,  \eqref{eqn:example3} is strictly larger than one, as expected.
\end{example} 
Example \ref{exm:IIDGtilde} shows that, when the \ac{mmse} estimate has i.i.d. entries, the excess average \ac{mse} of hardware-limited quantization with large-scale inputs scales with respect to the quantization rate $\Rate$ proportionally to the optimal vector quantizer. This indicates that the proposed hardware-limited quantization system can approach the optimal performance with an average \ac{mse} gap that becomes negligible as $\AsymDist \Opt$ approaches the average \ac{mmse} $\AsymDist \MMSE$.   A similar relation to \eqref{eqn:example3} can be obtained for any distribution using the upper bound on the distortion-rate function in \cite[Eq. (6)]{Gibson:17}.

Although Example \ref{exm:IIDGtilde} focuses on the case where  the \ac{mmse} estimate has i.i.d. entries,
in the simulations study in Section \ref{sec:Simulations} we demonstrate that the hardware-limited system of Theorem \ref{thm:OptimalDes} can also approach the optimal \ac{mse} of  Theorem  \ref{thm:OptVecQuant} in massive \ac{mimo} channel estimation with quantized measurements, where the entries of the \ac{mmse} estimate are correlated. The application of our results to such setups is described in the following section.

\vspace{-0.2cm}
\section{Application: Massive \ac{mimo} Channel Estimation}
\label{sec:MIMO}
\vspace{-0.1cm}
An important application of our study on task-based quantization with large-scale inputs  is channel estimation in massive \ac{mimo} communications networks. 
Specifically, in massive \ac{mimo} systems, there is a strong need to operate with simple low-resolution quantizers, as increasing quantization rate results in a sharp increase in power consumption and memory usage. The problem of channel estimation from quantized measurements has received considerable attention, most notably in massive \ac{mimo} systems with large-scale inputs \cite{Mo:18,Li:17,Choi:16,Jacobsson:17}, but also for finite-scale inputs \cite{Zeitler:12,Dabeer:10,Stien:18}. 
As discussed in the introduction, previous works on massive \ac{mimo} channel estimation focus only on the digital processing, while hybrid architectures utilizing analog combiners were designed assuming \ac{csi}  \cite{Mo:17,Choi:17,Stein:17}. By applying the analysis of Section \ref{sec:MSE}, we are able to jointly optimize both the analog and the digital processing to improve the channel estimation performance under a given quantization rate constraint.

In the following we first present the massive \ac{mimo} system model in Subsection \ref{subsec:MIMO_Model}. Then, we discuss the fundamental limits of massive \ac{mimo} channel estimation without quantization in Subsection \ref{subsec:MIMO_NoQuant}. Finally, in Subsection \ref{subsec:MIMO_VecQuant} we show how the results of Section \ref{sec:MSE} can be applied to characterize the achievable performance and design the corresponding massive \ac{mimo} channel estimators.

\vspace{-0.2cm}
\subsection{Massive \ac{mimo} System Model}
\label{subsec:MIMO_Model}
\vspace{-0.1cm} 
We consider pilot-aided channel estimation in a multi-cell multi-user \ac{mimo} system with $\Ncells$ cells. 
In each cell, a \ac{bs} equipped with an array of equally-spaced $\Nantennas$ antennas serves $\Nusers$ single-antenna \acp{ut}. 
The antennas are not necessarily half-wavelength spaced, hence, the channel outputs can be spatially correlated. 
We focus on the {\em massive \ac{mimo} regime}, namely, the number of antennas $\Nantennas$ is sufficiently large to carry out large-scale (asymptotic) analysis.  

The massive \ac{mimo} channel follows a block-fading model \cite{Marzetta:10}. To formulate the model, let $\Dmat_{l,m}$ be a $\Nusers \times \Nusers$ diagonal matrix with positive diagonal entries $\{\dcoeff_{l,m,u} \}_{u=1}^{\Nusers}$ representing the attenuation between the $u$th \ac{ut} of the $m$th cell and the $l$th \ac{bs}, $l,m \in \{1,\ldots,\Ncells\} \triangleq\NcellsSet$. 
Without loss of generality, we assume that for each $l \in \NcellsSet$, the coefficients $\{\dcoeff_{l,l,u} \}_{u=1}^{\Nusers}$ are arranged in descending order. 
Furthermore, let $\Hmat_{l,m}\in\mySet{C}^{\Nantennas\times\Nusers}$ be a random proper-complex zero-mean Gaussian matrix with i.i.d. entries of unit variance, representing the instantaneous channel response between the \acp{ut} of the $m$th cell and the $l$th \ac{bs},  $l,m \in \NcellsSet$. 
For each $(l_1,m_1) \ne (l_2,m_2)$, $\Hmat_{l_1,m_1}$ and $\Hmat_{l_2,m_2}$ are mutually independent, and we assume a block-fading model for $\{\Hmat_{l,m} \}_{l,m \in \NcellsSet}$. 
To account for coupling induced by antenna spacing, we use $\CorrMat[l] \in \mySet{C}^{\Nantennas \times \Nantennas}$  to model the receive side correlation, i.e., $\big(\CorrMat[l] \big)_{k_1,k_2}$ represents the correlation between the antennas of indexes $k_1$ and $k_2$. Following conventional models for antenna coupling, e.g., Jakes model \cite{Jakes:93}, the fact that the antennas are equally-spaced implies that $\CorrMat[l]$ is a Toeplitz matrix with unit diagonal entries, and we  write $\Acorr{l}[k_1-k_2] = \big(\CorrMat[l] \big)_{k_1,k_2}$, and set $\Psd{l}(\cdot)$ to be the \ac{dtft} of $\Acorr{l}[\tau]$.  
The overall random channel matrix from the \acp{ut} in the $m$th cell to the $l$th \ac{bs} is given by $\Gmat_{l,m} =\CorrMat[l]^{1/2} \Hmat_{l,m} \Dmat_{l,m}$.  
Let $\myW_l[i]\in \mySet{C}^{\Nantennas}$, $l \in \NcellsSet$, be an i.i.d. zero-mean proper-complex Gaussian  signal  representing the additive channel noise at the $l$th \ac{bs}. Due to the antenna coupling at the \ac{bs}, the noise is also spatially correlated, and its covariance matrix is $\SigW \CorrMat[l]$, with $\SigW > 0$.  

\begin{figure}
	\centering
	\includefig{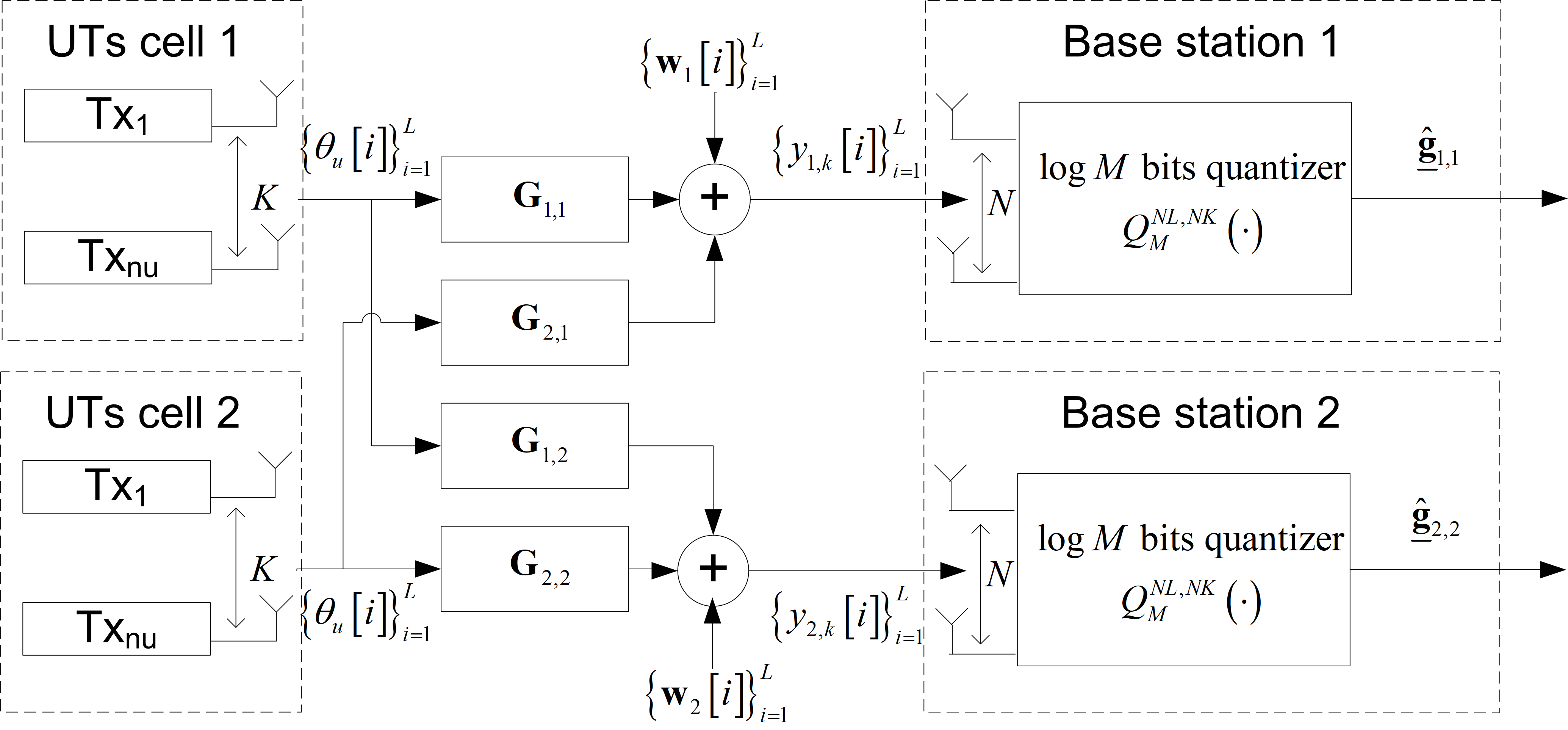}
	\caption{Massive MIMO channel estimation with $\Ncells = 2$ cells.}
	\vspace{-0.4cm}
	\label{fig:ChannelSetup1}
\end{figure}

Channel estimation is carried out in a \ac{tdd} fashion. Each \ac{ut} sends a deterministic orthogonal \ac{ps} consisting of $\Tpilots$ symbols, where the \acp{ps} are the same in all cells and known to the \acp{bs}. 
The \acp{bs} use the knowledge of the \acp{ps} to estimate the channel.
Let $\myTheta_u[i]$ be the $i$th pilot symbol of the $u$th user in each cell, $u \in \{1,\ldots,\Nusers\} \triangleq \NusersSet$, $i \in \{1,\ldots, \Tpilots\} \triangleq \TpilotsSet$. 
The channel output at the $k$th antenna of the $l$th \ac{bs} at time instance $i \in \TpilotsSet$  is 
\vspace{-0.1cm}
\begin{equation}
\label{eqn:Channel_Est1}
y_{l,k}[i] = \sum\limits_{m=1}^{\Ncells}\sum\limits_{u=1}^{\Nusers}\left( \Gmat_{l,m}\right)_{k,u} \myTheta_u[i] + \left( \myW_l[i]\right)_k.
\vspace{-0.1cm}
\end{equation} 
The orthogonality of the \acp{ps} implies that for all $l,m \in \NusersSet$,
$\sum\limits_{i=1}^{\Tpilots}\myTheta_l[i]\myTheta_m^*[i] = \Tpilots \cdot \delta_{m,k}$. Furthermore, the \ac{ps} length, $\Tpilots$, must not be smaller than the number of \acp{ut}, $\Nusers$ \cite[Sec. III-A]{Marzetta:10}.
Each \ac{bs} uses up to $\log M$ bits to represent the received signal $\{y_{l,k}[i] \}$, from which an estimate of the corresponding channel in vector  ${\Gvec}_{l,l} \triangleq {\rm vec}\left( {\Gmat}_{l,l}\right)$, denoted $\hat{\Gvec}_{l,l}$, is produced.
An illustration of the considered setup with $\Ncells = 2$ cells is depicted in Fig. \ref{fig:ChannelSetup1}.

Our goal is to derive the achievable average \ac{mse} in estimating the channel matrix at a given cell with index $l \in \NcellsSet$, and to characterize the corresponding quantization scheme.
\label{txt:Apriknow}
\textcolor{NewColor}{ 
 As common in the massive \ac{mimo} literature, see, e.g., \cite{Marzetta:10, Hoydis:13, Shlezinger:17}, we assume that the \ac{bs} knows:  $1)$ the pilot symbols; $2)$ the  channel input-output relationship, i.e., that the channel output are obtained from the \ac{ps} via \eqref{eqn:Channel_Est1}; and $3)$ the statistical model of the channel and the noise. This knowledge is utilized in the design of the quantization system to facilitate the estimation of each realization of the channel.} 
In our analysis, we fix the quantization rate, defined here as $\Rate \triangleq \frac{1}{\Nantennas \cdot \Tpilots} \log M$, and derive the achievable  \ac{mse} in the large number of antennas limit, 
$\AsymDist_l \triangleq \mathop{\lim}\limits_{\Nantennas \rightarrow \infty} \frac{1}{\Nantennas \cdot \Nusers} \E \{\|\Gvec_{l,l} - \hat{\Gvec}_{l,l} \|^2 \}$.

\vspace{-0.2cm}
\subsection{Achievable \ac{mse} without Quantization Constraints}
\label{subsec:MIMO_NoQuant}
\vspace{-0.1cm} 
As a preliminary step, we characterize   the average \ac{mse} without quantization, namely, the average \ac{mmse}. 
As stated in the previous subsection, the \acp{bs} use the orthogonal \acp{ps}   to produce the \ac{mmse} estimate of their corresponding channel responses.  
Define the $\Nantennas \times \Tpilots$ random matrices $\myYmat_l$ and $\myWmat_l $, such that $\left( \myYmat_l\right)_{k,i} = y_{l,k}[i]$ and $\left( \myWmat_l\right)_{k,i} = \left( \myW_l[i]\right)_k$, as well as the $\Nusers \times \Tpilots$ deterministic matrix  $\mySmat$ with entries $\left( \mySmat\right)_{u,i} = \myTheta_u[i]$. 
From \eqref{eqn:Channel_Est1} we have that for all $l \in \NcellsSet$: 
\vspace{-0.15cm}
\begin{equation}
\label{eqn:Channel_Training_MarForm}
\myYmat_l = \sum\limits_{m=1}^{\Ncells}\Gmat_{l,m} \mySmat + \myWmat_l,
\vspace{-0.2cm}
\end{equation}
or, alternatively, by writing $\myYvec_l \triangleq {\rm vec}(\myYmat_l)$, $\Gvec_{l,m} \triangleq {\rm vec}(\Gmat_{l,m})$, and $\myWvec_l  \triangleq {\rm vec}(\myWmat_l)$, \eqref{eqn:Channel_Training_MarForm} can be written as
\vspace{-0.15cm}
\begin{equation}
\label{eqn:Channel_Training_VecForm}
\myYvec_l = \sum\limits_{m=1}^{\Ncells}\left( \mySmat^T \otimes \myI_{\Nantennas}\right) \Gvec_{l,m}  + \myWvec_l.
\vspace{-0.2cm}
\end{equation}
Since the \acp{ps} are orthogonal it holds that $\mySmat \mySmat^H = \Tpilots \cdot \myI_{\Nusers}$. 
The covariance matrix of $\myYvec_l$ is given by $\CovMat{\myYvec_l} = \CovYtag\otimes \CorrMat[l]$, where
\vspace{-0.15cm}
\begin{equation}
\label{eqn:CovVecForm}
\CovYtag \triangleq  \sum\limits_{m=1}^{\Ncells} \mySmat^T \Dmat_{l,m}^2\mySmat^* + \SigW \myI_{\Tpilots}.
\vspace{-0.15cm}
\end{equation}

Next,  define the coefficients $\phicoeff_{l,u} \triangleq \sqrt{\bcoeff_{l,u}} \dcoeff_{l,l,u}$ where 
\vspace{-0.15cm}
\begin{equation}
\bcoeff_{l,u} \triangleq \frac{\Tpilots \dcoeff_{l,l,u}^2}{\SigW + \Tpilots\sum\limits_{m=1}^{\Ncells} \dcoeff_{l,m,u}^2 }, \quad l \in \NcellsSet, u \in \NusersSet,
\label{eqn:BmatDef}
\vspace{-0.1cm}
\end{equation}
as well as the $\Nusers \times \Nusers$ diagonal matrices  $\{\Phimat_{l}\}_{l \in \NcellsSet}$ and $\{\Bmat_{l}\}_{l \in \NcellsSet}$ with diagonal entries $\{\phicoeff_{l,u} \}_{u=1}^{\Nusers}$ and $\{\bcoeff_{l,u} \}_{u=1}^{\Nusers}$, respectively. The \ac{mmse} channel estimate  and its statistical characterization are stated in the following lemma:
\begin{lemma}
	\label{lem:ChEstLem}
	The \ac{mmse} estimate of $\tilde{\Gvec}_{l,l} \triangleq {\rm vec}\big( \tilde{\Gmat}_{l,l}\big)$ from $ \myYvec_l $   is given by
	\vspace{-0.1cm}
	\begin{equation}
	\label{eqn:ChEstLem1}
	\tilde{\Gvec}_{l,l} = \Tpilots^{-1}\left( \Bmat_{l} \mySmat^* \otimes \myI_{\Nantennas}\right)    \myYvec_l.
	\vspace{-0.1cm}
	\end{equation}
	Furthermore,  the vector form of the \ac{mmse} estimate $\tilde{\Gvec}_{l,l} \triangleq {\rm vec}\big( \tilde{\Gmat}_{l,l}\big)$  is a zero-mean $\Nantennas \cdot \Nusers \times 1$ Gaussian random vector with covariance matrix $\E\{\tilde{\Gvec}_{l,l}\tilde{\Gvec}_{l,l}^H\} = \left( \Phimat_{l}^2  \otimes \myI_{\Nantennas} \right)$.
\end{lemma}

\begin{IEEEproof}
The lemma follows from \cite[Lem. 1]{Shlezinger:17}, thus its proof is omitted for brevity.
\end{IEEEproof} 
%

\smallskip
Lemma \ref{lem:ChEstLem} can be used to obtain the average \ac{mmse} in the limit $\Nantennas \rightarrow \infty$, as stated in the following corollary:
\begin{corollary}
	\label{cor:mmse}
	The average \ac{mmse} in estimating ${\Gvec}_{l,l}$ is
	\vspace{-0.15cm}
	\begin{align}
	\AsymDist_l\MMSE 
	=  \frac{1}{\Nusers} \sum\limits_{u=1}^{\Nusers} \left( \dcoeff_{l,l,u}^2 - \phicoeff_{l,u}^2\right) . 
	\label{eqn:corMMMSE}
	\vspace{-0.15cm}
	\end{align}	
\end{corollary} 

\begin{IEEEproof}
	The corollary follows since the covariance matrix of ${\Gvec}_{l,l}$ {is $ \Dmat_{l,l}^2  \otimes \CorrMat[l] $. Thus, letting $\Nantennas \rightarrow \infty$, it follows from Szego's theorem \cite{Gray:06} combined with Lemma \ref{lem:ChEstLem} and the fact that $\Acorr{l}[0]=1$} that the asymptotic average \ac{mmse} is given by \eqref{eqn:corMMMSE}. 
\end{IEEEproof}

Having characterized the \ac{mmse} channel estimate for the massive \ac{mimo} setup without quantization, we are now ready to introduce quantization, and apply the results of Section \ref{sec:MSE}. 

\vspace{-0.2cm}
\subsection{Achievable \ac{mse}  with Quantized Channel Outputs}
\label{subsec:MIMO_VecQuant}
\vspace{-0.1cm} 
We now show how Theorems \ref{thm:OptVecQuant}-\ref{thm:OptimalDes} can be used to characterize the achievable average \ac{mse} for massive \ac{mimo} channel estimation with quantization constraints. 
 
To see that the massive \ac{mimo} system model detailed in Subsection \ref{subsec:MIMO_Model} is a special case of the general model described in Subsection \ref{subsec:Pre_Problem}, we note that  by writing $\myObstag_i = \left[ y_{l,i}[1] , \ldots,  y_{l,i}[\Tpilots]  \right]^T$, it holds that the set $\{\myObstag_i\}_{i=1}^{\Nantennas}$ {consists of  $\Tpilots \times 1$ zero-mean Gaussian random vectors with autocorrelation $\E\{\myObstag_{i_1}\myObstag_{i_2}^H\} =\CovYtag\Acorr{l}[i_1-i_2]$. Similarly, by letting $\mySOItag_i$ be the $i$th row of $\Gmat_{l,l}$, it holds that  $\{\mySOItag_i\}_{i=1}^{\Nantennas}$ are $\Nusers\times 1$ zero-mean Gaussian random vectors with autocorrelation $\E\{\mySOItag_{i_1}\mySOItag_{i_2}^H\} =\Dmat_{l,l}^2\Acorr{l}[i_1-i_2]$. Finally, by Lemma \ref{lem:ChEstLem} it holds that the \ac{mmse} estimate of $\Gmat_{l,l}$ from the channel output $\myYvec_l$ is given by the set of  \ac{mmse} estimates of $\mySOItag_i$ from $\myObstag_i$ for each $i \in \{1,\ldots,\Nantennas\}$,} which can be written as $\mySOIEsttag_i = \LmmseMattag \myObstag_i$ with $\LmmseMattag = \Tpilots^{-1} \Bmat_{l}\mySmat^* $. We thus conclude that the massive \ac{mimo} channel estimation setup is a special case of the general problem formulation stated in Subsection \ref{subsec:Pre_Problem}.

%
%
In the following, we first show how Theorems \ref{thm:OptVecQuant}-\ref{thm:TIVecQuant} characterize the achievable average \ac{mse} when the \ac{bs} uses vector quantizers. 
Then, we apply Theorem \ref{thm:OptimalDes} to obtain the minimal achievable average \ac{mse} when the \ac{bs} uses hardware-limited quantizers. Finally, we note that in massive \ac{mimo} systems, the \ac{bs} may be able to linearly combine only channel outputs received at the same time instance. By incorporating this constraint into the structure hardware-limited systems, we derive the minimal achievable average \ac{mse} and the resulting quantization system for this form of restricted hardware-limited quantization.

%

\subsubsection{Vector Quantization} 
In Subsection \ref{subsec:Pre_Problem} we discussed two vector quantization systems: the optimal vector quantizer, which is designed to recover the unknown channel $\Gvec_{l,l}$, and the task-ignorant vector quantizer, which represents the observed signal $\myYvec_l$ separately from the task of estimating the channel. 

Applying Theorem \ref{thm:OptVecQuant}, we obtain the minimal achievable average \ac{mse} of any quantization system operating with quantization rate $\Rate$, as stated in the following proposition:
\begin{proposition}
	\label{cor:OptVecQuant}
	The average \ac{mse} of the optimal vector quantizer for massive \ac{mimo} channel estimation is given by
	\begin{equation}
	\label{eqn:OptVecQuant1}
	\AsymDist_l\Opt = \AsymDist_l\MMSE + \frac{1}{\Nusers}\DistG\left(\frac{\Tpilots}{\Nusers}\cdot \Rate, \Phimat_{l}^2, 1\right), 
	\end{equation}
	where $\DistG(\cdot)$ is defined in \eqref{eqn:exmGaussian1}.
\end{proposition}
\begin{IEEEproof}
	The proposition follows directly from Theorem \ref{thm:OptVecQuant} by noting that in the limit $\Nantennas \rightarrow \infty$, the \ac{mmse} estimate $\mySOIEst_{l,l}$ can be represented as an $\Tpilots \times 1$ Gaussian source with  multivariate \ac{psd} $\PSD{\mySOIEst}(\omega)=\Phimat_{l}^2$ for each $\omega \in [0,2\pi]$ by Lemma \ref{lem:ChEstLem}.
\end{IEEEproof}	

\smallskip
Using Theorem \ref{thm:TIVecQuant}, we characterize the achievable average \ac{mse} with vector quantization carried out separately from the task  {for the case when $\{\myObstag_i\}$ are i.i.d., namely, $\Acorr{l}[\tau] = \delta_{\tau}$.} This is stated in the following proposition:
\begin{proposition}
	\label{cor:TIVecQuant}
	 {When $\Acorr{l}[\tau] = \delta_{\tau}$,} the average \ac{mse} of the task ignorant vector quantizer for massive \ac{mimo} channel estimation is given by
	\begin{equation}
	\label{eqn:TIVecQuant2}
	\hspace{-0.2cm}
	\AsymDist_l\Ign \!=\! \AsymDist_l\MMSE \!+\! \frac{1}{\Nusers \!\cdot\! \Tpilots^2} {\rm Tr}\left( \!\mySmat^T\!\Bmat_{l}^2\mySmat^* \!\left( \CovYtag \!-\! \CovMat{\myY_l, G}(\Rate)  \right) \! \right), 
	\end{equation}
	where $\CovYtag$ is defined in \eqref{eqn:CovVecForm}, and $\CovMat{\myY_l, G}(\Rate)$ is the covariance matrix of the optimal marginal distribution which achieves the distortion-rate function $\DistG\big(\Rate, \CovYtag, 1 \big)$, defined in \eqref{eqn:exmGaussian1}.
\end{proposition}
\begin{IEEEproof}
	The proposition is a result of  Theorem \ref{thm:TIVecQuant}, obtained by substituting $\LmmseMattag = \Tpilots^{-1} \Bmat_{l}\mySmat^* $ in \eqref{eqn:TIVecQuant}, as  $\{\myObstag_i\}$ are i.i.d.  Gaussian with covariance matrix $\CovYtag$. Therefore, \eqref{eqn:TIVecQuant} becomes
\ifsingle		
	\begin{align}
	\AsymDist_l\Ign &= \AsymDist_l\MMSE + \frac{1}{\Nusers}  {\rm Tr}\left( \left( \Tpilots^{-1} \Bmat_{l}\mySmat^* \right)^H \left(\Tpilots^{-1} \Bmat_{l}\mySmat^*  \right) \left( \CovMat{\myObstag} - \CovMat{\myObstag, D}(\Rate)  \right)  \right) \notag \\
	&= \AsymDist_l\MMSE + \frac{1}{\Nusers \cdot \Tpilots^2}  {\rm Tr}\left(  \mySmat^T \Bmat_{l}^H  \Bmat_{l}\mySmat^*  \left( \CovYtag - \CovMat{\myYvec_l', G}(\Rate)  \right)   \right).
	\label{eqn:TIVecQuant2Proof}
	\end{align}
\else
		\begin{align}
	&\AsymDist_l\Ign = \AsymDist_l\MMSE  \notag \\
	&\qquad  + \frac{1}{\Nusers}  {\rm Tr}\left( \left( \Tpilots^{-1} \Bmat_{l}\mySmat^* \right)^H \left(\Tpilots^{-1} \Bmat_{l}\mySmat^*  \right) \left( \CovMat{\myObstag} \!-\! \CovMat{\myObstag, D}(\Rate)  \right)  \right) \notag \\
	&\stackrel{(a)}{=} \AsymDist_l\MMSE 
	\!+\! \frac{1}{\Nusers\! \cdot\! \Tpilots^2}  {\rm Tr}\left(  \mySmat^T \Bmat_{l}^2 \mySmat^*\!  \left( \CovYtag \!-\! \CovMat{\myYvec_l', G}(\Rate)  \right)\!   \right),
	\label{eqn:TIVecQuant2Proof}
\end{align}
\fi
	where $(a)$ holds since $\Bmat_{l}$ is diagonal  with non-negative diagonal entries.
\end{IEEEproof}	
Note that since $\myYvec_l$ is Gaussian, $\CovMat{\myY_l, G}(\Rate) $ can be obtained using the inverse waterfilling algorithm \cite[Ch. 10.3]{Cover:06}.

\subsubsection{Hardware-Limited Quantization} 
\begin{figure}
	\centering
	{\includefig{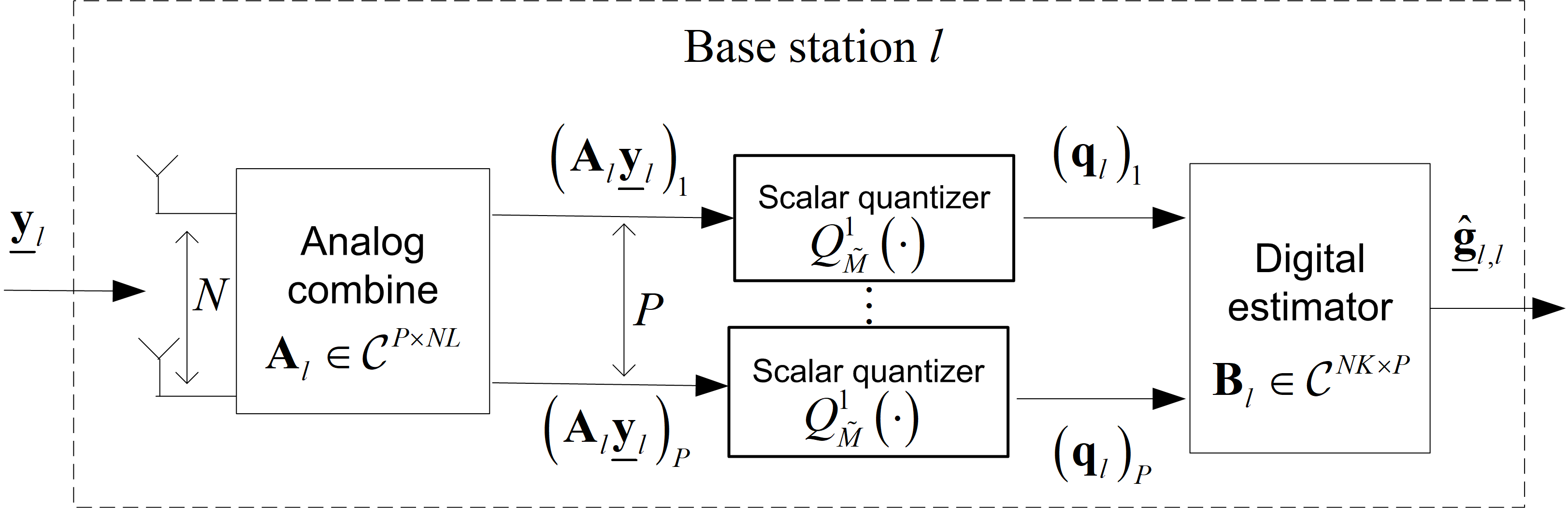}}
	\caption{Massive \ac{mimo} channel estimation with scalar \acp{adc}.}
	\label{fig:GenSetup2}		
	\vspace{-0.6cm}
\end{figure}

Utilizing vector quantization in massive \ac{mimo} systems is likely to be infeasible due to its extremely high complexity for large-scale inputs. It is thus desirable to utilize serial scalar uniform \acp{adc}, corresponding to the hardware-limited quantization setup described in Subsection \ref{subsec:Pre_Problem}. 
\label{txt:Complexity}
\textcolor{NewColor}
{Here, the linear mapping carried out in the analog domain can be implemented using a fully connected network with complex gains, as considered in \cite{Karamalis:06, Nsenga:11,Gong:19}. In some cases, networks with controllable gains may be complex to implement, and more restricted linear structures are desirable. Constrained analog combiners can represent common practical architectures such as  phase shifter networks \cite{Rial:16}, antenna selection structures \cite{Choi:18}, discrete cosine beamforming \cite{Kim:15}, and Lorentzian constrained phase combiners, which are encountered when using metasurface antennas  \cite{Shlezinger:DMA}.  For such scenarios, our analysis constitutes a lower bound on the achievable \ac{mse}, and can be used to facilitate the design of restricted analog combiners by approximating the resulting complex gain combiner matrix using a feasible structure, see, e.g., \cite{Cuba:17, Stein:17,Shlezinger:DMA}.} 
 An illustration of a receiver, representing the $l$th \ac{bs} in a massive \ac{mimo} network, applying channel estimation with hardware-limited quantization is depicted in Fig. \ref{fig:GenSetup2}.

We note that by setting the analog combining matrix $\myA_l$ to be the identity matrix, the resulting system specializes the standard model for \ac{mimo} channel estimation with quantized measurements, as in, e.g., \cite{Li:17,Choi:16,Jacobsson:17}. 
Consequently, the ability to jointly optimize the analog combining, which represents the linear processing of $\myYvec_l$ carried out in analog, along with the setting of the support and the digital processing, is the main difference between task-based quantization and previously proposed quantizers. In Section \ref{sec:Simulations} we numerically illustrate that jointly designing the quantization system components significantly improves the estimation accuracy over previously proposed schemes, and that the resulting hardware-limited system can  approach the optimal performance achievable with vector quantizers.  

Using Theorem \ref{thm:OptimalDes}, we next characterize  the minimal achievable average \ac{mse} in estimating massive \ac{mimo} channels using hardware-limited quantizers. 
To that aim, let $\{\eigT{l,u}\}$ be the singular values of $\Tpilots^{-1} \Bmat_{l}\mySmat^* \CovYtag^{1/2} \otimes \CorrMat[l]$ arranged in descending order.
The resulting optimal hardware-limited quantization system for a fixed quantization rate $\Rate$ and  analog combining ratio $\Ratio$, is stated in the following proposition: 
\begin{proposition}
	\label{cor:OptimalDes}
	\begin{subequations}
		\label{eqn:MOptimalDes}
			In the hardware-limited quantization system which minimizes the average \ac{mse}, 		
		the analog combining matrix $\myA_l\op$ is given by
		\ifsingle	
		\begin{equation*}
		\myA_l\op = \myMat{U}_{\myA} \myMat{\Lambda}_{\myA} \left( \myMat{V}_{\myA}^H \CovYtag^{-1/2} \otimes \CorrMat[l]^{-1/2}\right) ,
		\end{equation*}
		\else	
		$\myA_l\op = \myMat{U}_{\myA} \myMat{\Lambda}_{\myA} \left( \myMat{V}_{\myA}^H \CovYtag^{-1/2} \otimes \CorrMat[l]^{-1/2}\right) $,
		\fi		 
		where
		\begin{itemize}
			\item $\myMat{V}_{\myA} \in \mySet{C}^{\Tpilots \times \Tpilots}$ is the right singular vectors matrix of  $\Tpilots^{-1} \Bmat_{l}\mySmat^* \CovYtag^{1/2}$.
			\item  $\myMat{\Lambda}_{\myA} \in \mySet{C}^{\lenZ \times \Tpilots\cdot\Nantennas}$ is a diagonal matrix  with diagonal entries  
			\begin{equation}
			\label{eqn:MOptimalDesA}
			\left( \myMat{\Lambda}_{\myA}\right)_{u,u}^2 = \frac{4 \MyKappa}{{3\TilM^2} \cdot \Ratio} 
			\Plevel(\Wlevel\cdot \eigT{l,u}), 
			\end{equation}
			where $\Wlevel$ is set such that $\frac{4 \MyKappa}{{3\TilM^2} \cdot \lenZ} \sum\limits_{u=1}^{\lenZ}\Plevel(\Wlevel \cdot \eigT{l,u}) = 1$. 
			\item $\myMat{U}_{\myA} \in \mySet{C}^{\lenZ \times \lenZ}$ is a unitary matrix which guarantees that  $\myMat{U}_{\myA}\myMat{\Lambda}_{\myA}\myMat{\Lambda}_{\myA}^H\myMat{U}_{\myA}^H$ has identical diagonal entries. 
		\end{itemize}
		The support of the \ac{adc} is given by $	\DynRange^2   =  \frac{ \MyKappa}{\Ratio}$,
		and the digital processing matrix is  
		\ifsingle		
		\begin{equation}
		\label{eqn:MOptimalDesB}
		\myB_l\op  =  \left( \Dmat_{l,l}^2 \mySmat^* \otimes \CorrMat[l]\right)    \left( \myA_l\op\right) ^H\left( \myA_l\op \left( \CovYtag \otimes \CorrMat[l]\right)  \left( \myA_l\op\right) ^H + \frac{{4{\DynRange^2}}}{{3\TilM^2}}{\myI_\lenZ} \right)^{ - 1}.
		\end{equation}
		\else
		\vspace{-0.15cm}
		\begin{align}
		\myB_l\op  &=  \left( \Dmat_{l,l}^2 \mySmat^* \otimes \CorrMat[l]\right)    \left( \myA_l\op\right) ^H \notag \\
		&\qquad\times \left( \myA_l\op \left( \CovYtag \otimes \CorrMat[l]\right)  \left( \myA_l\op\right) ^H + \frac{{4{\DynRange^2}}}{{3\TilM^2}}{\myI_\lenZ} \right)^{ - 1}.
		\label{eqn:MOptimalDesB}
		\vspace{-0.15cm}
		\end{align}
		\fi 
		The corresponding achievable average \ac{mse} in the limit $\Nantennas \rightarrow \infty$ when $\lenZn \ge {\rm rank}\left( \Phimat_{l}\right)$  is given by
		\begin{align}
&\AsymDist_l \ADC
\!=\!  \AsymDist \MMSE \!+ \! \frac{1}{2\pi\lenStag}\int_{0}^{2\pi}\!  \sum\limits_{u=1}^{\lenStag }   \frac{  \phicoeff_{l,u}^2{\Psd{l}(\omega)}  } { \Glevel(\Wlevel \phicoeff_{l,u}\sqrt{\!\Psd{l}(\omega)}) \! + \! 1} d\omega.
\vspace{-0.2cm}
\label{eqn:MOptimalMSE2}
\end{align}		
	Furthermore, when $\Acorr{l}[\tau]=\delta_{\tau}$, the asymptotic average \ac{mse} for each $\lenZn\ge 0$ is given by	
\ifsingle		
		\begin{align}
		\AsymDist_l \ADC &= \AsymDist \MMSE + \frac{1}{\Nusers}\sum\limits_{u=1}^{\min(\Nusers, \lenZn)} \frac{\phicoeff_{l,u}^2}{\Plevel(\Wlevel\cdot \phicoeff_{l,u})  +1} \notag \\
		&+\frac{ \delta_{(\lenZn < \Nusers)}}{\Nusers}\left(\sum\limits_{u=\lenZn + 1}^{\Nusers} {\phicoeff_{l,u}^2}  - \left(\Ratio \cdot \Tpilots - \lenZn \right)\frac{\Plevel(\Wlevel\cdot \phicoeff_{l,\lenZn + 1 }) \phicoeff_{l,\lenZn + 1 }^2}{\Plevel(\Wlevel\cdot \phicoeff_{l,\lenZn + 1 }) +1}   \right).
		\label{eqn:MOptimalMSE}
		\end{align}
\else
\vspace{-0.15cm}
		\begin{align}
		&\AsymDist_l \ADC = \AsymDist \MMSE + \frac{1}{\Nusers}\sum\limits_{u=1}^{\min(\Nusers, \lenZn)} \frac{\phicoeff_{l,u}^2}{\Plevel(\Wlevel\cdot \phicoeff_{l,u}) +1} + \frac{\delta_{(\lenZn < \Nusers)}}{\Nusers}\notag \\
		&\times\Bigg(\sum\limits_{u=\lenZn \! + \! 1}^{\Nusers} {\phicoeff_{l,u}^2} \! -\! \left(\Ratio  \Tpilots \! - \! \lenZn \right)  \frac{\Plevel(\Wlevel \phicoeff_{l,\lenZn + 1 }) \phicoeff_{l,\lenZn \! + \! 1 }^2}{\Plevel(\Wlevel \phicoeff_{l,\lenZn + 1 }) \! + \!1}   \Bigg).
		\label{eqn:MOptimalMSE}
		\vspace{-0.15cm}
		\end{align}
\fi 
	\end{subequations}
\end{proposition}

\begin{IEEEproof}
	The proposition is a result of Theorem \ref{thm:OptimalDes}. In particular, here $\LmmseMattag\CovMat{\myObstag} \LmmseMattag^H = \Phimat_{l}^2$.  Setting this in Theorem \ref{thm:OptimalDes} proves \eqref{eqn:MOptimalDesA}, \eqref{eqn:MOptimalMSE2}, and \eqref{eqn:MOptimalMSE}. 
	Finally, \eqref{eqn:MOptimalDesB} is obtained from \eqref{eqn:OptimalDesB} by noting that for the massive \ac{mimo} setup, 
	\vspace{-0.15cm}
	\begin{align}
	&\LmmseMattag \CovMat{\myObstag} 
	= \Tpilots^{-1} \Bmat_{l}\mySmat^*\left( \sum\limits_{m=1}^{\Ncells} \mySmat^T \Dmat_{l,m}^2\mySmat^*  \! +\! \SigW \myI_{ \Tpilots}\right) \notag \\
	&\stackrel{(a)}{=}  \Tpilots^{-1} \Bmat_{l}\left(\Tpilots \sum\limits_{m=1}^{\Ncells}   \Dmat_{l,m}^2 \! +\! \SigW \myI_{ \Nusers} \right) \mySmat^* 
	\stackrel{(b)}{=} 	\Dmat_{l,l}^2 \mySmat^*, 	
	 \label{eqn:LMMSECov}
	 	\vspace{-0.15cm}
	\end{align}  
	where $(a)$ follows since $\mySmat \mySmat^H = \Tpilots \cdot \myI_{\Nusers}$, and $(b)$ follows from the definition of $\Bmat_{l}$ in \eqref{eqn:BmatDef}. 
\end{IEEEproof}	

\smallskip
We note that the matrix $\myA_l\op$ in Proposition \ref{cor:OptimalDes} linearly combines the vector $\myYvec_l$, which represents the channel outputs received over the entire channel estimation period. Thus, $\myA_l\op$  can linearly combine samples taken from different antennas, i.e., spatial combining, and at different time instances, i.e., temporal combining. While spatial combining can be implemented using simple hardware, see, e.g., \cite{Stein:17}, temporal combining requires storing samples for different durations in analog, which may be difficult when the number of training symbols $\Tpilots$ is large. 
 Consequently, we next characterize the optimal system   when  $\myA_l$ is restricted to implement only spatial combining. 


\subsubsection{Spatial Analog Combining} 
\label{subsec:MSE_SpaAnalog} 
In Proposition \ref{cor:OptimalDes} we characterized the achievable average \ac{mse} when the input to the scalar \acp{adc} can be written as any linear transformation of all the channel outputs, $\myYvec_l$. Consequently, we allowed  samples from different time instances and different receive antennas to be jointly combined. 
In fact, it follows from Corollary \ref{cor:OptimalCor} that if $\lenZ$ is an integer multiple of $\Nantennas$  {and the channel outputs are spatially uncorrelated, i.e., $\lenZ = \lenZn \cdot \Nantennas$ and $\Acorr{l}[\tau] = \delta_{\tau}$,} then the optimal analog combining matrix is $\myA_l\op = \myA_l' \otimes \myI_{\Nantennas}$, for some $\myA_l' \in \mySet{C}^{\lenZn \times \Tpilots}$. Namely, the optimal matrix $\myA_l\op$ implements {\em only temporal combining}, and does not utilize spatial combining. 
Since in some cases it may be preferable not to combine samples received at different time instances in the analog domain to avoid the need to store data in analog, in the following we restrict the analog combining matrix to operate only on samples received at the same time instance. It should be noted that this is the model used in previous works on analog combining design for \ac{mimo} systems \cite{Mo:17, Choi:17,Stein:17}, which assumed full \ac{csi} and fixed quantizers. 

\label{txt:Spatial1}
To formulate the resulting setup,  {we use $\lenZT$ to denote the number of samples quantized at each time instance, i.e., the number of RF chains, and} let $\myAT_l  \in \mySet{C}^{\Nantennas \times \lenZT}$ represent the analog combining, applied to each received channel output. Here, at each time index $i \in \TpilotsSet$, the vector $\myAT_l \myY_l[i]$ is quantized using $\lenZT$ identical scalar quantizers. As the overall number of quantization levels is fixed to $M$, each scalar quantizer has resolution $\TilM = \lfloor M^{1/(2\Tpilots\cdot \lenZT)} \rfloor$. 

The considered setup is a special case of the model illustrated in Fig. \ref{fig:GenSetup2}, with analog combining matrix $\myA_l = \myI_{\Tpilots} \otimes \myAT_l$ and $\lenZ = \lenZT \cdot \Tpilots$. The analog combining ratio is thus $\Ratio = \frac{\lenZ}{\Tpilots \cdot \Nantennas} = \frac{\lenZT}{\Nantennas} $. Since  $\Ratio$ is fixed and positive, letting $\Nantennas$ grow arbitrarily large implies that $\lenZT$ grows proportionally.
Let $\maxDiag$ be the maximal diagonal entry of $\CovYtag$, namely, $\maxDiag \triangleq \mathop{\max}\limits_{i=1,\ldots,\Tpilots} \big(\CovYtag\big)_{i,i}$. 
 Under this setting, the optimal system and the corresponding average \ac{mse} are stated in the following proposition:
\begin{proposition}
	\label{thm:SpatialDes}
	\begin{subequations}
		\label{eqn:SpatialDes}
		In the hardware-limited quantization system with spatial analog combining which minimizes the average \ac{mse}, the analog combining matrix $\myAT_l$ is given by  $\myAT_l =  \myMat{U}_{\myAT} \myMat{\Lambda}_{\myAT}\myMat{V}_{\myAT}^H\CorrMat[l]^{-1/2}$, where  $\myMat{U}_{\myAT}$ guarantees that $ \myMat{U}_{\myAT} \myMat{\Lambda}_{\myAT} \myMat{\Lambda}_{\myAT}^H \myMat{U}_{\myAT}^H$ has identical diagonal entires \cite[Alg. 2.2]{Palomar:07}; 
		 $\myMat{V}_{\myAT}^H$ is the eigenmatrix of $\CorrMat[l]$; 
		 and $\myMat{\Lambda}_{\myAT} $ is diagonal with diagonal entries  $\{\bar{a}_i\}$, which are the solution to the convex optimization problem:
\vspace{-0.2cm}
\begin{align}
\label{eqn:AnalogProblemSpat2}
 \{\bar{a}_i\}_{i=1}^{\lenZT } =& \mathop{\arg\max}\limits_{\{a_i\}_{i=1}^{\lenZT }} \sum\limits_{i=1}^{\lenZT} \sum\limits_{u=1}^{\Nusers} \frac{\Tpilots \cdot\phicoeff_{l,u}^4 \cdot a_i^2 \cdot \eig{\CorrMat[l],i}}{ \Tpilots \cdot\phicoeff_{l,u}^2  \cdot a_i^2 + \bcoeff_{l,u}^2 }
\\
& {\text{subject to }}\frac{{4{\MyKappa\cdot \maxDiag}}}{{3\TilM^2 \cdot \lenZT}} \sum\limits_{i=1}^{\lenZT} a_i^2= ,. \notag 
\vspace{-0.2cm}
\end{align} 
where $\eig{\CorrMat[l],i}$ is the $i$-th largest eigenvalue of $\CorrMat[l]$.	
	The support of the \ac{adc} is $	\DynRange^2   =   \frac{3\TilM^2}{4}$,
	and the digital processing matrix is  
	\ifsingle		
	\begin{equation}
	\label{eqn:SpatialDesB}
	\myBT_l\op  =  \left( \Dmat_{l,l}^2 \mySmat^* \otimes\CorrMat[l]\myAT_l ^H\right)    \left( \left( \CovYtag \otimes \myAT_l\CorrMat[l]\myAT_l ^H\right) +   \myI_{\Tpilots\cdot \lenZT} \right)^{ - 1}.
	\end{equation}
	The corresponding achievable average \ac{mse} in the limit $\Nantennas \rightarrow \infty$ is given by
\begin{align}
\AsymDist_l\sADC &= \AsymDist_l\MMSE + \frac{1}{\Nusers}\sum\limits_{u=1}^{\Nusers}\phicoeff_{l,u}^2  -   \frac{\Ratio}{\Nusers} \sum\limits_{u=1}^{\Nusers}  \mathop{\lim}\limits_{\lenZT \rightarrow\infty}\frac{1}{\lenZT} \sum\limits_{i=1}^{\lenZT}\frac{\Tpilots \cdot\phicoeff_{l,u}^4 \cdot \bar{a}_i^2 \cdot \eig{\CorrMat[l],i}}{ \Tpilots \cdot\phicoeff_{l,u}^2  \cdot \bar{a}_i + \bcoeff_{l,u}^2 }.
\label{eqn:SpatialMSE2}
\end{align}		
	\else
	\vspace{-0.1cm}
	\begin{align}
	\hspace{-0.2cm}\myBT_l\op \! = \! \left(\Dmat_{l,l}^2 \mySmat^*\! \!\otimes \! \CorrMat[l]\myAT_l^H\!\right) \! 
	\left( \!\left( \!\CovYtag\! \otimes\! \myAT_l\CorrMat[l] \myAT_l^H\!\right)\! +\!  \frac{{4{\DynRange^2}}}{{3\TilM^2}} \myI_{\Tpilots \lenZT} \right)^{\! -\! 1}\!\!.
	\label{eqn:SpatialDesB}
	\vspace{-0.1cm}		
	\end{align}	
	The corresponding achievable average \ac{mse} in the limit $\Nantennas \rightarrow \infty$ is given by
\begin{align}
\AsymDist_l\sADC &= \AsymDist_l\MMSE + \frac{1}{\Nusers}\sum\limits_{u=1}^{\Nusers}\phicoeff_{l,u}^2 \notag \\
&  -   \frac{\Ratio}{\Nusers} \sum\limits_{u=1}^{\Nusers}  \mathop{\lim}\limits_{\lenZT \rightarrow\infty}\frac{1}{\lenZT} \sum\limits_{i=1}^{\lenZT}\frac{\Tpilots \cdot\phicoeff_{l,u}^4 \cdot \bar{a}_i^2 \cdot \eig{\CorrMat[l],i}}{ \Tpilots \cdot\phicoeff_{l,u}^2  \cdot \bar{a}_i^2 + \bcoeff_{l,u}^2 }.
\label{eqn:SpatialMSE2}
\end{align}		
\fi 
	\end{subequations}
\end{proposition}  

{\em Proof:}
See Appendix \ref{app:ProofThmSpaDes}.

\smallskip
  {
The asymptotic average \ac{mse} in \eqref{eqn:SpatialMSE2} can be numerically evaluated by considering a large fixed value of $\Nantennas$, for which the set $\{\bar{a}_i\}_{i=1}^{\lenZT }$ can be computed by solving the concave optimization problem in \eqref{eqn:AnalogProblemSpat2}. When the \ac{bs} antennas are not coupled, i.e., $\Acorr{l}[\tau] = \delta_{\tau}$,  \eqref{eqn:SpatialMSE2} can be obtained in closed-form,} as stated in the following corollary:
\begin{corollary}
	\label{cor:SpatialIID}
	  {When $\Acorr{l}[\tau] = \delta_{\tau}$,} the asymptotic achievable average \ac{mse} using spatial analog combining is given by 
\ifsingle			
		\begin{align}
		\AsymDist_l\sADC &= \AsymDist_l\MMSE + \frac{1}{\Nusers}\sum\limits_{u=1}^{\Nusers}\phicoeff_{l,u}^2  -   \frac{\Ratio}{\Nusers} \sum\limits_{u=1}^{\Nusers}  \frac{ \phicoeff_{l,u}^4  }{  \phicoeff_{l,u}^2    + \frac{{4{\MyKappa\cdot \maxDiag}}}{{3\TilM^2 \cdot \Tpilots }} \cdot\bcoeff_{l,u}^2 }.
		\label{eqn:SpatialMSE}
		\end{align}		
\else
\vspace{-0.15cm}
		\begin{align}
		\hspace{-0.2cm}
		\AsymDist_l\sADC \!&=\! \AsymDist_l\MMSE\! +\! \frac{1}{\Nusers}\sum\limits_{u=1}^{\Nusers}\bigg( \phicoeff_{l,u}^2  
		\!-\! \frac{ \Ratio \cdot \phicoeff_{l,u}^4  }{  \phicoeff_{l,u}^2 \!   +\! \frac{{4{\MyKappa\cdot \maxDiag}}}{{3\TilM^2 \cdot \Tpilots }} \cdot\bcoeff_{l,u}^2 }\bigg) .
		\label{eqn:SpatialMSE}
\vspace{-0.15cm}		
		\end{align}		
\fi 
\end{corollary}
 
\begin{IEEEproof}
	For  $\Acorr{l}[\tau] = \delta_{\tau}$ it holds that $\eigT{l,i} = 1$ for each $i$. Thus, as the mapping   $\xi(x) \triangleq  \sum\limits_{u=1}^{\Nusers} \frac{\Tpilots \cdot\phicoeff_{l,u}^4 \cdot x}{ \Tpilots \cdot\phicoeff_{l,u}^2  \cdot x + \bcoeff_{l,u}^2 }$ is concave \cite[3.2.1]{Boyd:04}, we have 
	\begin{align}
	\frac{1}{\lenZT}\sum\limits_{i=1}^{\lenZT} \xi(a_i) 
	&\le \xi\left( \frac{1}{\lenZT}\sum\limits_{i=1}^{\lenZT} a_i\right) = \xi\left(\frac{{3\TilM[\lenZT \cdot \Tpilots]^2 }}{{4{\MyKappa[\lenZT \cdot \Tpilots]\cdot \maxDiag}}} \right),
	\label{eqn:OptSol} 
	\end{align}
	so that setting $a_i = \frac{{3\TilM^2 }}{{4{\MyKappa\cdot \maxDiag}}}$ maximizes \eqref{eqn:AnalogProblemSpat2}. Substituting  into Proposition \ref{thm:SpatialDes} proves the corollary.
\end{IEEEproof} 

\smallskip
The channel output model in \eqref{eqn:Channel_Est1} implies that, when $\Acorr{l}[\tau] = \delta_{\tau}$, the channel outputs received at different antennas for each time instance $i \in \TpilotsSet$, $\{y_{l,k}[i]\}_{k=1}^{\Nantennas}$, are i.i.d.. Therefore, intuitively, combining $\{y_{l,k}[i]\}_{k=1}^{\Nantennas}$ into a smaller set may result in an inaccurate estimation. This is also demonstrated in the numerical study in Subsection \ref{subsec:Sim_SelP}, where it is shown that  when the antennas are uncorrelated, the proposed quantizer performs better with increased analog combining ratio $\Ratio$ (unlike the hardware-limited quantizer with general analog combining, which, as noted in Corollary \ref{cor:OptimalP}, performs best when $\Ratio \le \frac{\Nusers}{\Tpilots}$).  
Furthermore, it follows from the proof of Corollary \ref{cor:SpatialIID} that for uncorrelated antennas, the optimal analog spatial combining matrix ${\myAT}_l$  multiplies each input by a constant, whose purpose is to guarantee that the quantized entries are within the support of the uniform scalar quantizers. This combining is different from conventional hybrid beamforming, which is typically designed assuming full \ac{csi}  to better capture the energy of the transmitted signal \cite{Mo:17, Stein:17}. Consequently, when the channel outputs are not spatially correlated and the quantization system cannot combine samples received at different time instances in the analog domain, most of the performance gain is a result of the processing in the digital domain. This insight is in agreement with a similar conclusion  in \cite{Cuba:17}, which considered only spatial analog combining. 

Finally, we note that even though the quantizer of  Corollary~\ref{cor:SpatialIID} may not reduce the dimensionality of the quantized signal, it does not operate only in digital, as it sets the support based on the statistics of the input. 
Unlike previous channel estimators for massive \ac{mimo} with quantized channel outputs, e.g., \cite{Li:17,Mo:18,Jacobsson:17}, which operated only in the digital domain,  the proposed quantizer reduces the quantization error by properly setting the support and scaling the channel output. 




\vspace{-0.2cm}
\section{Numerical Results and Discussion}
\label{sec:Simulations}
\vspace{-0.1cm}
In this section we numerically evaluate the performance of the quantization systems discussed in Section \ref{sec:MIMO} for  massive \ac{mimo} channel estimation. 
First, in Subsection \ref{subsec:Sim_SelP}, we focus on hardware-limited systems, and demonstrate how to set the number of scalar quantizers, dictated by the ratio $\Ratio$, by numerically computing the value which minimizes the average \ac{mse}. 
Then, in Subsection \ref{subsec:Sim_Sca}, we compare the performance of the hardware-limited quantizers to that achievable using  vector quantizers, illustrating their ability to approach optimality. 

\begin{figure}
	\centering
		\scalebox{0.48}{\includegraphics{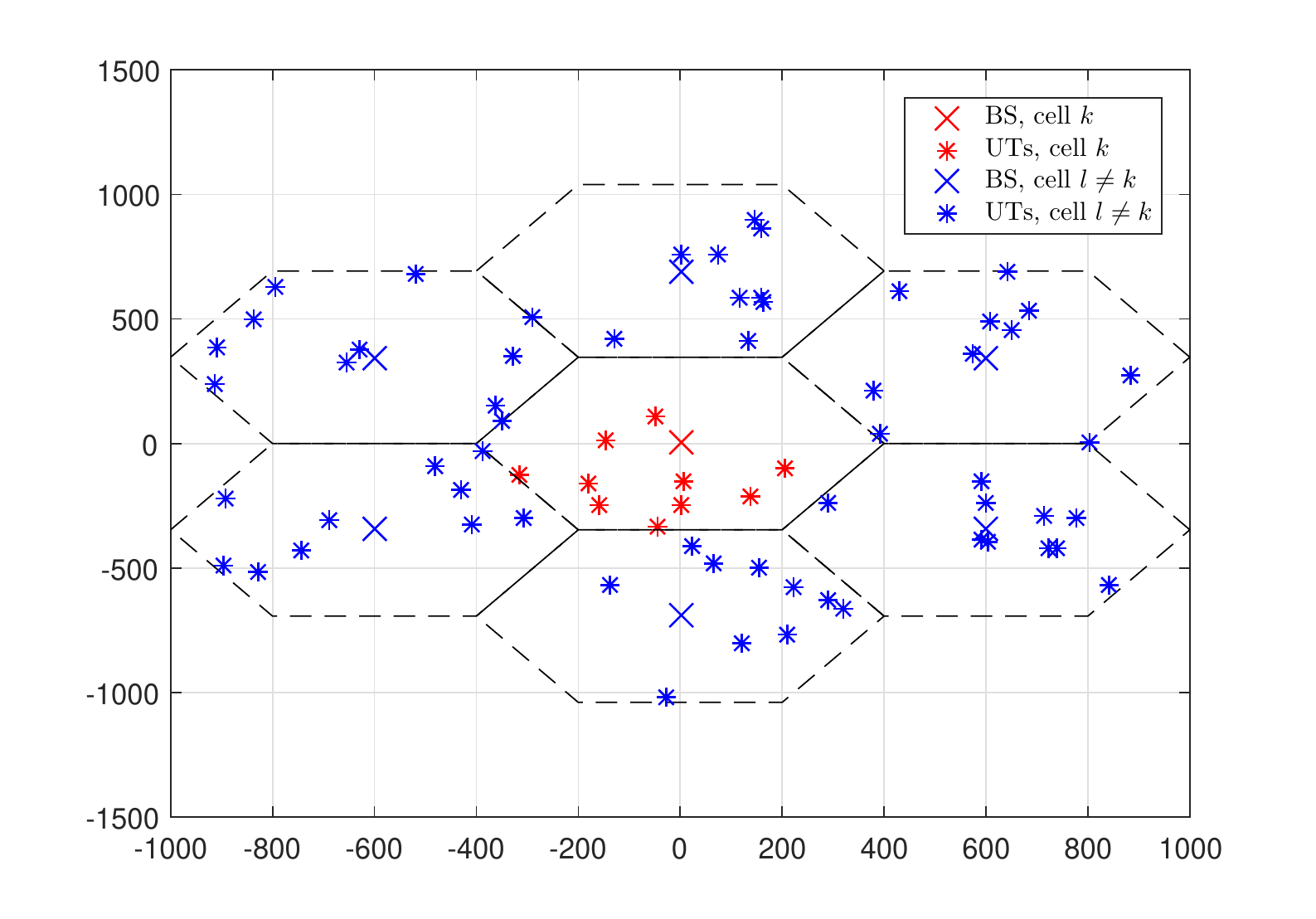}}
		\vspace{-0.4cm}
		\caption{Massive \ac{mimo} network illustration.}
		\label{fig:Network1}		
	\vspace{-0.4cm}
\end{figure}

We consider a massive \ac{mimo} network consisting of $\Ncells = 7$ hexagonal cells of radius $400$ m, with  $\Nusers = 10$ \acp{ut} in each cell. As in \cite{Marzetta:10},  the \acp{ut} are uniformly distributed in the cell, with the exception of a circle with radius $20$ m around the \ac{bs}. The attenuation coefficients  $\{\dcoeff_{l,m,u}\}_{u \in \NusersSet}$ are generated as $\big\{\frac{z_{l,m,u}}{\rho_{l,m,u}^2}\big\}_{m \in \NusersSet}$, where $\{z_{l,m,u}\}$ are the shadow fading coefficients, independently randomized from a log-normal distribution with standard deviation of $ 8$ dB, and  $\{\rho_{l,m,u}\}$ represent the range between the $u$th \ac{ut} of the $m$th cell and the $l$th \ac{bs}, $l,m \in \NcellsSet$, $u \in \NusersSet$ \cite[Sec. II-C]{Marzetta:10}. 
An illustration of such a network is given in Fig. \ref{fig:Network1}. We focus on the central cell in Fig. \ref{fig:Network1}, and thus drop the subscript $l$ indicating the cell index.

We use two models for the receive side correlation $\Acorr{l}[\tau]$: {\em Uncorrelated antennas}, namely,  $\Acorr{l}[\tau] = \delta_{\tau}$; and {\em Correlated antennas},  representing spatial correlation induced by antenna spacing of $0.4$ wavelength based on Jakes model $\Acorr{l}[\tau] = J_0\left(0.8\pi |\tau| \right)$, where $J_0(\cdot)$ is the zero-order Bessel function of the first type \cite{Jakes:93}.  
Following \cite[Sec. II-A]{Li:17}, the pilots matrix $\mySmat$ is  the first $\Nusers$ columns of the $\Tpilots \times \Tpilots$ discrete Fourier transform matrix. The noise power is $\SigW\! =\! 10^{-3}$, and for the scalar quantizers we fix $\myEta\! =\! 2$. 
In the following all hardware-limited quantization systems are simulated with dithered quantizers, with the exception of the channel estimator of \cite{Jacobsson:17},  used for comparison in Subsection \ref{subsec:Sim_Sca}, which is evaluated in the sequel with standard non-dithered uniform quantizers as derived in \cite{Jacobsson:17}.
 Our results are averaged over $10^3$ Monte-Carlo simulations.

\vspace{-0.2cm}
\subsection{Selecting the Analog Combining Ratio $\Ratio$}
\label{subsec:Sim_SelP}
\begin{figure}
	\centering
	\begin{minipage}{0.45\textwidth}
		\centering
		\scalebox{0.48}{\includegraphics{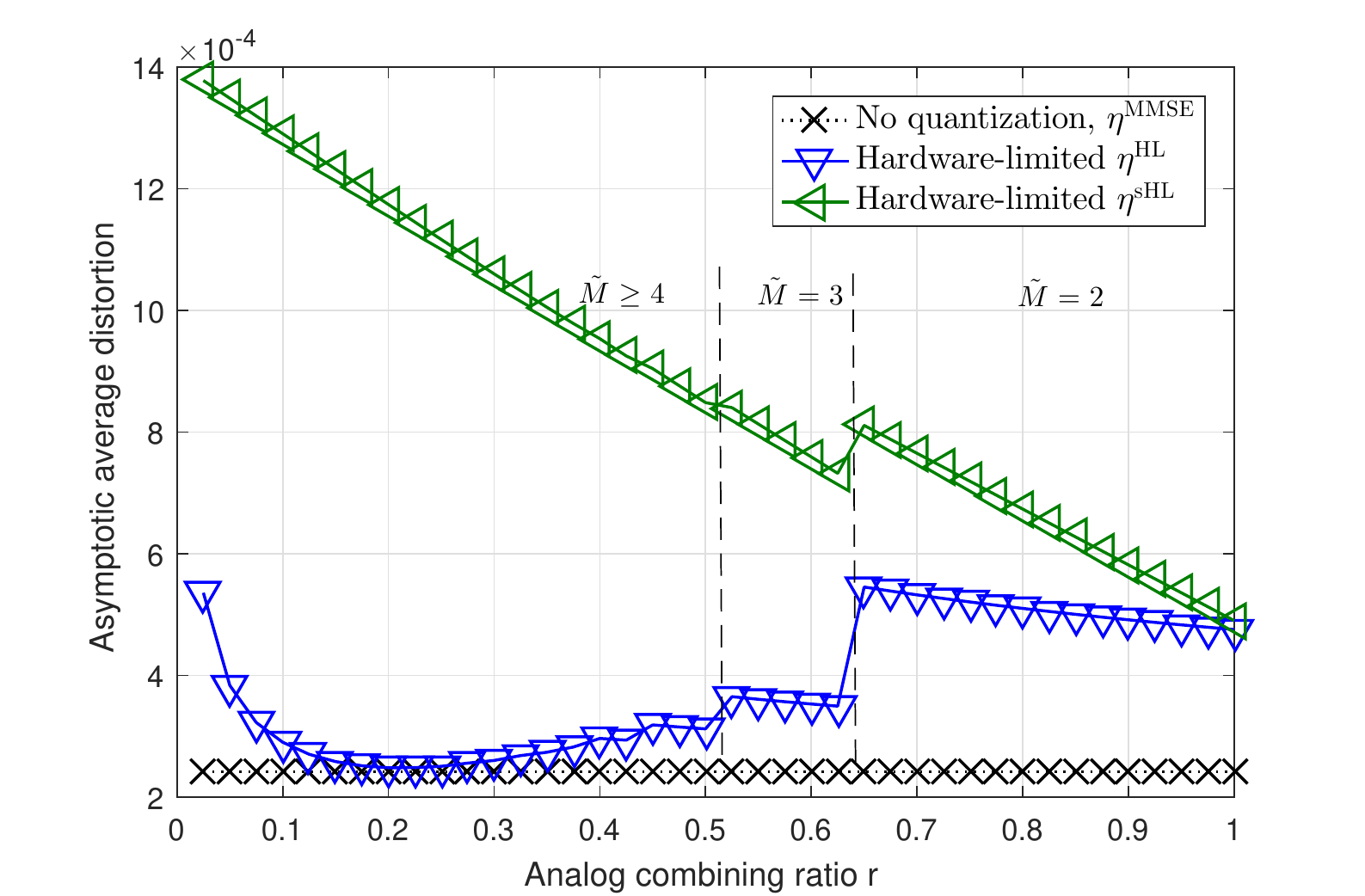}}
		\vspace{-0.4cm}
		\caption{Asymptotic average \ac{mse} vs. $\Ratio$ for $\Rate= 2$, uncorrelated antennas.}
		\label{fig:MassiveMIMO1}		
	\end{minipage}
	$\quad$
	\begin{minipage}{0.45\textwidth}
		\centering
		\scalebox{0.48}{\includegraphics{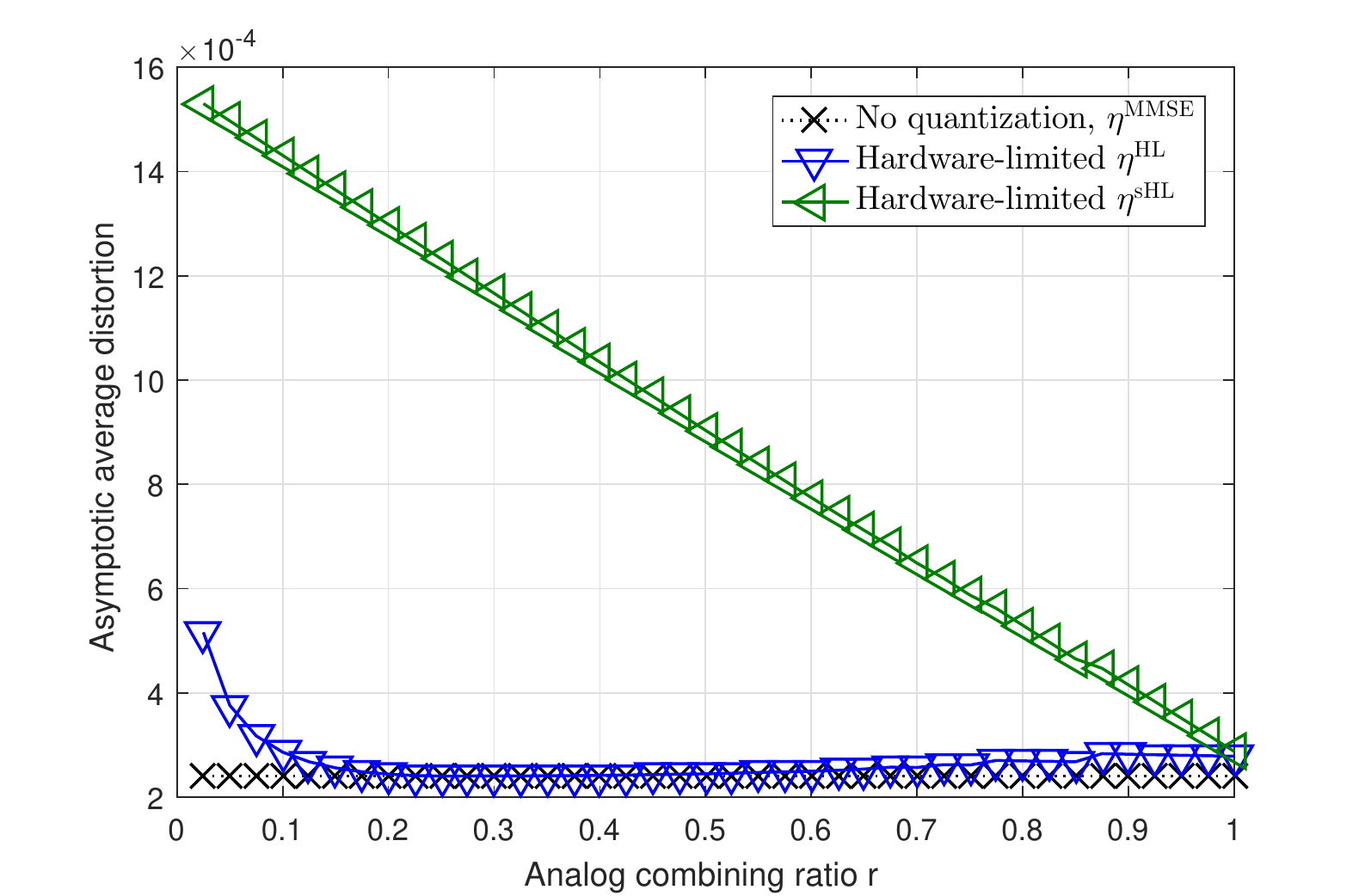}}
		\vspace{-0.4cm}
		\caption{Average \ac{mse} vs. $\Ratio$ for $\Rate= 4$, uncorrelated antennas.}
		\label{fig:MassiveMIMO2}
	\end{minipage}
	\vspace{-0.6cm}
\end{figure}

\begin{figure}
		\centering
		\scalebox{0.48}{\includegraphics{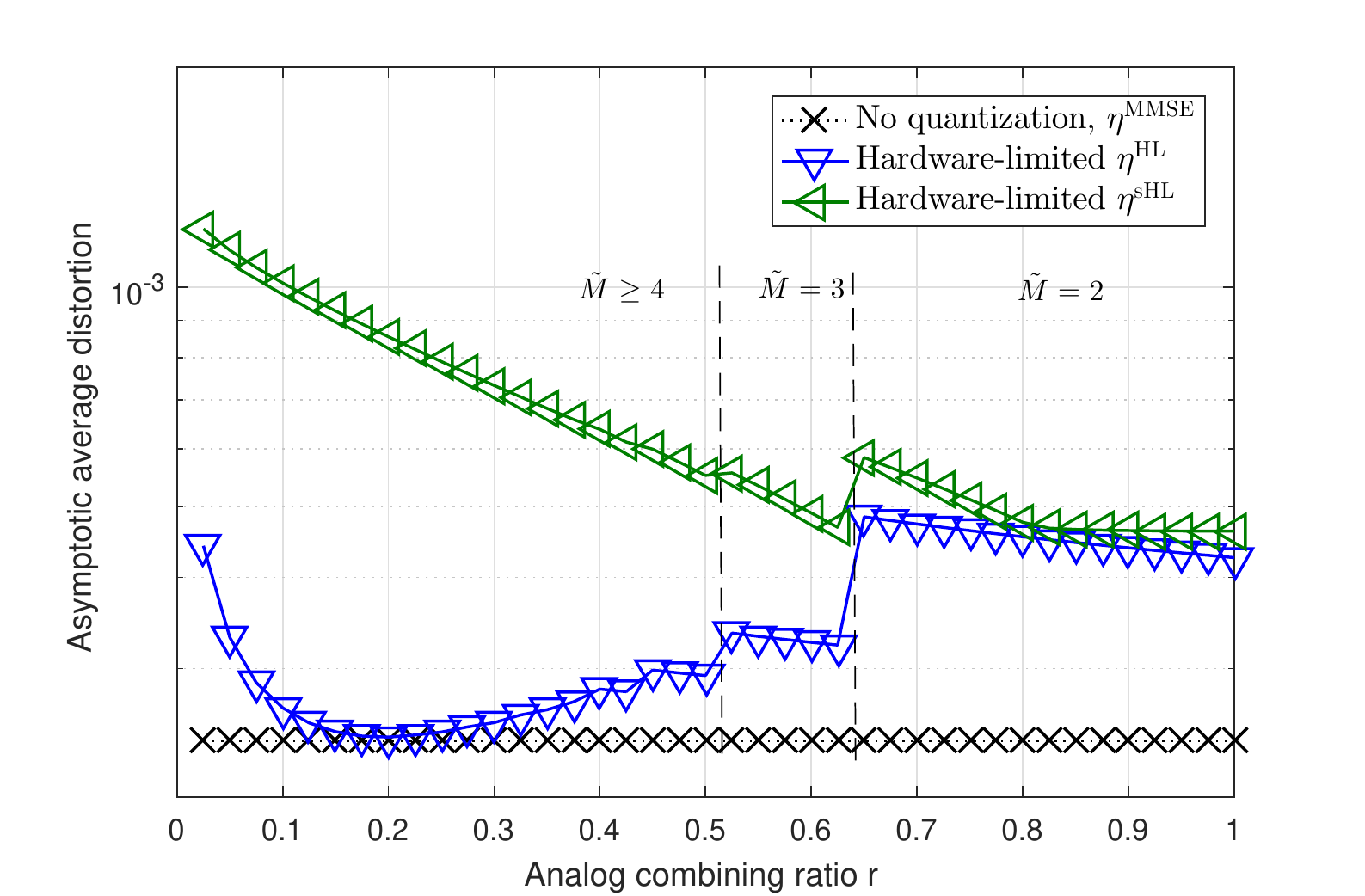}}
		\vspace{-0.4cm}
		\caption{Average \ac{mse} vs. $\Ratio$ for $\Rate= 2$, correlated antennas.}
		\label{fig:MassiveMIMO2a}
		\vspace{-0.6cm}
\end{figure}

\vspace{-0.1cm}
We first numerically evaluate the number of scalar quantizers, dictated by the analog combining ratio $\Ratio = \frac{\lenZ}{\Nantennas \cdot \Tpilots}$, for which the achievable average \ac{mse} of the hardware-limited quantization systems studied in Section \ref{sec:MIMO} is minimized. 
To that aim, we fix $\Tpilots = 40$, and evaluate the achievable average \ac{mse} versus $\Ratio \in (0, 1]$ for general analog combining via Proposition \ref{cor:OptimalDes}, and for spatial analog combining via Proposition \ref{thm:SpatialDes}. 
When the asymptotic average \ac{mse} is given by a limit expression, e.g., \eqref{eqn:SpatialMSE2} with correlated antennas, we compute the \ac{mse} with $\Nantennas = 100$ antennas.
Note that for $\Ratio < \frac{\Nusers}{\Tpilots} = 0.25$, the number of quantized samples is smaller than the number of estimated parameters. The achievable average \acp{mse} for uncorrelated antennas quantization rates $\Rate = 2$ and $\Rate = 4$ are depicted in Figs. \ref{fig:MassiveMIMO1}-\ref{fig:MassiveMIMO2}, respectively,  {and for correlated antennas with quantization rate $\Rate = 2$ in Fig. \ref{fig:MassiveMIMO2a}.} In Figs. \ref{fig:MassiveMIMO1}-\ref{fig:MassiveMIMO2a} we also depict the minimal average \ac{mse} achievable without quantization, namely, the average \ac{mmse}, computed via Corollary \ref{cor:mmse}.

We first observe in Figs. \ref{fig:MassiveMIMO1}-\ref{fig:MassiveMIMO2a} that the analog combining ratio has a notable effect on the average \ac{mse} of the considered systems. In particular, for different values of $\Ratio$, the achievable average \ac{mse} with quantization rate $\Rate = 2$  {and uncorrelated antennas} varies from $5.3\cdot 10^{-4}$ to $2.4\cdot 10^{-4}$ for general analog combining and from $1.3\cdot 10^{-3}$ to $4.9\cdot 10^{-4}$ for spatial analog combining. 
Furthermore, we note that for hardware-limited quantizers with general analog combining, the analog combining ratio which minimizes the average \ac{mse} $\AsymDist\ADC$ is not larger than $\frac{\Nusers}{\Tpilots} = 0.25$, in agreement with Corollary \ref{cor:OptimalP}.
This follows since properly combining correlated samples from different time indexes results in an error which is negligible compared to that induced by the uniform quantizers, hence, hardware-limited quantizers with general analog combining operate best when the analog combining decreases the number of quantized samples to be not larger than the number of channel coefficients, i.e., $\Ratio \le \frac{\Nusers}{\Tpilots}$, allowing the quantization to be carried out with improved resolution.

 When the analog combining matrix is restricted to spatial combining, we observe in  Figs. \ref{fig:MassiveMIMO1}-\ref{fig:MassiveMIMO2} that for uncorrelated antennas, increasing the combining ratio, namely, increasing the number of scalar quantizers, improves the average \ac{mse} $\AsymDist\sADC$. This implies that combining only the independent samples received at the same time index induces a more dominant error compared to the quantization error which results from using quantizers with lower resolution. 
 However, when the antennas are correlated, the error induced by combining the correlated samples is less notable compared to the uncorrelated case, and thus setting an analog combining ratio smaller than one can minimize the overall average \ac{mse}. In particular, it is noted in Fig. \ref{fig:MassiveMIMO2a} that increasing the analog combining ratio from $\Ratio = 0.8$ to $\Ratio = 1$, for which the number of bits $\TilM = 2$ does not change, hardly affects the overall performance, even though more samples quantized at the same resolution are processed in the digital domain. 
 Additionally, as expected, for all values of $\Ratio$ and for all considered scenarios, the minimal \ac{mse} achievable with general analog combining is smaller than the special case where it is restricted to spatial combining.

Finally, recall that the number of quantization levels is $\TilM = \lfloor2^{\frac{\Rate}{2\Ratio}} \rfloor$, thus different values of $\Ratio$ may result in the same $\TilM$, most notably when $\Rate$ is small and $\Ratio$ is relatively large. Consequently, when increasing $\Ratio$ does not reduce  $\TilM$, the overall performance is typically improved by increasing $\Ratio$ as more samples are processed in digital. However, when increasing $\Ratio$ causes the \ac{adc} quantization to be less accurate, the average \ac{mse} typically increases. For example, in Figs. \ref{fig:MassiveMIMO1} and  \ref{fig:MassiveMIMO2a} we explicitly mark the regions of $\Ratio$ for which $\TilM = 2$ and $\TilM = 3$. Observing the average \acp{mse} in these regions, we note that for uncorrelated antennas with a fixed $\TilM$, $\AsymDist\sADC$ decreases quite sharply as $\Ratio$ increases, due to the relationship between $\AsymDist\sADC$ and $\Ratio$  in \eqref{eqn:SpatialMSE}. In both Figs. \ref{fig:MassiveMIMO1} and  \ref{fig:MassiveMIMO2a} we note that $\AsymDist\sADC$ increases substantially when switching from $\TilM= 3$ to $\TilM=2$.   For general analog combining, increasing $\Ratio$ for fixed $\TilM$ has a less notable effect on the average \ac{mse}, as in this case \eqref{eqn:MOptimalMSE} only depends on $\Ratio$ through the setting of $\Wlevel$.

The numerical study in Figs. \ref{fig:MassiveMIMO1}-\ref{fig:MassiveMIMO2a} can be used for determining the combining ratio $\Ratio$ when using hardware-limited quantizers. In particular, the insights gained in this study are used in the comparison of hardware-limited quantization to task-based vector quantization in the following subsection. 


\vspace{-0.2cm}
\subsection{Hardware-Limited vs. Vector Quantization}
\label{subsec:Sim_Sca}
\vspace{-0.1cm}
We now compare the  average \ac{mse} of hardware-limited quantization, which utilizes scalar \acp{adc}, to that achievable using vector quantizers. 
In particular, we compare the performance of the hardware-limited quantizers  to the optimal vector quantizer, computed via Proposition \ref{cor:OptVecQuant}; to the average \ac{mse} achievable using task-ignorant vector quantization, computed via Proposition \ref{cor:TIVecQuant}; and to the channel estimator of \cite{Jacobsson:17}, which extends the $1$-bit Bussgang-LMMSE estimator of \cite{Li:17} to multiple bits. The  Bussgang estimator of \cite{Jacobsson:17} is computed by setting the number of antennas to $\Nantennas = 100=10\Nusers$ and the support of the quantizers  to $\DynRange = 1$.  The performance of the estimator of \cite{Jacobsson:17} is numerically averaged over $10^3$ Monte Carlo simulations in which the estimator processes a uniform non-dithered quantized version of the channel output. Note that \cite{Jacobsson:17} considered a single cell thus we expect its channel estimation accuracy to be impaired due to the presence of intercell interference. Finally, we compute the achievable \ac{mse} of the linear \ac{mmse} digital estimator given in \eqref{eqn:ThmProof2b} with no analog combining and $\DynRange = 1$. Comparing this digital only estimator to $\AsymDist\sADC$ quantifies the gain of properly setting the support and the analog scaling in the spatial-only system of Proposition~\ref{thm:SpatialDes}. 

 Note that the analog combining ratio must satisfy $\Ratio \le \frac{\Rate}{2}$ in order to have $\log \TilM \ge 1$, i.e., to assign at least one bit for each scalar quantizer.
Combining this with the numerical study of the values of $\Ratio$ in Subsection \ref{subsec:Sim_SelP}, we set $\Ratio = \min\left( \frac{\Nusers}{\Tpilots},\frac{\Rate}{2}\right) $  when using the system with general analog combining, and $\Ratio = \min\left( 1,\frac{\Rate}{2}\right) $ when restricted to spatial analog combining  {and $\Acorr{l}[\tau] = \delta_{\tau}$}. 

\begin{figure}
	\centering
	\begin{minipage}{0.45\textwidth}
		\centering
		\scalebox{0.48}{\includegraphics{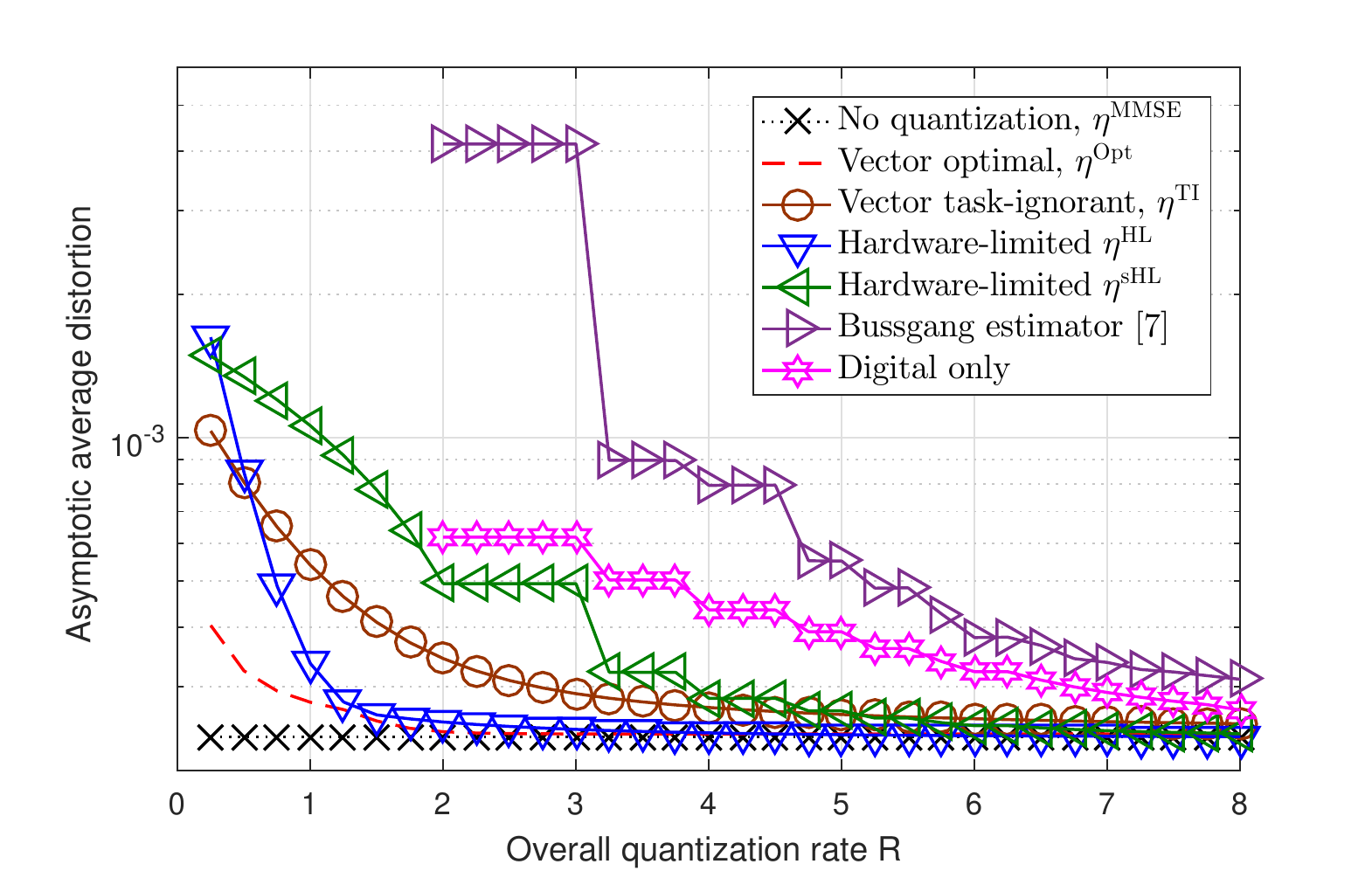}}
		\vspace{-0.4cm}
		\caption{Average \ac{mse} vs. $\Rate$, uncorrelated antennas.}
		\label{fig:MassiveMIMO3}		
	\end{minipage}
	$\quad$
	\begin{minipage}{0.45\textwidth}
		\centering
		\scalebox{0.48}{\includegraphics{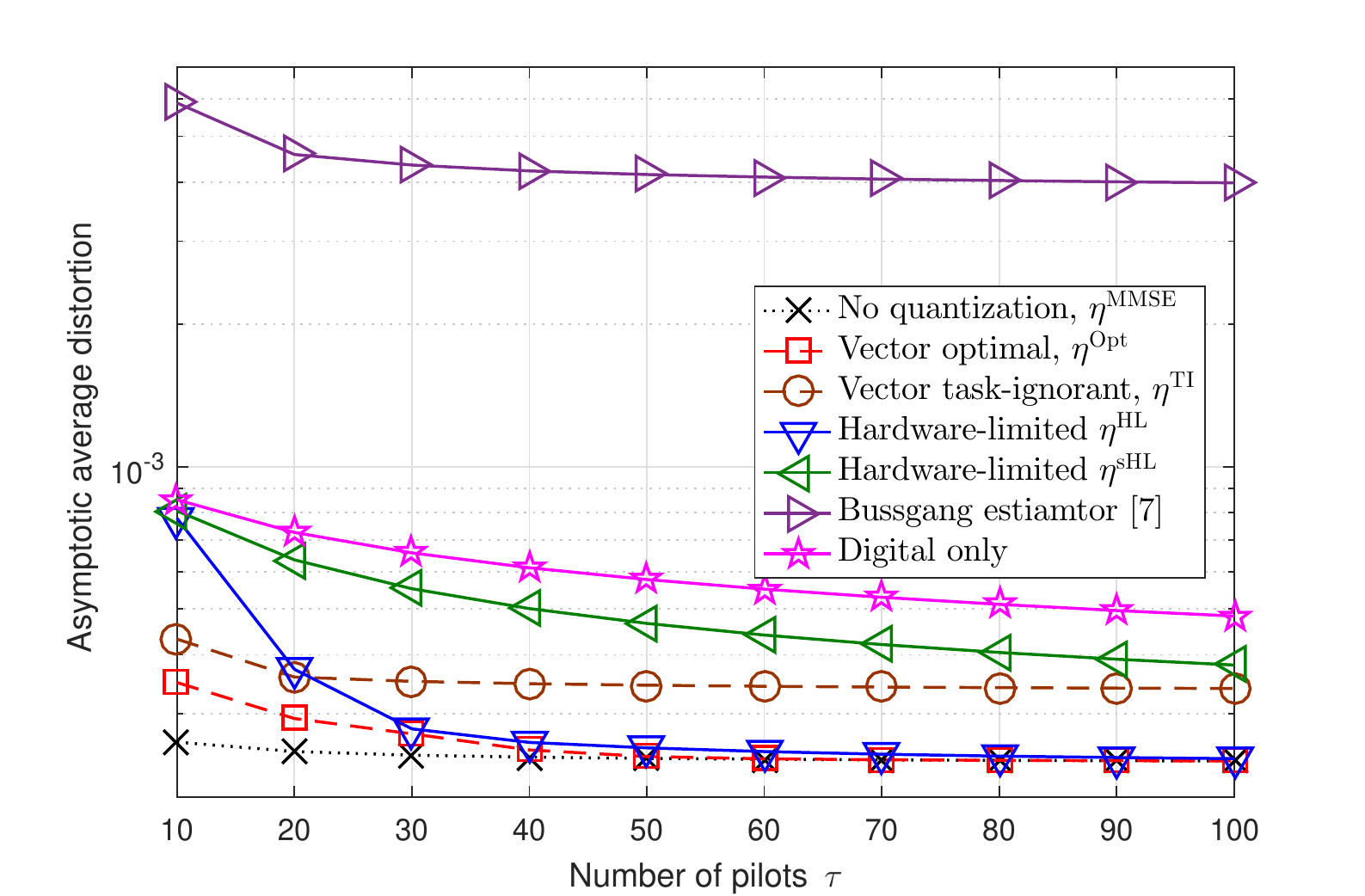}}
		\vspace{-0.4cm}
		\caption{Average \ac{mse} vs. $\Tpilots$ for $\Rate= 2$, uncorrelated antennas.}
		\label{fig:MassiveMIMO4}
	\end{minipage}
	\vspace{-0.6cm}
\end{figure}

In Fig. \ref{fig:MassiveMIMO3} we fix the number of pilot symbols to $\Tpilots = 40$, and evaluate the achievable average \ac{mse} versus $\Rate \in[0.5, 8]$  for uncorrelated antennas.   
Observing Fig. \ref{fig:MassiveMIMO3}, we note that the performance of the hardware-limited quantizer with general analog combining $\AsymDist\ADC$ approaches the optimal performance $\AsymDist\Opt$, achievable with vector quantizers, for quantization rates larger than $\Rate = 1.5$.
{ It is emphasized that while $\AsymDist\Opt$ is smaller than $\AsymDist\ADC$, both measures are within a gap which is negligible compared to the average \ac{mmse}, which constitutes the error floor.  The existence of this error floor is an inherent property of task-based quantization problems, in which, unlike standard quantization, the error cannot be made arbitrarily small by increasing the quantization rate, as it cannot be smaller than the average \ac{mmse}.
 Furthermore, the performance of the hardware-limited quantizer with spatial combining $\AsymDist\sADC$ also approaches $\AsymDist\Opt$ as $\Rate$ increases, and effectively coincides with the minimal achievable \ac{mse} for $\Rate > 5$. 
The estimator of \cite{Jacobsson:17}, which operates only in the digital domain and assumes no intercell interference,  is outperformed by our proposed systems for all considered quantization rates. The digital only estimator, which is designed for multiple cells yet operates only in the digital domain, is also outperformed by $\AsymDist\sADC$, especially at quantization rates $\Rate \in [3,6]$, where setting the support of the quantizers can notably reduce the quantization error.  Furthermore, even for $\Rate = 2$ where one-bit quantizers are used without analog combining, the \ac{mse} of the digital only estimator is still larger than $\AsymDist\sADC$. This follows since properly setting the support, as done in Proposition \ref{thm:SpatialDes}, is still beneficial here as it controls the energy of the dither signal.

These results indicate that properly designed quantization systems operating with scalar \acp{adc} can approach the optimal performance for channel estimation in massive \ac{mimo} systems.
Additionally, we note that for nearly all the considered quantization rates, our proposed hardware-limited system with general analog combining outperforms  vector quantization carried out separately from the channel estimation task. This demonstrates the clear benefits of taking the task of the system into account when designing quantizers for massive \ac{mimo} systems.

Next, we fix $\Rate = 2$. In this case, when no analog combining is applied, each complex sample is represented using two bits, and thus the real and imaginary part are quantized using one-bit sign quantizers. In Fig. \ref{fig:MassiveMIMO4}, we compare the achievable \acp{mse} versus $\Tpilots \in [10, 100]$ for uncorrelated antennas.
From Fig. \ref{fig:MassiveMIMO4} we note that as $\Tpilots$ increases, the hardware-limited quantizer with general analog combining approaches the optimal performance for a fixed quantization rate $\Rate$,  as its analog combining ratio $\frac{\Nusers}{\Tpilots}$ decreases. When this happens, uniform quantization can be carried out at more accurately for the same $\Rate$, reducing the quantization error.  
Furthermore, the quantizer with spatial analog combining, which, following the results of Subsection \ref{subsec:Sim_SelP}, does not decrease its combining ratio as $\Tpilots$ increases, also demonstrates a steady improvement in the average \ac{mse}. This behavior is in agreement with the fact that as $\Tpilots \rightarrow \infty$, $\AsymDist\sADC$ in \eqref{eqn:SpatialMSE} approaches $\AsymDist\MMSE$.

So far we have considered the case of uncorrelated antennas. In Fig. \ref{fig:MassiveMIMO5} we compare the achievable average \acp{mse} of the hardware-limited quantizers to the optimal vector quantizer and to the digital only quantizer for the correlated antennas setup. As in Fig. \ref{fig:MassiveMIMO3}, we compute the average \ac{mse} versus $\Rate \in[0.5, 8]$ when the number of pilot symbols is fixed to $\Tpilots = 40$. 
Based on  the numerical study of the values of $\Ratio$ in Subsection \ref{subsec:Sim_SelP}, we use here $\Ratio = \min\left( \frac{\Nusers}{\Tpilots},\frac{\Rate}{2}\right) $  when using the system with general analog combining, and $\Ratio = \min\left( 0.8,\frac{\Rate}{2}\right) $ when restricted to spatial analog combining.  
Recall that the asymptotic average \ac{mse} of the task-ignorant vector quantizer is given in Proposition \ref{cor:TIVecQuant} only for uncorrelated antennas, and is thus not evaluated in this correlated setup. 
Observing Fig. \ref{fig:MassiveMIMO5} we note that, similarly to the uncorrelated setup in Fig. \ref{fig:MassiveMIMO3}, $\AsymDist\ADC$ is within a very small gap from optimal performance $\AsymDist\Opt$ for quantization rates larger than $\Rate = 1.5$. 
The hardware-limited quantizer with spatial combining, which for the uncorrelated case required the quantization rate to be $\Rate > 5$ to approach  $\AsymDist\Opt$, is capable of achieving near-optimal performance for $\Rate > 3$ here, due to its  ability to exploit the spatial correlation. It is also observed that the average \ac{mse} of estimating the channel only in the digital domain is notably higher compared to $\AsymDist\sADC$. This indicates that, as noted in \cite{Ordonez:16}, spatial correlation in massive \ac{mimo} systems with quantized outputs can be exploited by combining the samples received at the same time instance, leading to more accurate recovery.

Finally, we note that our hardware-limited quantizers require accurate knowledge of the channel input-output statistical relationship, from which, e.g., the covariance matrix $\CovYtag$ is obtained. In practice, such a-priori knowledge may not be available, and one must utilize noisy estimates of the channel parameters instead of their actual value. In order to  evaluate the robustness of the proposed quantization systems to inaccurate knowledge of the underlying channel, we numerically compute the average \ac{mse} achieved when using a noisy estimate of the  \acp{ut} attenuation $\{\dcoeff_{l,m,u}\}$,  given by  $\dcoeff_{l,m,u} + \sigma_{d} \cdot w_{l,m,u}$, for each $m \in \NcellsSet$ and $u \in \NusersSet$, where $\{w_{l,m,u}\}$ are i.i.d. zero mean Gaussian \acp{rv} with unit variance. 
Inaccurate knowledge of $\{\dcoeff_{l,m,u}\}$ leads to a noisy estimation of  the covariance matrix $\CovYtag$ and the  matrix $\LmmseMattag$. For each simulated  realization of $\{\dcoeff_{l,m,u}\}$, we evaluate the average \ac{mse} over $40$ realizations of   $\{w_{l,m,u}\}$. We consider both correlated a well as uncorrelated antennas, recalling that the average \ac{mmse} in Corollary \ref{cor:mmse} is identical in both. 
In Fig. \ref{fig:MassiveMIMO6} we depict the computed average \acp{mse} of our proposed hardware-limited quantizers with $\Nantennas =100$ antennas and fixed quantization rate $\Rate = 2$ compared to the digital only estimator, versus the coefficients noise level  $\sigma_{d}^2 \in [0,0.2]$. 
Since the average \acp{mse} here are computed by simulating the proposed quantization systems, and not by computing an analytical expression, we do not simulate vector quantizers, which are very computationally complex to implement at large input sizes. 
Observing Fig. \ref{fig:MassiveMIMO6}, we note that while the performance of all considered quantizers degrades rapidly as $\sigma_{d}^2$ increases, the relative gain of our proposed quantizers compared to digital only estimation is maintained. This behavior is observed for both uncorrelated as well as correlated antennas. These results indicate that the benefits of the proposed hardware-limited quantizers hold also in the presence of inaccurate \ac{csi}. 
%
\begin{figure}
	\centering
	\begin{minipage}{0.45\textwidth}
		\centering
		\scalebox{0.48}{\includegraphics{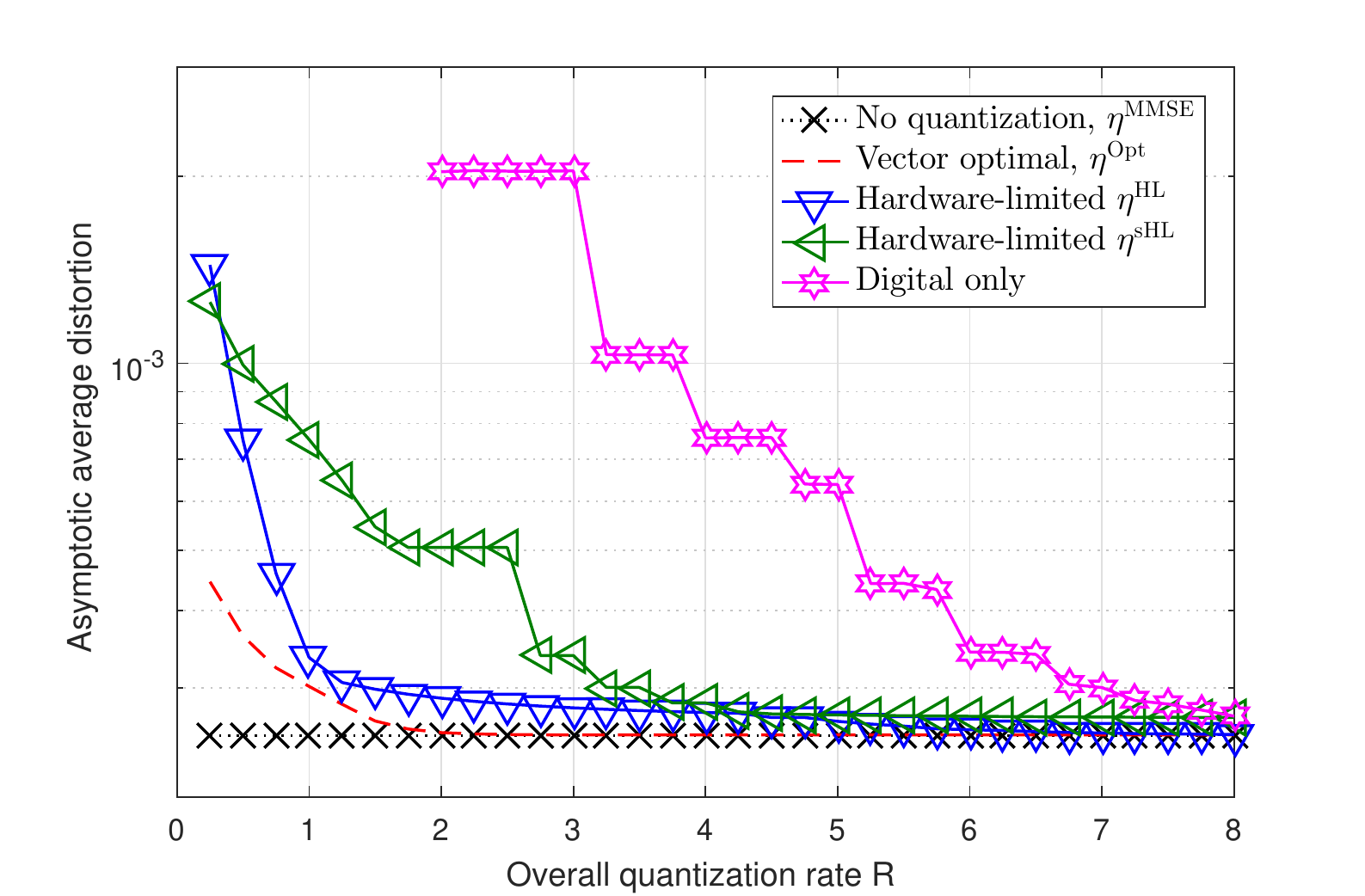}}
		\vspace{-0.4cm}
		\caption{Average \ac{mse} vs. $\Rate$, correlated antennas.}
		\label{fig:MassiveMIMO5}		
	\end{minipage}
$\quad$
	\begin{minipage}{0.45\textwidth}
		\centering
		\scalebox{0.48}{\includegraphics{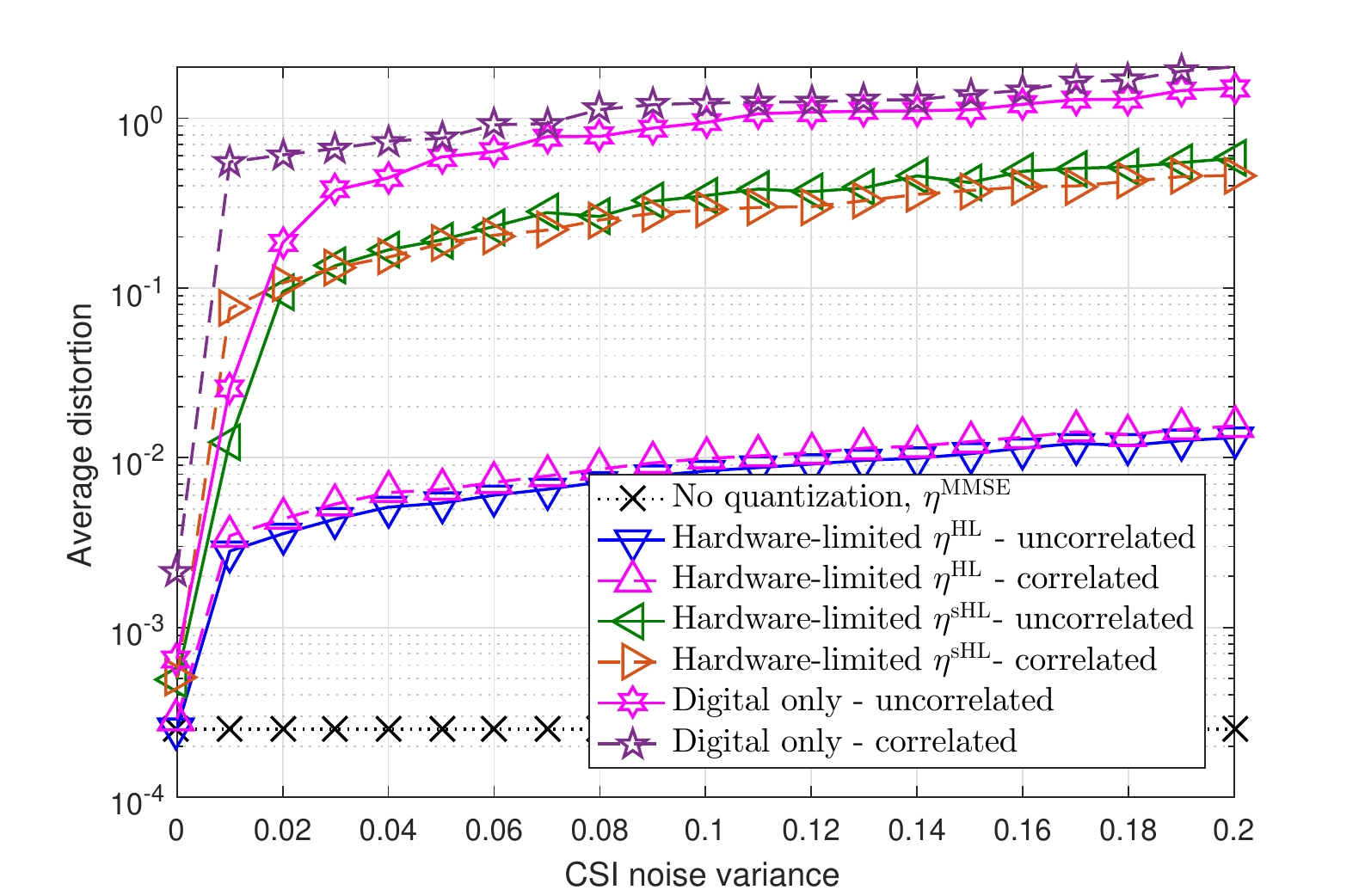}}
		\vspace{-0.4cm}
		\caption{Average \ac{mse} vs. $\sigma_{d}^2$ for $\Rate= 2$.}
		\label{fig:MassiveMIMO6}
	\end{minipage}
	\vspace{-0.6cm}
\end{figure}

The simulation results presented in this section demonstrate the fundamental performance limits of channel estimation in massive \ac{mimo} systems, and illustrate that properly designed hardware-limited quantization systems are capable of approaching these limits at relatively low quantization rates. 

\vspace{-0.2cm}
\section{Conclusions}
\label{sec:Conclusions}
\vspace{-0.1cm}
In this work we studied task-based quantization with large-scale inputs. We first derived the average achievable \ac{mse} when using vector quantization, and extended our earlier analysis of task-based quantization systems operating with scalar \acp{adc} to large-scale data. 
Then, we showed how these results can be applied to studying channel estimation in massive \ac{mimo} systems with quantized inputs. Our numerical results demonstrate that the minimal achievable average \ac{mse} in massive \ac{mimo} channel estimation can be approached by properly designed quantization systems utilizing  scalar low-resolution \acp{adc}, and that the proposed approach outperforms previous channel estimators operating only in the digital domain. 

 
\vspace{-0.2cm}
\begin{appendix}
\numberwithin{proposition}{subsection} 
\numberwithin{lemma}{subsection} 
\numberwithin{corollary}{subsection} 
\numberwithin{remark}{subsection} 
\numberwithin{equation}{subsection}	
%
\vspace{-0.2cm}
\subsection{Proof of Theorem \ref{thm:OptVecQuant}}
\label{app:ProofVecAsym}
\vspace{-0.1cm} 
 Recall  that  the optimal quantizer for finite $\lenAsym$ quantizes the \ac{mmse} estimate~\cite{Wolf:70}.  {Thus, using the notation $ \Quan{M}{\lenS}(\cdot) =  \Quan{M}{\lenS, \lenS}(\cdot)$, the minimal average \ac{mse} is  given by} 
\ifsingle 
  \begin{align}
   \frac{1}{\lenS}\mathop{\min}\limits_{\Quan{M}{\lenX, \lenS} (\cdot )} \E\left\{\left\|\mySOI -  \Quan{M}{\lenX, \lenS}\left( \myObs\right) \right\|^2  \right\}   
  &  = \frac{1}{\lenS} \E\left\{\left\|\mySOI  -  \mySOIEst  \right\|^2  \right\}  + \frac{1}{\lenS} \mathop{\min}\limits_{\Quan{M}{\lenS} (\cdot )}\E\left\{\left\| \mySOIEst -  \Quan{M}{\lenS}\left( \mySOIEst\right) \right\|^2  \right\} \notag \\
  & = \AsymDist \MMSE \!+\! \frac{1}{\lenS} \mathop{\min}\limits_{\Quan{M}{\lenS} (\cdot )}\E\left\{\left\|\mySOIEst \!-\!  \Quan{M}{\lenS}\left(\mySOIEst\right) \right\|^2  \right\}.
  \label{eqn:Proof3Eq1}  
  \end{align}
\else
\vspace{-0.2cm}
 	\begin{align}
	&\frac{1}{\lenS}\mathop{\min}\limits_{\Quan{M}{\lenX, \lenS} (\cdot )} \E\left\{\left\|\mySOI -  \Quan{M}{\lenX, \lenS}\left( \myObs\right) \right\|^2  \right\}   \notag \\
	& \quad = \AsymDist \MMSE + \frac{1}{\lenS} \mathop{\min}\limits_{\Quan{M}{\lenS} (\cdot )}\E\left\{\left\|\mySOIEst -  \Quan{M}{\lenS}\left(\mySOIEst\right) \right\|^2  \right\}.
	\label{eqn:Proof3Eq1}  
	\vspace{-0.2cm}
	\end{align}
\fi 
  The second summand in \eqref{eqn:Proof3Eq1} is the minimal average distortion in quantizing the \ac{mmse} estimate $\mySOIEst$ at  rate $\frac{1}{\lenS} \log M = \frac{\lenXtag}{\lenStag} \frac{1}{\lenX}\log M = \frac{\lenXtag}{\lenStag} \cdot \Rate$. 
  Since $\mySOIEst$  {consists of $\lenAsym$  zero-mean random vectors sampled from a stationary distribution, it follows from \cite[Ch. 5.9]{Han:03}} that for $\lenAsym \rightarrow \infty$, the minimal achievable distortion coincides with the distortion-rate function for  $\mySOIEsttag$, namely, 
  \begin{equation*}
	  \mathop{\lim}\limits_{\lenAsym \rightarrow \infty} \frac{1}{\lenAsym } \mathop{\min}\limits_{\Quan{M}{\lenS} (\cdot )}\E\left\{\left\| \mySOIEst -  \Quan{M}{\lenS}\left( \mySOIEst\right) \right\|^2  \right\}   
	  = D_{\mySOIEsttag}\left( \frac{\lenXtag}{\lenStag} \cdot \Rate \right). 
  \end{equation*}
  Substituting this in \eqref{eqn:Proof3Eq1} proves the theorem.
  \qed

\vspace{-0.2cm}
\subsection{Proof of Theorem \ref{thm:TIVecQuant}}
\label{app:ProofTIAsym}
\vspace{-0.1cm}  
To prove the theorem, we first express the excess distortion due to quantization. Then, we let $\lenAsym \rightarrow \infty$, and show that the excess distortion coincides with the second summand in \eqref{eqn:TIVecQuant}.

From the orthogonality principle,  the resulting distortion in estimating $\mySOIEst$ from the quantized $\myObs$ is given by
\ifsingle
	\begin{align} 
	\frac{1}{\lenS}\E\left\{ \left\| \mySOI- \E\left\{ \mySOI\big|  {\Quan{M}{\lenX}}\left(  \myObs\right) \right\}\right\|^2 \right\}  
	& = \frac{1}{\lenS}\E\left\{ \left\|  \mySOI- \mySOIEst \right\|^2 \right\}  
	+   \frac{1}{\lenS}\E\left\{ \left\| \mySOIEst  \! - \!
	 \E\left\{ \mySOI\big|  {\Quan{M}{\lenX}}\left(  \myObs\right) \right\} \right\|^2 \right\} \notag \\
	&  \stackrel{(a)}{=}  \AsymDist \MMSE +  
	\frac{1}{\lenS}\E\left\{ \left\| \mySOIEst  \! - \!
	\E\left\{ \mySOIEst\big|  {\Quan{M}{\lenX}}\left(  \myObs \right) \right\} \right\|^2 \right\},
	\vspace{-0.1cm}
	\label{eqn:Proof4Eq1}
	\end{align} 
\else
\vspace{-0.2cm}
	\begin{align} 
	&\frac{1}{\lenS}\E\left\{ \left\| \mySOI- \E\left\{ \mySOI\big|  {\Quan{M}{\lenX}}\left(  \myObs\right) \right\}\right\|^2 \right\}  \notag \\
	& = \frac{1}{\lenS}\E\left\{ \left\|  \mySOI- \mySOIEst \right\|^2 \right\}  
	+   \frac{1}{\lenS}\E\left\{ \left\| \mySOIEst  \! - \!
	\E\left\{ \mySOI\big|  {\Quan{M}{\lenX}}\left(  \myObs\right) \right\} \right\|^2 \right\} \notag \\
	&  \stackrel{(a)}{=}  \AsymDist \MMSE +  
	\frac{1}{\lenS}\E\left\{ \left\| \mySOIEst  \! - \!
	\E\left\{ \mySOIEst\big|  {\Quan{M}{\lenX}}\left(  \myObs \right) \right\} \right\|^2 \right\},
	\vspace{-0.1cm}
	\label{eqn:Proof4Eq1}
	\vspace{-0.2cm}
	\end{align} 
\fi 
where  $(a)$ follows since $\mySOI \mapsto \myObs  \mapsto  {\Quan{M}{\lenX}}\big(  \myObs \big) $ form a Markov chain, thus, by \cite[Prop. 4]{Rioul:10}, 
$ \E\left\{ \mySOI \big|  {\Quan{M}{\lenX}}\big( \myObs\big) \right\}
=  \E\left\{ \mySOIEst\big|  {\Quan{M}{\lenX}}\big(  \myObs \big) \right\}$. 

Next, we note that   $\mySOIEst =  \left( \LmmseMattag \otimes \myI_{\lenAsym}\right) \myObs$, it thus follows that
%
\ifsingle
\begin{align}
&\E\left\{ \left\| \mySOIEst  \! - \!
\E\left\{ \mySOIEst\big|  {\Quan{M}{\lenX}}\left(  \myObs \right) \right\} \right\|^2 \right\} 
= \E\left\{ \left\| \left( \LmmseMattag \otimes \myI_{\lenAsym}\right)\left(   \myObs  \! - \!
\E\left\{\myObs\big|  {\Quan{M}{\lenX}}\left( \myObs\right) \right\} \right) \right\|^2 \right\} \notag \\
&\qquad = {\rm Tr}\left(\left(  \left( \LmmseMattag\right)^H \LmmseMattag \otimes \myI_{\lenAsym}\right) \E \left\{ \left(   \myObs \! - \!
\E\left\{\myObs\big|  {\Quan{M}{\lenX}}\left(  \myObs\right) \right\} \right) 
\left(   \myObs \! - \!
\E\left\{\myObs\big|  {\Quan{M}{\lenX}}\left(  \myObs\right) \right\} \right)^H \right\}  \right)  \notag \\
&\qquad \stackrel{(a)}{=}    {\rm Tr}\left( \left( \LmmseMattag ^H \LmmseMattag \otimes \myI_{\lenAsym}\right)\left(\CovMat{\myObs} - \CovMat{{\Quan{M}{\lenX}}\left(  \myObs \right)  } \right)  \right), 
\label{eqn:Proof4Eq2}
\end{align}
\else
\vspace{-0.2cm}
\begin{align}
&\E\left\{ \left\| \mySOIEst  \! - \!
\E\left\{ \mySOIEst\big|  {\Quan{M}{\lenX}}\left(  \myObs \right) \right\} \right\|^2 \right\}  \notag \\
&\quad= \E\left\{ \left\| \left( \LmmseMattag \otimes \myI_{\lenAsym}\right)\left(   \myObs  \! - \!
\E\left\{\myObs\big|  {\Quan{M}{\lenX}}\left( \myObs\right) \right\} \right) \right\|^2 \right\} \notag \\
&\quad \stackrel{(a)}{=}    {\rm Tr}\left( \left(  \LmmseMattag ^H \LmmseMattag \otimes \myI_{\lenAsym}\right)\left(\CovMat{\myObs} - \CovMat{{\Quan{M}{\lenX}}\left(  \myObs \right)  } \right)  \right), 
\label{eqn:Proof4Eq2}
\vspace{-0.2cm}
\end{align}
\fi 
where $(a)$ holds as the optimal quantizer output is uncorrelated with the quantization error \cite[Sec. III]{Gray:98}. 
Since  $\myObs$   consists here of $\lenAsym$ i.i.d. $\lenXtag \times 1$  random vectors distributed as $\myObstag$, it follows from  \cite[Ch. 23.2]{Polyanskiy:15} that in the limit $\lenAsym \rightarrow \infty$, the output of the optimal quantizer consists of $\lenAsym$ i.i.d. $\lenXtag \times 1$ random vectors whose distribution is the marginal distortion-rate distribution which achieves $D_{\myObstag}\left(\Rate  \right)$, i.e.,  $ \CovMat{{\Quan{M}{\lenX}}\left(\myObs\right)  } = \CovMat{\myObstag, D}(\Rate) \otimes \myI_{\lenAsym}$. Plugging this into \eqref{eqn:Proof4Eq2} and letting $\lenAsym \rightarrow \infty$ yields
\ifsingle
\begin{align}
& \mathop{\lim}\limits_{\lenAsym \rightarrow \infty}
\frac{1}{\lenStag \cdot \lenAsym}\E\left\{ \left\|  \mySOIEst \! - \!
\E\left\{ \mySOIEst\big|  {\Quan{M}{\lenX}}\left(  \myObs \right) \right\} \right\|^2 \right\}   =  \frac{1}{\lenStag} {\rm Tr}\left(   \LmmseMattag ^H \LmmseMattag\left( \CovMat{\myObstag} - \CovMat{\myObstag, D}(\Rate)  \right)  \right).
\label{eqn:Proof4Eq3}
\end{align}
\else
\vspace{-0.2cm}
\begin{align}
& \mathop{\lim}\limits_{\lenAsym \rightarrow \infty}
\frac{1}{\lenStag \cdot \lenAsym}\E\left\{ \left\|  \mySOIEst \! - \!
\E\left\{ \mySOIEst\big|  {\Quan{M}{\lenX}}\left(  \myObs \right) \right\} \right\|^2 \right\}   \notag \\
&\quad =  \frac{1}{\lenStag} {\rm Tr}\left(  \LmmseMattag ^H \LmmseMattag\left( \CovMat{\myObstag} - \CovMat{\myObstag, D}(\Rate)  \right)  \right).
\label{eqn:Proof4Eq3}
\vspace{-0.2cm}
\end{align}
\fi 
Combining \eqref{eqn:Proof4Eq3} and \eqref{eqn:Proof4Eq1} proves the theorem. 
\qed

\vspace{-0.2cm}
\subsection{Proof of Theorem \ref{thm:OptimalDes}}
\label{app:ProofThmDes}
\vspace{-0.1cm}
For a finite $\lenAsym$, the optimal system and the resulting \ac{mse} for the considered setup can be obtained from \cite{Shlezinger:18}. 
Consequently, in the following we formulate the results of \cite{Shlezinger:18} (adapted to complex-valued signals), and then let $\lenAsym$ grow to infinity, obtaining  Theorem~\ref{thm:OptimalDes}.
In particular, under the model detailed in Subsection \ref{subsec:Pre_Problem},  the optimal digital processing in \eqref{eqn:OptimalDesB} is obtained from \cite[Lem. 1]{Shlezinger:18}.  
 The  analog combining of \cite[Thm. 1]{Shlezinger:18}  is given by
  {
  $\myA\op = \myMat{U}_{\myA} \myMat{\Lambda}_{\myA}\left( \myMat{V}_{\myA}^H \CovMat{\myObstag}^{-1/2} \otimes\CorrMat^{-1/2} \right) $,} where $\left( \myMat{\Lambda}_{\myA}\right)_{l,l}^2 = 	\frac{4\MyKappa}{3 \TilM^2  \cdot \Ratio}\Glevel(\Wlevel\cdot \eigT{l})$.  
The waterfilling parameter $\Wlevel> 0 $ is set such that  $\frac{4\MyKappa}{3 \TilM^2 \cdot \Ratio}\sum\limits_{l=1}^{\lenZ} \Glevel(\Wlevel\cdot \eigT{l}) = \lenX$, which can be written as 
$\frac{4\MyKappa}{3 \TilM^2 \cdot \lenZ}\sum\limits_{l=1}^{\lenZ} \Glevel(\Wlevel\cdot \eigT{l}) = 1$.  
The support is set to satisfy 
\vspace{-0.2cm}
\begin{align}
\DynRange^2  = \MyKappa \mathop{\max}\limits_{l=1,\ldots,\lenZ} \E \left\{  \left|\left( \myA\op  \myObs\right)_l \right| ^2 \right\}, 
\label{eqn:DynRange}
\vspace{-0.2cm}
\end{align}
and is thus given by 
$\DynRange^2   =   \frac{ \MyKappa}{\lenZ} {\rm Tr}\left( \myMat{\Lambda}_{\myA}\myMat{\Lambda}_{\myA}^H\right) = \frac{\MyKappa}{\Ratio}$.

The resulting optimal average excess \ac{mse} compared to the \ac{mmse} in \cite[Thm. 1]{Shlezinger:18} under this setting can be written as
\vspace{-0.2cm}
\begin{align}
\hspace{-0.2cm}
{\rm MSE}_{\lenAsym}\!\!\left(\myA\op \right)
\!= \!\frac{1}{\lenS } \sum\limits_{l=1}^{\lenS }    \eigT{l}^2 \! - \!
\frac{1}{\lenS}\!\! \!\!\sum\limits_{l=1}^{\min(\lenS,\lenZ)}\!\!  \frac{ \Glevel(\Wlevel\cdot \eigT{l}) \cdot \eigT{l}^2  } { \Glevel(\Wlevel\cdot \eigT{l}) \! + \! 1}.
\vspace{-0.2cm}
\label{eqn:ProofMSE2}
\end{align} 
When both sums in \eqref{eqn:ProofMSE2} have the same number of summands, i.e., $\lenZn \ge {\rm rank}(\LmmseMatT\CovMat{\myObstag}\LmmseMatT^H)$, \eqref{eqn:ProofMSE2} yields
\vspace{-0.2cm}
\begin{align}
&{\rm MSE}_{\lenAsym}\left(\myA\op \right)
= \frac{1}{\lenS } \sum\limits_{l=1}^{\lenS }    \frac{  \eigT{l}^2  } { \Glevel(\Wlevel\cdot \eigT{l}) \! + \! 1}.
\vspace{-0.2cm}
\label{eqn:ProofMSE22}
\end{align}
By letting $\eig{\CorrMat,k}$ be the $k$-th largest eigenvalue of $\CorrMat$, it follows that each singular value $\eigT{l}$ can be written as $\eigT{l} = \phicoeff_i\sqrt{\eig{\CorrMat,k}}$ for some pair of indexes $i \in \{1,\ldots,\lenStag\}$ and $k \in \{1,\ldots,\lenAsym\}$, where each $l$ corresponds to a different $(i,k)$ pair.
 The average \ac{mse} in \eqref{eqn:ProofMSE22} can thus be written as 
\vspace{-0.2cm}
\begin{align}
&{\rm MSE}_{\lenAsym}\left(\myA\op \right)
= \frac{1}{\lenStag } \sum\limits_{i=1}^{\lenStag }  \frac{1}{\lenAsym } \sum\limits_{k=1}^{\lenAsym }   \frac{  \phicoeff_i^2{\eig{\CorrMat,k}}  } { \Glevel(\Wlevel\cdot \phicoeff_i\sqrt{\eig{\CorrMat,k}}) \! + \! 1}.
\vspace{-0.2cm}
\label{eqn:ProofMSE23}
\end{align}
Since the mapping $f(x) \triangleq \frac{x}{\Glevel({\Wlevel \cdot \sqrt{x}}) + 1}$ is continuous over $\mySet{R}^+$ and since the rows of $\CorrMat$ are absolutely summable, it follows from Szego's theorem \cite[Eq. (1.6)]{Gray:06} that in the limit $\lenAsym \rightarrow \infty$, \eqref{eqn:ProofMSE23} becomes 
\begin{align}
&{\rm MSE}\left(\myA\op \right)
= \frac{1}{\lenStag } \sum\limits_{i=1}^{\lenStag } \frac{1}{2\pi}\int_{0}^{2\pi}   \frac{  \phicoeff_i^2{\Psd{}(\omega)}  } { \Glevel(\Wlevel\cdot \phicoeff_i\sqrt{\Psd{}(\omega)}) \! + \! 1} d\omega,
\vspace{-0.2cm}
\label{eqn:ProofMSE24}
\end{align}
thus proving \eqref{eqn:OptimalMSE2}.  

Now,  {when $\Acorr{}[l] = \delta_l$, then $\eigT{l} = \phicoeff_{ \left\langle l \right\rangle_{\lenAsym}}$ and $\Psd{ }(\omega) \equiv 1$. In this case,} we can write \eqref{eqn:ProofMSE2} for any setting of $\lenZ$ as
\vspace{-0.2cm}
\begin{equation}
\hspace{-0.2cm}
{\rm MSE}_{\lenAsym}\!\left(\myA\op \right) 
\!= \! \frac{1}{\lenS  }\! \sum\limits_{l=1}^{\lenZ } \!  \frac{\phicoeff_{ \left\langle l \right\rangle_{\lenAsym}}^2 } { \Glevel(\Wlevel\cdot \phicoeff_{ \left\langle l \right\rangle_{\lenAsym}}) \! + \! 1} \! + \! \frac{1}{\lenS  }\! \sum\limits_{l=\lenZ\! + \!1}^{\lenS } \!\phicoeff_{ \left\langle l \right\rangle_{\lenAsym}}^2.
\label{eqn:ProofMSE3}
\vspace{-0.2cm}
\end{equation}
In order to express \eqref{eqn:ProofMSE3} in the limit $\lenAsym \rightarrow \infty$, we recall that by  \eqref{eqn:lenZDefa}, $\lenZ < \lenS$ implies that $\lenZn < \lenStag$, thus, \eqref{eqn:ProofMSE3} becomes
\ifsingle
\begin{align}
&{\rm MSE}_{\lenAsym}\left(\myA\op \right) 
=  \frac{1}{\lenS  } \sum\limits_{l=1}^{\lenZn \cdot \lenAsym }   \frac{\phicoeff_{  \left\langle l \right\rangle_{\lenAsym}}^2 } { \Glevel(\Wlevel \cdot \phicoeff_{  \left\langle l \right\rangle_{\lenAsym}}) \! + \! 1}
 \! + \! \frac{1}{\lenS  } \sum\limits_{l=\lenZn \cdot \lenAsym  \! + \! 1}^{\lenZn \cdot \lenAsym \! + \! \lenZq }   \frac{\phicoeff_{  \left\langle l \right\rangle_{\lenAsym}}^2 } { \Glevel(\Wlevel \cdot \phicoeff_{  \left\langle l \right\rangle_{\lenAsym}})  \! + \! 1}\notag \\
&\qquad  \qquad\! + \! \frac{1}{\lenS  } \sum\limits_{l=\lenZn \cdot \lenAsym \! + \! \lenZq \! + \! 1}^{(\lenZn \! + \! 1 )\cdot \lenAsym } \phicoeff_{  \left\langle l \right\rangle_{\lenAsym}}^2
 \! + \! \frac{1}{\lenS  } \sum\limits_{l=(\lenZn \! + \! 1 )\cdot \lenAsym \! + \!1}^{\lenStag \cdot \lenAsym } \phicoeff_{  \left\langle l \right\rangle_{\lenAsym}}^2 \notag \\
 &=  \frac{1}{\lenStag  } \sum\limits_{i=1}^{\lenZn}   \frac{\phicoeff_{  i}^2 } {\Glevel(\Wlevel\cdot \phicoeff_{  i}) \! + \! 1} 
 \! + \! \frac{\lenZq}{\lenS  }   \frac{\phicoeff_{  (\lenZn\! + \!1)}^2 } {\Glevel(\Wlevel\cdot \phicoeff_{  (\lenZn\! + \!1)}) \! + \! 1} 
 \! + \! \frac{\lenAsym - \lenZq}{\lenS  } \phicoeff_{  (\lenZn \! + \! 1 )}^2
 \! + \! \frac{1}{\lenStag  } \sum\limits_{i=\lenZn \! + \! 2 }^{\lenStag } \phicoeff_{  i}^2 \notag \\
 &=\frac{1}{\lenStag  } \sum\limits_{i=1}^{\lenZn}   \frac{\eig{\LmmseMatT, i}^2 } {\Glevel(\Wlevel\cdot \phicoeff_{  i}) \! + \! 1} 
 \! + \!  \frac{1}{\lenStag  } \sum\limits_{i=\lenZn \! + \! 1 }^{\lenStag } \phicoeff_{  i}^2 
 - \frac{\lenZq}{\lenS  } \frac{\phicoeff_{  (\lenZn\! + \!1)}^2 \Glevel(\Wlevel\cdot \phicoeff_{  (\lenZn\! + \!1)}) } {\Glevel_{\lenZn\! + \!1}(\Wlevel\cdot \phicoeff_{  (\lenZn\! + \!1)})  \! + \! 1}.
\label{eqn:ProofMSE3b}
\end{align}
\else
\vspace{-0.2cm}
\begin{align*}
&{\rm MSE}_{\lenAsym}\left(\myA\op \right) 
\!=\!  \frac{1}{\lenS  } \!\sum\limits_{l=1}^{\lenZn \cdot \lenAsym } \!  \frac{\phicoeff_{  \left\langle l \right\rangle_{\lenAsym}}^2 } { \Glevel(\Wlevel\cdot \phicoeff_{  \left\langle l \right\rangle_{\lenAsym}}) \! + \! 1} \! + \! \!\frac{1}{\lenS  } \!\!\sum\limits_{l=(\lenZn \! + \! 1 )\cdot \lenAsym \! + \!1}^{\lenStag \cdot \lenAsym }\!\!\!\! \phicoeff_{  \left\langle l \right\rangle_{\lenAsym}}^2\notag \\
&\qquad \! + \! \frac{1}{\lenS  } \!\sum\limits_{l=\lenZn \cdot \lenAsym  \! + \! 1}^{\lenZn \cdot \lenAsym \! + \! \lenZq }\!   \frac{\phicoeff_{  \left\langle l \right\rangle_{\lenAsym}}^2 } { \Glevel(\Wlevel\cdot \phicoeff_{  \left\langle l \right\rangle_{\lenAsym}})  \! + \! 1} \! + \! \frac{1}{\lenS  } \!\!\sum\limits_{l=\lenZn \cdot \lenAsym \! + \! \lenZq \! + \! 1}^{(\lenZn \! + \! 1 )\cdot \lenAsym }\!\!\phicoeff_{ \left\langle l \right\rangle_{\lenAsym}}^2\notag \\
&=\frac{1}{\lenStag  }\! \sum\limits_{i=1}^{\lenZn}   \frac{\phicoeff_{  i}^2 } {\Glevel(\Wlevel\cdot \phicoeff_{  i}) \! + \! 1} 
\! + \!  \frac{1}{\lenStag  }\! \!\sum\limits_{i=\lenZn \! + \! 1 }^{\lenStag }\! \!\phicoeff_{  i}^2 
\!-\! \frac{\lenZq}{\lenS  } \frac{\phicoeff_{  (\lenZn\! + \!1)}^2 \Glevel(\Wlevel\!\cdot\! \phicoeff_{  (\lenZn\! + \!1)}) } {\Glevel(\Wlevel\!\cdot\!\phicoeff_{  (\lenZn\! + \!1)})  \! + \! 1}.
\vspace{-0.2cm}
\end{align*}
\fi 
Writing $\frac{\lenZq}{\lenS  } = \Ratio \cdot \lenXtag - \lenZn$ yields  an expression which does not depend on $\lenAsym$, and thus holds for $\lenAsym \rightarrow \infty$.
Combining this with \eqref{eqn:ProofMSE24}  {while setting $\Psd{ }(\omega) \equiv 1$} proves \eqref{eqn:OptimalMSE}. 
\qed

\vspace{-0.2cm}
\subsection{Proof of Proposition \ref{thm:SpatialDes}}
\label{app:ProofThmSpaDes}
\vspace{-0.1cm}
To prove the proposition, we first characterize the achievable average \ac{mse} for a fixed $\myAT_l$  
using \cite[Lem. 1]{Shlezinger:18}. Then, as in \cite[Appendix C]{Shlezinger:18},  
we derive the optimal unitary rotation for a given $\myAT_l$, and obtain the analog combining matrix as well as the resulting average \ac{mse}. We characterize the average excess \ac{mse} compared to the average \ac{mmse}, from which the overall average \ac{mse} can be obtained by adding   $\AsymDist_l\MMSE$.

Note that spatial analog combining can be written as a special case of the hardware-limited setup by fixing $\myA  = \myI_{\Tpilots} \otimes \myAT_l$ and $\lenZ = \lenZT \cdot \Tpilots$.  
Under this setting, it can be shown that for a given $\myAT_l$,  the achievable average \ac{mse} for fixed $\Nantennas$ when setting the digital processing $\myBT$ to the linear \ac{mmse} estimator is given by
\begin{align}
\hspace{-0.2cm}
{\rm MSE}_\Nantennas\!\left(\myAT_l\right)  
&\!=	\!  
\frac{1}{\Nusers}{\rm Tr}\!\left( \Phimat_{l}^2\right)   \! -\!
\frac{1}{\Nantennas\Nusers} {\rm Tr}\Bigg(  \!\left( \mySmat^T\!\Dmat_{l,l}^4 \mySmat^*\! \otimes  \myAT_l\CorrMat[l]^2\myAT_l^H\right) 
\notag \\
& \hspace{-0.2cm} \times \bigg( \left(  \CovYtag \otimes \myAT_l\CorrMat[l]\myAT_l^H\right)  \! + \!\frac{{4{\DynRange^2}}}{{3\TilM^2}}\myI_{\lenZT \cdot \Tpilots} \bigg)^{ - 1} \Bigg). 
\label{eqn:ThmProof2a}
\end{align}

Similarly, the optimal digital processing matrix is given by  
\ifsingle
\begin{equation}
\myB_l\op \left( \myAT_l\right)  = \left( \Dmat_{l,l}^2\mySmat^* \otimes \CorrMat[l]\myAT_l^H\right) \!\left( \left( \CovYtag \otimes \myAT_l\CorrMat[l]\myAT_l^H\right)  \! + \!\frac{{4{\DynRange^2}}}{{3\TilM^2}}\myI_{\lenZT \cdot \Tpilots} \right)^{ - 1} .
\label{eqn:ThmProof2b}
\end{equation}
\else
\begin{align}
\myB_l\op \left( \myAT_l\right) & = \left( \Dmat_{l,l}^2\mySmat^* \otimes \CorrMat[l]\myAT_l^H\right) \notag \\
&\times \left( \left( \CovYtag \otimes \myAT_l\CorrMat[l]\myAT_l^H\right)  \! + \!\frac{{4{\DynRange^2}}}{{3\TilM^2}}\myI_{\lenZT \cdot \Tpilots} \right)^{ - 1} .
\label{eqn:ThmProof2b}
\end{align}
\fi 
 
Next, recall that   $\DynRange$ is set to $\myEta$ times the maximal standard deviation of the quantizer input. Thus, by \eqref{eqn:DynRange}, 
\begin{align}
\DynRange^2  
&= \MyKappa \mathop{\max}\limits_{i=1,\ldots,\lenZT \cdot \Tpilots} \E \left\{  \left|\left(  \left( \myI_{\Tpilots} \otimes \myAT_l\right) \myYvec_l\right)_l \right| ^2 \right\} \notag \\
&\stackrel{(a)}{=} \MyKappa\cdot \maxDiag \cdot \mathop{\max}\limits_{i=1,\ldots,\lenZT}   \left( \myAT_l\CorrMat[l]\myAT_l^H\right)_{i,i}  ^2 ,
\label{eqn:DynRange2}
\end{align}
where $(a)$ holds by writing the covariance of $ \myYvec_l$ and as   the maximal diagonal entry of a Kronecker product of positive semi-definite matrices is the product of the maximal diagonal entries \cite[Ch. 7.8]{Meyer:00}. 
Defining $\myAB \triangleq \myAT_l\CorrMat[l]^{1/2}$ and substituting \eqref{eqn:DynRange2} in \eqref{eqn:ThmProof2a} results in
\ifsingle
 \begin{align}
 {\rm MSE}_\Nantennas\left(\myAB\right) =  \frac{1}{\Nusers}{\rm Tr}\!\left( \Phimat_{l}^2\right)   -   \frac{1}{\Nusers\cdot \Nantennas} {\rm Tr} &\Bigg(  \left( \mySmat^T\Dmat_{l,l}^4 \mySmat^* \otimes  \myAB\CorrMat[l]\myAB^H\right)   \!\bigg( \left(  \CovYtag \otimes \myAB\myAB^H\right)\notag \\
 &  \! + \!\frac{{4{\MyKappa\cdot \maxDiag}}}{{3\TilM^2}} \mathop{\max}\limits_{i=1,\ldots,\lenZT}     \left( \myAB\myAB^H\right)_{i,i}  ^2 \myI_{\lenZT \cdot \Tpilots} 
 \bigg)^{ - 1} 
 \Bigg). 
 \label{eqn:ThmProof34}
 \end{align}
 \else
  \begin{align}
 &{\rm MSE}_{\Nantennas}\left(\myAB\right) =  \frac{1}{\Nusers}{\rm Tr}\!\left( \Phimat_{l}^2\right)   \notag \\
 &-   \frac{1}{\Nusers\cdot \Nantennas} {\rm Tr} \Bigg(  \left( \mySmat^T\Dmat_{l,l}^4 \mySmat^* \otimes  \myAB\CorrMat[l]\myAB^H\right)   \!\bigg( \left(  \CovYtag \otimes \myAB\myAB^H\right)\notag \\
 & \qquad \! + \!\frac{{4{\MyKappa\cdot \maxDiag}}}{{3\TilM^2}} \mathop{\max}\limits_{i=1,\ldots,\lenZT}    \left( \myAB\myAB^H\right)_{i,i}  ^2 \myI_{\lenZT \cdot \Tpilots} 
 \bigg)^{ - 1} 
 \Bigg). 
 \label{eqn:ThmProof34}
 \end{align}
 \fi 
 Using \eqref{eqn:ThmProof34}, we can now characterize the optimal unitary rotation for any given $\myAB$, as stated in the following lemma:
 \begin{lemma}
  \label{lem:OptRotation2}
  	For every matrix $\myAB \in \mySet{C}^{\lenZT \times   \Nantennas}$ there exists a unitary matrix $\myMat{U}_{\myAT} \in \mySet{C}^{\lenZT \times \lenZT}$ such that  
\ifsingle  	
  	\begin{align}
  	{\rm MSE}_{\lenAsym}\left( \myAB\right) \ge  {\rm MSE}_{\lenAsym}\left( \myMat{U}_{\myAT}\myAB\right)& =\frac{1}{\Nusers}{\rm Tr}\!\left( \Phimat_{l}^2\right) -   \frac{1}{\Nusers\cdot \Nantennas}{\rm Tr} \Bigg(  \left( \mySmat^T\Dmat_{l,l}^4 \mySmat^* \otimes \myAB\CorrMat[l]\myAB^H\right) 
  	 \notag \\
  	&  \times   \!\bigg( \left( \CovYtag \otimes \myAB\myAB^H\right)  \! + \!\frac{{4{\MyKappa\cdot \maxDiag}}}{{3\TilM^2 \cdot \lenZT}}  {\rm Tr} \left( \myAB\myAB^H\right)  \myI_{\lenZT \cdot \Tpilots} 
  	\bigg)^{ - 1} 
  	\Bigg). 
  	\label{eqn:OptRotation2}
  	\end{align}
\else
  	\begin{align}
 & 	\hspace{-0.4cm}
{\rm MSE}\left( \myAT_l\right) \ge  {\rm MSE}\left( \myMat{U}_{\myAT} \myAT_l\right) =\frac{1}{\Nusers}{\rm Tr}\!\left( \Phimat_{l}^2\right) \notag \\
&  \hspace{-0.4cm}-   \frac{1}{\Nusers\cdot \Nantennas} {\rm Tr} \Bigg(  \left( \mySmat^T\Dmat_{l,l}^4 \mySmat^* \otimes  \myAB\CorrMat[l]\myAB^H\right)   \notag \\
& \hspace{-0.4cm}\times \! \bigg(\! \left( \CovYtag \otimes \myAB\myAB^H\right)  \! + \!\frac{{4{\MyKappa\!\cdot\! \maxDiag}}}{{3\TilM^2 \!\cdot\! \lenZT}}  {\rm Tr} \left( \myAB^H\right) \! \myI_{\lenZT \cdot \Tpilots} \!
\bigg)^{\!  \!- 1} 
\Bigg). 
\label{eqn:OptRotation2}
\end{align}
\fi  	 
  	The unitary matrix $\myMat{U}_{\myAT}$ is a set such that $\myMat{U}_{\myAT}\myAB\myAB^H \myMat{U}_{\myAT}^H$ is weakly majorized by all possible rotations of $\myAB\myAB^H$.
 \end{lemma}
 
 \begin{IEEEproof}
 The lemma is obtained by repeating the arguments in  
 \cite[Lem. C.1]{Shlezinger:18}, 
 thus its proof is omitted for brevity.
\end{IEEEproof}

 We can now characterize the optimal $\myAB$ as the matrix which minimizes \eqref{eqn:OptRotation2}.   
 Note that the right hand side of \eqref{eqn:OptRotation2} is invariant to replacing $\myAB$  with $\alpha \cdot \myMat{U} \myAB$ for any $\alpha > 0$ and for any unitary $\myMat{U}$. Consequently, we can fix $\frac{{4{\MyKappa[\lenZT \cdot \Tpilots]\cdot \maxDiag}}}{{3\TilM[\lenZT \cdot \Tpilots]^2 \cdot \lenZT}}{\rm Tr}\left(  \myAB\myAB^H\right) = 1$,. 
 and thus, minimizing \eqref{eqn:OptRotation2} reduces to solving
\begin{align} 
& \!\!\mathop{\arg\max}\limits_{\myAB}  {\rm Tr} \Bigg(\! \! \left( \mySmat^T\!\Dmat_{l,l}^4 \mySmat^*\! \otimes\!  \myAB\CorrMat[l]\myAB^H\!\right)   \!\bigg(\! \!\left( \! \CovYtag \!\otimes\! \myAB\myAB^H\!\right)  \! +  \!  \myI_{\lenZT  \Tpilots} \!
\bigg)^{\!\! \!- \!1} \!
\Bigg), \notag 
 \\
& {\text{subject to }}\frac{{4{\MyKappa\!\cdot \maxDiag}}}{{3\TilM^2 \cdot \lenZT}}{\rm Tr}\left(  \myAB\myAB^H\right) \!=\! 1. 
\label{eqn:AnalogProblemSpat1}
\end{align} 
By \eqref{eqn:DynRange2}, the support is now 
$\DynRange^2   =   \frac{  \MyKappa\cdot \maxDiag }{\lenZT} {\rm Tr}\left( \myAB\myAB^H \right) =  \frac{3\TilM[\lenZT \cdot \Tpilots]^2}{4}$.
Plugging the resulting $\DynRange$ into \eqref{eqn:ThmProof2b} proves \eqref{eqn:SpatialDesB}.

In order to solve \eqref{eqn:AnalogProblemSpat1}, we define the matrix 
\ifsingle
\begin{align}
\myMat{M} &\triangleq  \left( \CovYtag \otimes \myAB\myAB^H\right)  \! +    \myI_{\lenZT \cdot \Tpilots} 
= \left( \left(\mySmat^T \sum\limits_{m=1}^{\Ncells}\Dmat_{l,m}^2 \mySmat^* + \SigW\myI_{\Tpilots}  \right) \otimes \myAB\myAB^H\right)  \! +    \myI_{\lenZT \cdot \Tpilots} \notag \\
&=\left(  \myI_{\Tpilots} \otimes \left(\myI_{\lenZT} + \SigW\myAB\myAB^H \right)\right)  + \left(\mySmat^T \otimes  \myI_{\lenZT}\right) \left(\sum\limits_{m=1}^{\Ncells}\Dmat_{l,m}^2 \otimes \myAB\myAB^H\right) \left(\mySmat^* \otimes  \myI_{\lenZT}\right).
\label{eqn:Mmat1} 
\end{align}
\else
\begin{align}
\myMat{M} &\triangleq  \left( \CovYtag \otimes \myAB\myAB^H\right)  \! +    \myI_{\lenZT \cdot \Tpilots} 
=\left(  \myI_{\Tpilots} \otimes \left(\myI_{\lenZT} + \SigW\myAB\myAB^H \right)\right) \notag \\
& + \left(\mySmat^T \otimes  \myI_{\lenZT}\right) \left(\sum\limits_{m=1}^{\Ncells}\Dmat_{l,m}^2 \otimes \myAB\myAB^H\right) \left(\mySmat^* \otimes  \myI_{\lenZT}\right).
\label{eqn:Mmat1} 
\end{align}
\fi
Applying the matrix inversion lemma to \eqref{eqn:Mmat1}, recalling that $\mySmat\mySmat^H = \Tpilots \cdot \myI_{\Nusers}$ results in
\ifsingle
\begin{align} 
&{\rm Tr} \left(   \left( \mySmat^T\Dmat_{l,l}^4 \mySmat^* \otimes  \myAB\myAB^H\right)   \! \myMat{M}^{-1}
\right) 
=  {\rm Tr} \bigg(\left( \Tpilots \Dmat_{l,l}^4 \otimes \left( \left(\myI_{\lenZT} \!+\! \SigW\myAB\myAB^H \right)^{-1}  \myAB\CorrMat[l]\myAB^H\right) \right)  \notag \\
&\times \bigg(\Big(\Tpilots\sum_{m=1}^{\Ncells}\Dmat_{l,m}^2  \otimes  \left(\myI_{\lenZT} \!+\! \SigW\myAB\myAB^H \right)^{-1}  \myAB\myAB^H \Big) +  \myI_{\Nusers \cdot \lenZT} \bigg)^{-1}   \bigg).
\label{eqn:Mmat3} 
\vspace{-0.2cm}
\end{align}
\else
\begin{align} 
& \hspace{-0.2cm}
{\rm Tr} \left(   \left( \mySmat^T\Dmat_{l,l}^4 \mySmat^* \otimes  \myAB\myAB^H\right)   \! \myMat{M}^{-1}
\right) \notag \\
&\hspace{-0.2cm}=  {\rm Tr} \Bigg(\left( \Tpilots \Dmat_{l,l}^4 \otimes \left( \left(\myI_{\lenZT} \!+\! \SigW\myAB\myAB^H \right)^{-1}  \myAB\CorrMat[l]\myAB^H\right) \right)  \notag \\
&\hspace{-0.2cm}\times\!\! \bigg(\!\Big(\!\Tpilots\sum_{m\!=\!1}^{\Ncells}\!\Dmat_{l,m}^2 \!\! \otimes\!  \left(\myI_{\lenZT} \!+\! \SigW\myAB\myAB^H\! \right)^{\!-\!1} \!\! \myAB\myAB^H\! \Big) \!+ \! \myI_{\Nusers  \lenZT}\! \bigg)^{\!-\!1} \!  \Bigg).
\label{eqn:Mmat3} 
\vspace{-0.2cm}
\end{align}
\fi
We note that \eqref{eqn:Mmat3} is invariant to replacing $\myAB$ with $\alpha \cdot \myMat{U} \myAB$, we henceforth set $ \myAB = \myMat{\Lambda}\myMat{V}^H$, where $\myMat{\Lambda} \in \mySet{C}^{\lenZT \times \Nantennas}$ is diagonal with diagonal entries arranged in descending magnitude order, and $\myMat{V} \in \mySet{C}^{\Nantennas\times \Nantennas}$ is unitary.  Substituting this in \eqref{eqn:Mmat3} and using the invariance of the trace operator to cyclic permutations results in
\ifsingle
\begin{align} 
&{\rm Tr} \left(   \left( \mySmat^T\Dmat_{l,l}^4 \mySmat^* \otimes  \myAB\myAB^H\right)   \! \myMat{M}^{-1}
\right) 
=  {\rm Tr} \bigg(\left( \Tpilots \Dmat_{l,l}^4 \otimes \left(\myMat{V}^H\CorrMat[l]\myMat{V}\right) \right)  \notag \\
&\times 
\left( \myI_{ \Nusers} \otimes \myMat{\Lambda} \myMat{\Lambda}^H \right)  \left(\left( \Tpilots\sum_{m=1}^{\Ncells}\Dmat_{l,m}^2  \otimes  \myMat{\Lambda}\myMat{\Lambda}^H\right) + \left( \myI_{ \Nusers} \otimes  \left(\myI_{\lenZT} \!+\!  \myMat{\Lambda}\myMat{\Lambda}^H\right) \right)  \right)^{-1} \bigg).
\label{eqn:Mmat3a} 
\vspace{-0.2cm}
\end{align}
\else
\begin{align} 
&\hspace{-0.2cm}{\rm Tr} \left(   \left( \mySmat^T\Dmat_{l,l}^4 \mySmat^* \otimes  \myAB\myAB^H\right)   \! \myMat{M}^{-1}
\right) \notag \\
&\hspace{-0.2cm}=  {\rm Tr} \Bigg(\left( \Tpilots \Dmat_{l,l}^4 \otimes \left(\myMat{V}^H\CorrMat[l]\myMat{V}\right) \right)  
\left( \myI_{ \Nusers} \otimes \myMat{\Lambda} \myMat{\Lambda}^H \right) \notag \\
&\hspace{-0.2cm}\times \! \!\left(\!\left(\! \Tpilots\sum_{m=1}^{\Ncells}\!\Dmat_{l,m}^2 \! \otimes \! \myMat{\Lambda}\myMat{\Lambda}^H\!\right) \!+\! \left( \myI_{ \Nusers}\! \otimes \! \left(\myI_{\lenZT} \!+\!  \myMat{\Lambda}\myMat{\Lambda}^H\!\right) \right)  \!\right)^{\!-\!1}\! \Bigg).
\label{eqn:Mmat3a} 
\vspace{-0.2cm}
\end{align}
\fi
Note that the matrix $\left( \myI_{ \Nusers} \otimes \myMat{\Lambda} \myMat{\Lambda}^H \right)  \Big(\Big( \Tpilots\sum\limits_{m=1}^{\Ncells}\Dmat_{l,m}^2  \otimes  \myMat{\Lambda}\myMat{\Lambda}^H\Big) + \Big( \myI_{ \Nusers} \otimes  \left(\myI_{\lenZT} \!+\!  \myMat{\Lambda}\myMat{\Lambda}^H\right) \Big)  \Big)^{-1} $ is diagonal with non-negative diagonal entries arranged in descending order. Therefore, it follows from \cite[Thm. II.1]{Lassare:95} that \eqref{eqn:Mmat3a} is maximized by setting $\myMat{V}$ to be the eigenmatrix of $\CorrMat[l]$. Thus, by letting  $a_i$ be the diagonal entries of $\myMat{\Lambda}$, the objective \eqref{eqn:Mmat3a} can be written as
\ifsingle
\begin{align}
\vspace{-0.2cm}
{\rm Tr} \left(   \left( \mySmat^T\Dmat_{l,l}^4 \mySmat^* \otimes   \myAB\myAB^H \right)   \! \myMat{M}^{-1}
\right) 
&= \sum\limits_{u=1}^{\Nusers}\sum\limits_{i=1}^{\lenZT} \frac{\Tpilots \cdot\dcoeff_{l,l,u}^4 \cdot a_i^2 \cdot \eigT{l,i}}{1 +\left( \SigW + \Tpilots \sum\limits_{u=1}^{\Ncells}\dcoeff_{l,m,u}^2 \right)  a_i^2 } \notag \\
&\stackrel{(a)}{=} \sum\limits_{u=1}^{\Nusers}\sum\limits_{i=1}^{\lenZT} \frac{\Tpilots \cdot\phicoeff_{l,u}^4 \cdot a_i^2\cdot \eigT{l,i}}{ \Tpilots \cdot\phicoeff_{l,u}^2  \cdot a_i^2 + \bcoeff_{l,u}^2 }, 
\label{eqn:Mmat4} 
\vspace{-0.2cm}
\end{align}
\else
\vspace{-0.2cm}
\begin{align}
\vspace{-0.2cm}
&{\rm Tr} \left(   \left( \mySmat^T\Dmat_{l,l}^4 \mySmat^* \otimes   \myAB\myAB^H \right)   \! \myMat{M}^{-1}
\right) \notag \\
&\qquad= \sum\limits_{u=1}^{\Nusers}\sum\limits_{i=1}^{\lenZT} \frac{\Tpilots \cdot\dcoeff_{l,l,u}^4 \cdot a_i^2 \cdot \eigT{l,i}}{1 +\left( \SigW + \Tpilots \sum\limits_{u=1}^{\Ncells}\dcoeff_{l,m,u}^2 \right)  a_i^2 } \notag \\ 
&\qquad\stackrel{(a)}{=} \sum\limits_{u=1}^{\Nusers}\sum\limits_{i=1}^{\lenZT} \frac{\Tpilots \cdot\phicoeff_{l,u}^4 \cdot a_i^2\cdot \eigT{l,i}}{ \Tpilots \cdot\phicoeff_{l,u}^2  \cdot a_i^2 + \bcoeff_{l,u}^2 }, 
\label{eqn:Mmat4} 
\vspace{-0.2cm}
\end{align}
\fi 
where $(a)$ follows from the definition of $\bcoeff_{l,u}$ in \eqref{eqn:BmatDef}, and  since $\phicoeff_{l,u}^2 = \bcoeff_{l,u}\dcoeff_{l,l,u}^2$.
By combining \eqref{eqn:Mmat4} and \eqref{eqn:AnalogProblemSpat1} it holds that  the analog combining matrix which minimizes the average \ac{mse} is given by $\myMat{U}_{\myAT} \myMat{\Lambda}_{\myAT}\myMat{V}_{\myAT}^H$, where $\myMat{U}_{\myAT}$ is given in Lemma \ref{lem:OptRotation2}, $\myMat{V}_{\myAT}^H$ is the eigenmatrix of $\CorrMat[l]$, and $\myMat{\Lambda}_{\myAT} $ is diagonal with diagonal entries  $\{\bar{a}_i\}$, which are the solution to
\vspace{-0.2cm}
\begin{align}
& \{\bar{a}_i\}_{i=1}^{\lenZT } = \mathop{\arg\max}\limits_{\{a_i\}_{i=1}^{\lenZT }} \sum\limits_{i=1}^{\lenZT} \sum\limits_{u=1}^{\Nusers} \frac{\Tpilots \cdot\phicoeff_{l,u}^4 \cdot a_i^2 \cdot \eigT{l,i}}{ \Tpilots \cdot\phicoeff_{l,u}^2  \cdot a_i^2 + \bcoeff_{l,u}^2 }\notag
 \\
& {\text{subject to }}\frac{{4{\MyKappa\cdot \maxDiag}}}{{3\TilM^2 \cdot \lenZT}} \sum\limits_{i=1}^{\lenZT} a_i^2= 1. 
\label{eqn:AnalogProblemSpat2a}
\vspace{-0.2cm}
\end{align} 
The concavity of the objective in \eqref{eqn:AnalogProblemSpat2a} stems from the concavity of the mapping $x \mapsto  \frac{\Tpilots \cdot\phicoeff_{l,u}^4\cdot\eigT{l,i} \cdot x}{ \Tpilots \cdot\phicoeff_{l,u}^2  \cdot x + \bcoeff_{l,u}^2 }$ over $\mySet{R}^+$.  

Combining \eqref{eqn:ThmProof34} and \eqref{eqn:AnalogProblemSpat2a}, noting that $\Nantennas \rightarrow \infty$ implies that $\lenZT \rightarrow \infty$, proves \eqref{eqn:SpatialMSE2}, thus concluding the proof.
\qed

\end{appendix}  


\end{document}